\documentclass[a4paper,11pt]{article}
\pdfoutput=1 
\usepackage{jheppub}  
\usepackage{graphicx,color}
\usepackage{amsmath}
\usepackage{autobreak}
\allowdisplaybreaks
\usepackage{floatrow}
\usepackage[latin1]{inputenc}
\DeclareFloatFont{tiny}{\tiny}% "scriptsize" is defined by floatrow, "tiny" not
\newcommand{\ben}{\begin{enumerate}}
\newcommand{\een}{\end{enumerate}}
\newcommand{\beq}{\begin{equation}}
\newcommand{\eeq}{\end{equation}}
\newcommand{\bal}{\begin{align}}
\newcommand{\eal}{\end{align}}

\newcommand{\bea}{\begin{eqnarray}}
\newcommand{\eea}{\end{eqnarray}}
\newcommand{\nn}{\nonumber}
\allowdisplaybreaks

\usepackage{hyperref}

\usepackage{array}
\usepackage{ulem}
\newcolumntype{P}[1]{>{\centering\arraybackslash}p{#1}}
\allowdisplaybreaks
\RequirePackage{ifpdf}
\usepackage{amsmath}
\usepackage{mathtools}

\usepackage{autobreak}

\usepackage[final]{pdfpages}
\usepackage{ifpdf}
\usepackage{slashed}

\usepackage{hyperref}

\usepackage{color}
\usepackage{graphics}

\usepackage{etoolbox}
\usepackage{fixmath}

\usepackage{caption}
\usepackage{subcaption}
\usepackage{amsfonts}
\usepackage{multirow}
\usepackage{epstopdf}

\usepackage{relsize}

\usepackage{float}

\usepackage[table]{xcolor}
\usepackage{tabularx}

\allowdisplaybreaks

\newcommand*\oline[1]{%
   \vbox{%
     \hrule height 0.5pt%                  % Line above with certain width
     \kern0.25ex%                          % Distance between line and content
     \hbox{%
       \kern-0.2em%                        % Distance between content and left side of box, negative values for lines shorter than content
       \ifmmode#1\else\ensuremath{#1}\fi%  % The content, typeset in dependence of mode
       \kern-0.05em%                        % Distance between content and left side of box, negative values for lines shorter than content
     }% end of hbox
   }% end of vbox
}

\newcommand{\Nb}{\overline{N}}

\newcommand{\as}{a_s}

\def\Dm1{{{\delta(1-z)}}}

\def\g0#1DY{{g_{0#1}^{DY}}}

\def\LogmW1{{{\ln (1-\omega)}}}

\title{Resummed Higgs boson cross section at next-to SV to NNLO+$ \rm \overline {\textbf{NNLL}}$}
\author{A. H. Ajjath,}
\author{Pooja Mukherjee,}
\author{V. Ravindran,}
\author{Aparna Sankar,}
\author{Surabhi Tiwari,}

\affiliation{The Institute of Mathematical Sciences, HBNI, IV Cross Road, Taramani, Chennai 600113, India}
\emailAdd{ajjathah@imsc.res.in}
\emailAdd{poojamukherjee@imsc.res.in}
\emailAdd{ravindra@imsc.res.in}
\emailAdd{aparnas@imsc.res.in}
\emailAdd{surabhit@imsc.res.in}
\abstract{We present the resummed predictions for inclusive cross section for the production of Higgs boson 
at next-to-next-to leading logarithmic ($\rm \overline {NNLL}$) accuracy taking into account both soft-virtual ($\rm SV$) and next-to SV ($\rm NSV$) threshold logarithms. We derive the $N$-dependent coefficients  and the  $N$-independent constants in Mellin-$N$ space for our study. Using the minimal prescription we perform the inverse Mellin transformation and match it with the corresponding fixed order results. We report in detail the numerical impact of $N$-independent part of resummed result and explore the ambiguity involved in exponentiating them. By studying the K factors at different logarithmic accuracy, 
we find that  the perturbative expansion shows better convergence improving the reliability of the prediction at $\rm NNLO + \overline{NNLL}$ accuracy. For instance, the cross-section at $\rm NNLO + \overline{NNLL}$ accuracy reduces by $3.15\%$ as compared to the $\rm NNLO$ result for the central scale $\mu_R = \mu_F = m_H/2$ at 13 TeV LHC. We also observe that the resummed $\rm SV + NSV$ result improves the renormalisation scale uncertainty at every order in perturbation theory. The uncertainty from the renormalisation scale $\mu_R$   ranges between $(+8.85\% ,-10.12\%)$ at $\rm NNLO$ whereas it goes down to $(+6.54\% , - 8.32\%)$ at $\rm NNLO + \overline{NNLL}$ accuracy. However, the factorisation scale uncertainty is worsened by the inclusion of these NSV logarithms  hinting the importance of resummation beyond $\rm NSV$ terms. We also present our predictions for $\rm SV + NSV$ resummed result at different collider energies.} 
\begin{document} 
%%%Preprint
\preprint{IMSc/2021/09/06}
\maketitle

%%%Introduction
\section{Introduction} \label{intro}

%%general intro
% {\color{red}The remarkable discovery of the Higgs boson at the Large Hadron Collider \cite{Aad:2012tfa,Chatrchyan:2012xdj}
 %is a mile stone in the particle physics.  This
% puts the Standard Model (SM) in a firm position to describe the dynamics of all known elementary particles.
% %of course[D[D[D[D[D[D[D[O[C[C[C[C[C[C[C[C there are several short comings in the SM which lead physicists to explore physics beyond the SM.
% There have been tremendous efforts in constructing models that address the short comings of SM and at the same time, search for new physics signatures. These studies
% demonstrate rich phenomenology that can be explored at present and future colliders.}  In addition, model independent
% studies using sophisticated effective field theory approach have also improved our understanding.  
% To improve the search strategies of  new physics signatures at the LHC, 
% predictions from SM as well
% as from BSM have to be precise and less sensitive to theoretical 
% uncertainties in order to arrive at reliable conclusions.   
% The combined efforts from the theorists and experimentalists in the precision studies 
% within the SM at the LHC have set stringent constraints on the parameters
% present in various beyond the SM (BSM) scenarios \cite{Englert:2014uua}.

{\color{black}The milestone discovery of the Higgs boson at the Large Hadron Collider (LHC) \cite{Aad:2012tfa,Chatrchyan:2012xdj}
puts the Standard Model (SM) on a firm position to describe the dynamics of all known elementary particles.
There have been tremendous efforts in constructing models that address the shortcomings of SM and, at the same time, search for new physics signatures. These studies
demonstrate rich phenomenology that can be explored at present and future colliders.  In addition, model-independent
studies using the sophisticated effective field theory approach have also aided to a better understanding of the underlying theory.  
To improve the search strategies of  new physics signatures at the LHC and thereby to arrive at sensible conclusions, we require
precise predictions from SM as well as beyond SM (BSM) scenarios with minimum theoretical uncertainties.} Such studies within the SM  by both theorists and experimentalists have set stringent constraints on the parameters
present in various BSM scenarios \cite{Englert:2014uua}. 
%{\bf \color{blue} The last sentence looks out of context? More to add to connect with next para?}

At the LHC, Higgs bosons are produced dominantly in the gluon fusion channel, and hence most of the
theoretical studies centers around this.   In addition, the radiative corrections from perturbative Quantum Chromodynamics (QCD) beyond the leading order (LO) were found to be sizeable. Hence efforts to include higher order effects became inevitable.
{\color{black} As the accuracy in Higgs boson measurements improved, the level of precision in the predictions was also increased by including contributions from other partonic channels, the radiative effects from the electroweak (EW) sector of the SM, and other tiny effects resulting from finite quark masses.}   

In hadronic collisions, the Higgs bosons are dominantly produced in the gluon fusion subprocess through a top quark loop.  
Beyond next to LO (NLO), QCD corrections were obtained in the Higgs effective field theory (HEFT) where top quark degrees of freedom are
integrated out \cite{Dawson:1990zj,Spira:1995rr,Chetyrkin:1997un}
and this approach turned out to be the most advantageous technically.       
By multiplying higher order corrections computed in HEFT with the exact leading order cross section, one can retrieve most of the 
top-quark mass effects.  
See 
\cite{Georgi:1977gs,Graudenz:1992pv,Djouadi:1991tka,Spira:1995rr,Catani:2001ic,Harlander:2001is,Anastasiou:2002yz,Harlander:2002wh,Catani:2003zt,Ravindran:2003um,Moch:2005ky,Ravindran:2006cg,deFlorian:2012za,Bonvini:2014jma,deFlorian:2014vta,Anastasiou:2014vaa,Li:2014afw,Anastasiou:2015yha,Anastasiou:2015vya,Das:2020adl} for a complete list of results for the Higgs boson production in gluon fusion 
both in exact and in HEFT up to NNLO level and also for threshold effects beyond NNLO in QCD.  
The first step towards complete  N$^3$LO  QCD corrections was obtained by expanding the partonic contributions near the threshold \cite{Anastasiou:2015vya}.
The electroweak corrections to this channel were computed in \cite{Anastasiou:2008tj,Actis:2008ug,Bonetti:2017ovy, Anastasiou:2018adr,Bonetti:2018ukf, Becchetti:2020wof}.
A complete study taking into account all these effects was done in \cite{Anastasiou:2016cez,deFlorian:2016spz,Harlander:2016hcx}.
In \cite{Mistlberger:2018etf,Duhr:2019kwi} the full $z$-dependence for all the partonic channels was recovered at N$^3$LO.  To estimate the error resulting from finite top quark mass in HEFT,  the top quark mass effects were also systematically studied.  In particular, they were included at  
large partonic center of mass energy as well as from  $1/m_t$ terms 
\cite{Marzani:2008az,Harlander:2009my,Harlander:2009mq,Pak:2009dg} and they amount to $1\%$ correction.
Recently, in \cite{Czakon:2021yub}, the exact top quark effects were included at NNLO for all the channels and this
results in a decrease in cross section compared to HEFT to $-0.26\%$ at $13$ TeV and $-0.1\%$ at $8$ TeV.

{\color{black}While these fixed order predictions often provide reliable estimates of the radiative corrections, there are certain kinematical
regions in phase space, where these estimates can go wrong due to the presence of very large logarithms at every order in perturbative series.}  In the inclusive reactions, this happens in the threshold regions where most of the partonic center of mass energy goes in producing the Higgs boson and at the
same time the Parton distribution functions (PDFs) that are driving these partonic reactions dominates.  These large logarithms can be systematically resummed using standard threshold resummation \cite{Sterman:1986aj,Catani:1989ne}. 
For the Higgs boson production, in \cite{Bonvini:2016frm} the results up to N$^3$LL accuracy 
\cite{Bonvini:2014joa,Catani:2014uta,Bonvini:2014tea} 
were used to predict matched results at  N$^3$LO+N$^3$LL for $\mu_R=\mu_F=m_H$ which resulted in $4\%$  error from the
missing higher effects at $13$ TeV LHC.

The successive terms in the threshold expansion of various partonic channels are found to be numerically important   
and in many cases, or they are even larger than the leading contributions.  In particular, the subleading threshold logarithms, 
called the next-to soft virtual (NSV) terms in the inclusive production of Higgs boson dominate over the leading ones as pointed out in \cite{Anastasiou:2014lda}. 
There have already been several studies \cite{Laenen:2008ux,Grunberg:2009yi,Moch:2009hr,Laenen:2010kp,Laenen:2010uz,Bonocore:2014wua,deFlorian:2014vta,Bonocore:2015esa,Beneke:2019oqx,Beneke:2019mua,Bonocore:2016awd,DelDuca:2017twk,Das:2020adl} 
to understand the structure of NSV terms.
Recently, in \cite{Ajjath:2020ulr}, we set up a formalism that can resum these NSV logarithms in Mellin space 
and in \cite{Ajjath:2021lvg}, we studied their numerical significance for the production of a pair of leptons in Drell-Yan process.
In this article, we extend the study to the production of Higgs boson in gluon fusion at the LHC.  

The plan of the paper is as follows.  Sec.\ref{theory} is devoted to provide a theoretical overview of our formalism for completeness.  Here,
we first describe the importance of threshold expansion as well as the soft gluon exponentiation and then review the formalism given in \cite{Ajjath:2020ulr} to include NSV terms for the production of Higgs boson at the LHC.  We also discuss different resummation prescriptions to explore the ambiguity involved in exponentiating the N-independent constants. In Sec. \ref{pheno}, we study the numerical impact of our NSV resummed result in detail and present our findings.  Finally we conclude in Sec.\ref{concl}.

\section{Theoretical overview} \label{theory}
In QCD improved parton model, the inclusive cross section for producing Higgs boson
in the hadronic collisions can be written as 
a convolution of perturbatively calculable coefficient
functions (CFs), $\Delta_{ab}$,  and non-perturbative
flux $\tilde \Phi_{ab}$:
\begin{eqnarray}
\sigma^H(m_H^2,\tau) = \tau ~ {\pi G_B^2(\mu_R^2) \over 8 (N^2-1)} \sum_{a,b=g,q,\overline q}\int_\tau^1 {d z \over z} \tilde \Phi_{ab}\left({\tau \over z},\mu_F^2\right) 
\Delta_{ab}(z,m_H^2,\mu_R^2,\mu_F^2)
\end{eqnarray}
where $N=3$ in QCD and $\tau=m_H^2/S$ with $m_H$ being the Higgs boson mass and $S$ the square of hadronic center of mass energy. \textcolor{black}{The partonic scaling variable  $z=\frac{m_H^2}{\hat s}$ is given in terms of the partonic center of mass energy, $\sqrt{\hat s}$.
%As we work in the HEFT framework, the CFs computed in HEFT are multiplied with the  factor $G_B$ which is the product of born cross section computed in the full theory with finite quark masses and the square of the Wilson coefficient in the HEFT. 
We work in the HEFT framework, where the CFs are multiplied with the factor $G_B$ which is the product of born cross section computed in full theory with finite quark masses and the square of Wilson coefficient
in HEFT. }
The factor $G_B$ explicitly depends on both $m_H$ and $m_t$ and is renormalised at the renormalisation scale $\mu_R$ (see \cite{Ravindran:2003um}
for the details).  
The flux, renormalised at the factorisation scale $\mu_F$, is the convolution of parton distribution functions $f_a,f_b$ of incoming
partons $a,b$ respectively and is given by
\begin{eqnarray}
\tilde \Phi_{ab}\left({\tau \over z},\mu_F^2\right) = \int_{\tau \over z}^1 {d y \over y } f_a(y,\mu_F^2) f_b\left({\tau \over zy},
\mu_F^2\right)\,.
\end{eqnarray}
The CFs in HEFT are perturbatively calculable in powers of $a_s = g_s^2/16 \pi^2$ where $g_s$ is the QCD strong coupling constant:
\begin{eqnarray}
\Delta_{ab}(\mu_R^2,\mu_F^2,z) = \sum_{i=0}^\infty a_s^i(\mu_R^2) \Delta_{ab}^{(i)}(\mu_R^2,\mu_F^2,z)\,,
\end{eqnarray}
with the normalisation $\Delta_{ab}^{(0)} = \delta(1-z)$.

Beyond leading order, the general structure of CFs can be decomposed into,   
\textcolor{black}{\begin{eqnarray}
\Delta_{ab}(\mu_R^2,\mu_F^2,z) =  \delta_{b \bar a} \Delta^{SV}_{a \bar{a}}(\mu_R^2,\mu_F^2,z) + \Delta_{ab}^{reg}(\mu_R^2,\mu_F^2,z)\,,
\end{eqnarray}}
where $\Delta_{a\overline a}^{SV}$ denotes the soft-virtual (SV)
corrections {\color{black} while $\Delta_{ab}^{reg}$ constitutes to the regular part of CFs.} Here, the SV corrections resulting from pure virtual contributions in the born process $g + g \rightarrow H$ and
also the leading contributions from gluon gluon initiated  partonic channels with at least one emission of on-shell parton.
While the pure virtual contributions are proportional to only $\delta(1-z)$, the latter ones contain additionally the distributions of the form %to $\delta(1-z)$ terms the so called $+$ distributions, denoted by 
${\cal D}_k(z)$:
\begin{eqnarray}
{\cal D}_k(z) = \left( {\ln^k(1-z) \over 1-z} \right)_+ \,.
\end{eqnarray}
These distributions satisfy 
{\color{black}
\begin{eqnarray}
\int_0^1 dz ~f(z)~\left( {\ln^k(1-z) \over 1-z} \right)_+  = \int_0^1 dz ~\big(f(z)-f(1)\big) \left( {\ln^k(1-z) \over 1-z} \right)
\end{eqnarray}}
for any regular function $f(z)$.  
Further decomposing the SV CFs \textcolor{black}{order by order}, we get 
\textcolor{black}{\begin{align}
\Delta^{SV}_{a\bar a}(z) &=\sum_{i=0}^{\infty}a_s^i~\Delta^{SV,(i)}_{a\bar a}(z)\nonumber\\
&= \sum_{i=0}^{\infty}a_s^i~\left( \Delta^{(i)}_{\delta} \delta(1-z)+ \sum_{k=0}^{2i-1}  \Delta^{(i)}_{{\cal D}_k} {\cal D}_k(z) \right)\, ,
\label{delsvexp}
\end{align}
with $\Delta^{(i)}_{\delta}$ refers to the coefficients proportional to $\delta(1-z)$ and $\Delta^{(i)}_{{\cal D}_k}$ are those corresponds to ${{\cal D}_k}$}.

The regular part of CFs, denoted by $\Delta_{ab}^{reg}$, contain 
functions \textcolor{black}{of the form $(1-z)^{m}\ln^k(1-z),m,k=0,1,\cdots \infty$}. It can be seen that, the computation of the SV part of the CFs are relatively easy compared to the regular part.
For instance, the Feynman integrals that contribute to the
pure virtual corrections depend only on the hard scale $m_H$ 
\textcolor{black}{and those corresponds to the remaining soft part of SV contributions
%the corresponding soft part contains
proportional to the soft scale $m_H (1-z)$. 
However, the regular part depends on more than one scale through various powers of 
$m_H (1-z)^\alpha$, with $ \alpha \ge 1$ leading to their rich analytical structure.}
Already at NNLO level, the Feynman integrals in the perturbative expansions contain rich analytical
structure through the appearance of iterated integrals \cite{goncharov:1998ab}, which are often called generalised polylogarithms. 
Further the N$^3$LO \cite{Mistlberger:2018etf,Duhr:2019kwi} computation demonstrates the appearance of a new class of functions called elliptic functions (see for example \cite{Bloch:2013tra}) in the iterated integrals for the first time. 
%{\bf \color{blue} Need a connection to next section}

\subsection{Threshold expansion and soft gluon exponentiation} 
Owing to the complexity involved in computing CFs exactly in the variable $z$ {\color{black} at higher orders}, one resorts to the method of expansion near threshold region, namely around $z=1$.  Note that the scaling variable $z$ quantifies the
fraction of partonic center of mass energy that goes into producing on-shell Higgs boson in hadronic collisions.  
In the threshold region, the partons that accompany the Higgs boson are mostly soft and the Feynman integrals from
the real emission subprocesses in this region can be evaluated as a power series in $1-z$.  
These expansions show remarkably simple structure in terms of logarithms of $1-z$:
\textcolor{black}{\begin{align}
\Delta^{reg}_{ab}(z) &=\sum_{i=0}^{\infty}a_s^i~\Delta^{reg,(i)}_{ab}(z)\nonumber\\
&=\sum_{i=0}^{\infty}\sum_{k=0}^{2i-1} \sum_{l=0}^\infty  a_s^i~\Delta^{reg,(i)}_{ a b,l,k} ~(1-z)^l \ln^k(1-z) .
\label{delregexp}
\end{align}}
The above threshold expansion of CFs in partonic scaling variable $z$ will be useful provided
the partonic flux $\tilde \Phi_{ab}(\tau/z)$ that multiplies them to give hadronic cross sections also dominates in the
same region for a given hadronic scaling variable $\tau$.  It is interesting to note that the first NNLO result \cite{Harlander:2002wh}
for the inclusive Higgs boson production was achieved through the expansion of CFs around $z=1$. 
Similarly, at N$^3$LO level, in \cite{Anastasiou:2014lda}, next-to soft contributions from all the partonic channels were     
obtained for the Higgs production.  Later on, in \cite{Anastasiou:2015vya}, beyond next-to soft terms at N$^3$LO were 
obtained using threshold expansion.  In \cite{Mistlberger:2018etf}, the analytical results for all the channels were
reported for the first time and numerically they were found to be consistent with those obtained using threshold expansion
\cite{Anastasiou:2015vya}.
While the state-of-the-art predictions at N$^3$LO level is good enough for 
most of the phenomenological analysis, the results in perturbation theory
demonstrates rich analytical structure \textcolor{black}{which can be exploited to study their all-order behaviour.}
One such study was pioneered for the inclusive processes by Sterman \cite{Sterman:1986aj} and Catani and Trentedue \cite{Catani:1989ne}
who showed that the leading threshold contributions can be systematically resummed to all orders in a process
independent way.  This goes by the name of threshold resummation and the resummed results are organised 
in terms of certain exponents that are controlled by universal anomalous dimensions.    
It is convenient to perform resummation in Mellin $N$ space and the leading threshold logarithms in $z$ space translate to
large $\ln N$ terms in $N$ space.  These large $\ln N$s accompanied by small $a_s$ at every order
give order one terms, $\omega = 2 a_s(\mu_R^2) \beta_0 \ln N$, spoiling the reliability of the perturbative results.  The resummation of these order
one terms provide sensible expansion.  The successive terms in the expansion are arranged as leading logarithm (LL), next-to LL
(NLL) and so on.  For the Higgs boson production, the results up to N$^3$LL level is available
\cite{Bonvini:2014joa,Catani:2014uta,Bonvini:2014tea,Schmidt:2015cea} and in 
\cite{Bonvini:2016frm}, analysis at N$^3$LO+N$^3$LL at $\mu_R=\mu_F=m_H$ provided the estimate of $4\%$  error from the 
missing higher order effects at $13$ TeV LHC. 
In addition to standard QCD approach, 
the threshold effects in the large order perturbation theory can be studied using
soft collinear effective theory (SCET) which captures
the  soft and collinear modes at the Lagrangian level.  Using SCET \cite{Bauer:2000yr,Bauer:2001yt,Bauer:2002nz} one can conveniently 
compute the SV results order by order in perturbation theory.  The intrinsic scales that separates various
modes in theory can be used
to set up renormalisation group equations \cite{Ahrens:2008nc} whose solutions sum up large logarithms from threshold regions to all orders.
The numerical impact of these results provide reliable estimates of missing higher order effects and they are found to be very close
to standard QCD ones. 

%\subsection{Going beyond soft gluon exponentiation} 
The logarithms that are present in the threshold expansion of CFs, such as  
plus distributions ${\cal D}_k$ and the subleading logarithms from $\ln^k(1-z)$ show rich 
universal structure.
Their coefficients are mostly controlled by certain IR anomalous dimensions which are process independent.  
For example, the $+$ distributions are controlled by cusp and soft anomalous dimensions while   
the NSV logarithms by collinear anomalous dimensions.
The perturbative structure can be understood through IR evolution equations 
whose kernels depend on these IR anomalous dimensions.  
Thanks to several higher order results that are available,  there is a better understanding 
of IR structure of multi-loop amplitudes 
beyond two loops \cite{Catani:1998bh,Becher:2009cu,Becher:2009qa,
Gardi:2009qi,Catani:1998bh} (see \cite{Ajjath:2019vmf,H:2019nsw} for a QFT with mixed gauge groups).
Thanks to these results, SV results are available for several important observables.
For complete list, see \cite{Georgi:1977gs,Graudenz:1992pv,Djouadi:1991tka,Spira:1995rr,Catani:2001ic,Harlander:2001is,Anastasiou:2002yz,Harlander:2002wh,Catani:2003zt,Ravindran:2003um,Moch:2005ky,Ravindran:2006cg,deFlorian:2012za,Bonvini:2014jma,deFlorian:2014vta,Anastasiou:2014vaa,Idilbi:2005ni,Li:2014afw,Anastasiou:2015yha,Anastasiou:2015vya,Das:2020adl} for Higgs production in gluon fusion
and \cite{Altarelli:1978id,Altarelli:1979ub,Matsuura:1987wt,Matsuura:1988nd,Matsuura:1988sm,Matsuura:1990ba,Hamberg:1990np,vanNeerven:1991gh,Harlander:2002wh,Moch:2005ky,Ravindran:2006cg,deFlorian:2012za,Ahmed:2014cla,Ahmed:2015qda,Catani:2014uta,Li:2014afw,Duhr:2020seh} for Drell-Yan production.

\subsection{Going beyond soft gluon exponentiation}
Inclusive results that are available to third order for DY and Higgs productions suggest that
the subleading logarithms present in the regular part of CFs, $\Delta^{reg}_{ab}$ from various 
partonic channels also play important role in the precise predictions.  
In particular, the leading logarithmic terms in $\Delta^{reg}_{ab}(z)$ near $z=1$, namely
\begin{eqnarray}\label{NSV}
\Delta_{ab}^{NSV}(z) = \sum_{i=0}^\infty a_s^i(\mu_R^2) \Delta_{ab}^{NSV,(i)}(z)
,\, 
\end{eqnarray}
where $\Delta_{ab}^{NSV,(i)}(z)$ is defined by setting $l=0$ in \eqref{delregexp},i.e.,
\begin{eqnarray}
\label{DeltaR}
\Delta_{ab}^{NSV,(i)}(z) = \sum_{k=0}^{2 i-1} \Delta_{ab,0,k}^{reg,(i)} \ln^k(1-z) \,,
\end{eqnarray}
were shown  to spoil the approximations based on threshold expansions as the order of perturbation
increases \cite{Anastasiou:2014lda}.  One finds that these NSV contributions are larger than SV ones at the LHC energies.
Hence, these corrections as well as those from ${\cal O}(1-z)$ can not be ignored \cite{Anastasiou:2015vya}. 
Theoretically, NSV terms also show rich perturbative structure like the SV ones.
The NSV terms in  
the inclusive processes have been a topic of interest for more than a decade, see 
\cite{Laenen:2008ux,Laenen:2010kp,Laenen:2010uz,Bonocore:2014wua,Bonocore:2015esa,Beneke:2019oqx,Beneke:2019mua,Bonocore:2016awd,DelDuca:2017twk}.
In \cite{Grunberg:2009yi}, using physical evolution equations for the CFs of deep inelastic scattering and exploiting renormalisation group invariance,
the all order structure of the logarithms were studied.  In particular, by suitably modifying the kernels of these equations,
certain leading logarithms in NSV terms were correctly reproduced. 
In \cite{Moch:2009hr} (and \cite{deFlorian:2014vta,Das:2020adl}), a remarkable development was made by Moch and Vogt
by studying the logarithmic structure for the  physical evolution kernels (PEK).  
They  observed that there is an enhancement of single logarithms near threshold up to third
order obtained from the fixed order results of DIS, 
semi-inclusive $e^+ e^-$ annihilation and Drell-Yan production of
a pair of leptons in hadron collisions.
Based on this finding, they conjectured that it will hold true to all orders in $1-z$,
consequently they could predict certain NSV logarithms to all orders in $a_s$ like the way the SV terms can be predicted. Very recently, in \cite{Abele:2021nyo} the resummation of NSV terms has been addressed in the context of semi-inclusive DIS(SIDIS) process.
The NSV corrections to various inclusive processes were also studied 
in a series of papers \cite{Laenen:2008ux,Laenen:2010kp,Laenen:2010uz,
Bonocore:2014wua,Bonocore:2015esa,Beneke:2019oqx,Beneke:2019mua} and
lot of progress have been made leading to a better understanding of the underlying physics.
Recently, in \cite{Ajjath:2020ulr}, we investigated  the structure of NSV terms present in the 
quark anti quark initiated channels in the inclusive production of a pair of leptons in Drell-Yan process and 
gluon/bottom anti bottom initiated ones for production of Higgs boson exploiting 
the factorisation properties and renormalisation group invariance along with the certain universal structure
of real and virtual contributions obtained through Sudakov K+G equation. We also studied the all-order perturbative structure of the NSV logarithms in the CFs of deep inelastic scattering (DIS) and semi-inclusive e$^+$e$^-$ annihilation
(SIA) processes in \cite{Ajjath:2020sjk}. The formalism was later extended for studying the all-order behaviour of NSV terms in rapidity distributions of  Drell-Yan  and Higgs production through gluon fusion and bottom quark annihilation in \cite{Ajjath:2020lwb}. 

%%%%

% We restricted ourselves to diagonal channels namely quark anti quark initiated partonic channels for DY and
% gluon or bottom anti bottom initiated ones for Higgs boson production.  
The goal of \cite{Ajjath:2020ulr} was to set up a theoretical formalism to systematically include the NSV contributions 
in the diagonal channels of inclusive cross sections in general.
We used collinear factorisation and studied the structure of various building blocks that
constitute the CFs near threshold.  The building blocks include the form factors from $2 \rightarrow 1$ processes, 
the real emission contributions and the Altarelli-Parisi kernels.
For the diagonal channels, for SV and NSV terms,
the collinear factorisation of bare partonic channels remarkably simplifies.  For example, 
{\color{black}one observes that  only diagonal part of collinear singular Altarelli-Parisi kernels containing only diagonal splitting 
functions contribute}.   We then use a set of 
differential equations that govern these building blocks.  For instance, the form factors and real emission
contributions satisfy K+G equation while AP kernels satisfy AP evolution equations.  In addition,
each of these terms is renormalisation group invariant.  Note that
these differential equations are governed by set of universal anomalous dimensions and a few process dependent
constants.  
The solutions to these differential equations can be used to obtain an all order result
that encapsulates both SV and NSV terms.   The all order result can be displayed in a compact form through 
an integral representation in $z$ space.
The integral representation can be used  to resum order one terms in Mellin $N$ space 
to obtain reliable theoretical predictions at colliders.

For the Higgs boson production in gluon initiated channels, the sum of SV and NSV contributions is denoted by $\Delta_g$,
\begin{eqnarray}
\Delta_g(z) = \Delta_{g g}^{SV}(z) +
\Delta_{g g}^{NSV}(z)\,.
\end{eqnarray}
Following \cite{Ajjath:2020ulr} in dimensional regularisation ($n=4+\varepsilon$),we obtain
\begin{eqnarray}
\label{masterq}
\Delta_g(\mu_R^2,\mu_F^2,z) = 
\mathcal{C}\exp \bigg( \Psi^g\big(\mu_R^2,\mu_F^2,z,\varepsilon\big)\bigg)\bigg |_{\varepsilon=0} \,,
\end{eqnarray}
where $\Psi^g$ is finite in the limit $\varepsilon \rightarrow 0$
and is  given by
\begin{eqnarray}\label{Psi}
    \Psi^g\big(\mu_R^2,\mu_F^2,z,\varepsilon\big) = &\Bigg( \ln \bigg( Z_{UV,g}\big(\hat{a}_s,\mu^2,\mu_R^2,\varepsilon\big)\bigg)^2 +   \ln \big| \hat{F}_{g}\big(\hat{a}_s,\mu^2,- m^2_H,\varepsilon\big)\big|^2\Bigg) \delta\big(1-z\big)\nonumber \\
    &+2 \mathrm{\Phi}_g\big(\hat{a}_s,\mu^2,m^2_H,z,\varepsilon\big) - 2\mathcal{C} \ln \Gamma_{gg}\big(\hat{a}_s,\mu^2,\mu_F^2,z,\varepsilon\big) \,.
\end{eqnarray}
where $Z_{UV,g}$ is the overall renormalisation constant.
The constant $\hat a_s=\hat g_s^2/16$ is the bare strong coupling constant of QCD.  
The scale $\mu$ results from dimensional regularisation.  The form factor $\hat F_g$ is from virtual contributions 
to $g + g \rightarrow H$  while the soft-collinear function $\mathrm{\Phi}_g$ is obtained from real emission subprocesses
normalised by square of the form factor.  
The diagonal AP kernels are denoted by $\Gamma_{gg}$ which contain 
only diagonal splitting functions $P_{g g}$.

The symbol ``$\mathcal{C}$" refers to convolution which acts on any exponential of a function $f(z)$ takes the
following expansion:
\begin{eqnarray}
        \mathcal{C}e^{f(z)} = \delta(1-z) + \frac{1}{1!}f(z) + \frac{1}{2!}\big(f\otimes f\big)(z) + \cdots
\end{eqnarray}
We keep only  SV distributions namely $\delta(1-z)$,
${\cal D}_i(z)$ and NSV terms $\ln^i(1-z)$ with $i=0,1,\cdots$ resulting from the convolutions.
The solution to the K+G equation of the form factor is expressed in terms
of cusp ($A^g$), soft ($f^g$), collinear ($B^g$) anomalous dimensions and process dependent constants, 
while from the AP evolutions equations one finds the solution in terms of cusp ($A^g$), collinear ($B^g,C^g,D^g$) anomalous dimensions.  
The K+G equation of ${\mathrm \Phi_g}$ contains kernels that 
contain $\delta(1-z)$, $+$ distributions and $\ln(1-z)$ terms.
Due to the explicit $z$ dependence in $\mathrm \Phi_g$, one needs the knowledge of  
logarithmic structure of CFs, $\Delta_g$, from fixed order
perturbative results.  Thanks to CFs that are known to third order, 
we have successfully parametrised ${\mathrm \Phi_g}$ in terms of a set of specific functions of $z$ in dimensional 
regularisation. The function ${\mathrm \Phi_g}$ is also found to be dependent on the cusp ($A^g$), 
soft and the  collinear ($C^g$ and $D^g$)
anomalous dimensions.  Except the overall renormalisation constant the remaining terms are independently 
IR singular.  However, the sum of these contributions is finite and it takes the following integral representation 
in $z$ space: 
\begin{eqnarray}
\label{resumz}
\Delta_g(\mu_R^2,\mu_F^2,z)= C^g_0(\mu_R^2,\mu_F^2)
~~{\cal C} \exp \Bigg(2 \Psi^g_{\cal D} (\mu_F^2,z) \Bigg)\,,
\end{eqnarray}
where
\begin{eqnarray}
\label{phicint}
\Psi^g_{\cal D} (\mu_F^2,z) &=& {1 \over 2}
\int_{\mu_F^2}^{q^2 (1-z)^2} {d \lambda^2 \over \lambda^2}
        P^{\prime}_{gg} (a_s(\lambda^2),z)  
+\left({1 \over 1-z} \overline G^g_{SV}(a_s(q^2 (1-z)^2))\right)_+ 
\nonumber\\
&&+ \varphi_{f,g}(a_s(q^2(1-z)^2),z).
%
%+ {\cal Q}^q(a_s(q^2 (1-z)^2),z)\,,
\end{eqnarray}
%with
%\begin{eqnarray}
%\label{calQc}
%{\cal Q}^q (a_s(q^2(1-z)^2),z) &=&  \left({1 \over 1-z} \overline G^c_{SV}(a_s(q^2 (1-z)^2))\right)_+ + \varphi_{f,q}(a_s(q^2(1-z)^2),z).
%\end{eqnarray}
The $z$ independent coefficient $C_0^g$ is expanded in powers of $a_s(\mu_R^2)$ as
\begin{eqnarray}
\label{C0expand}
C_0^g(\mu_R^2,\mu_F^2) = \sum_{i=0}^\infty a_s^i(\mu_R^2) C_{0i}^g(\mu_R^2,\mu_F^2)\,.
\end{eqnarray}
The AP splitting function $P^\prime_{gg}$ is 
\begin{eqnarray}
        P^\prime_{gg}\big(z,a_s(\mu_F^2)\big) &=& 2 \Bigg(B^g(a_s(\mu_F^2)) \delta(1-z) + 
         A^g(a_s(\mu_F^2)) \left({ 1 \over 1-z} \right)_+
\nonumber\\&&
                      + C^g(a_s(\mu_F^2)) \ln(1-z) + D^g(a_s(\mu_F^2)) \Bigg)\,.
%\nonumber\\&&
%P^{\prime}_{qq}\big(z,a_s(\mu_F^2)\big)\,,
\end{eqnarray}
%where,
%\begin{eqnarray}
%        P^{\prime}_{qq}\big(z,a_s(\mu_F^2)\big) &=& 2  \Bigg[ A^q(a_s(\mu_F^2)) {\cal D}_0(z)
%%\nonumber\\&&
%                      + C^q(a_s(\mu_F^2)) \ln(1-z) + D^q(a_s(\mu_F^2)) \Bigg]
%%\nonumber
%\end{eqnarray}
The constants $C^g$ and $D^g$ 
are known to three loops in QCD \cite{Moch:2004pa,Vogt:2004mw}
********************************* BORN *************************************
(see \cite{GonzalezArroyo:1979df,Curci:1980uw,Furmanski:1980cm,Hamberg:1991qt,Ellis:1996nn,Moch:2004pa,Vogt:2004mw,Soar:2009yh,Ablinger:2017tan,Moch:2017uml} for the lower order ones).
The cusp, soft and the collinear anomalous dimensions 
%and  the constants $C^g$ and $D^g$
are all expanded
in powers of $a_s(\mu_F^2)$ as:
\begin{eqnarray}
X^g(a_s(\mu_F^2)) = \sum_{i=1}^\infty a_s^i(\mu_F^2) X_i^g, \quad \quad \textcolor{black}{X = A,f,B,C,D}
\end{eqnarray}
where $X^g_i$ to third order are available in \cite{Moch:2004pa,Vogt:2004mw} and are listed in appendix \ref{app:anodim}.

The SV function 
$\overline{G}^g_{SV}\big(a_s(q^2(1-z)^2),\epsilon\big)$ is related to the threshold exponent $\textbf{D}^g\big(a_s(q^2(1-z)^2)\big)$ via Eq.(46) of \cite{Ravindran:2006cg}.
The NSV function $\varphi_{f,g}$ is given by
\begin{eqnarray}
\label{varphiexp}
\varphi_{f,g}(a_s(q^2(1-z)^2),z) = \sum_{i=1}^\infty a_s^i(q^2(1-z)^2) \sum_{k=0}^i \varphi_{g,i}^{(k)} \ln^k(1-z) \,,
\end{eqnarray} 
where $q^2 = m^2_H$ and the coefficients $\varphi_{g,i}^{(k)}$ are known to third order and are listed below.
\begin{align}
\begin{autobreak} 
\varphi_{g,1}^{(0)} =
          4 C_A \,,
\end{autobreak} 
********************************* BORN *************************************
\\ 
\begin{autobreak} 
\varphi_{g,1}^{(1)} = 
     0\,,
\end{autobreak} 
\\ 
\begin{autobreak} 
\varphi_{g,2}^{(0)} =

        C_A n_f   \bigg\{
          - \frac{196}{27}
          + \frac{16}{3} \zeta_2
          \bigg\}

      + C_A^2   \bigg\{
          \frac{1306}{27}
          - \frac{208}{3} \zeta_2
          - 28 \zeta_3
          \bigg\}\,,
\end{autobreak} 
\\ 
\begin{autobreak} 
\varphi_{g,2}^{(1)} =

         C_A n_f   \bigg\{
          - \frac{2}{3}
          \bigg\}

      + C_A^2   \bigg\{
          \frac{2}{3}
          \bigg\}\,,
\end{autobreak} 
\\ 
\begin{autobreak} 
\varphi_{g,2}^{(2)} =

          -4 C_A^2 \,,
\end{autobreak} 
\\ 
\begin{autobreak} 
\varphi_{g,3}^{(0)} =

          C_A n_f^2   \bigg\{
          \frac{1568}{729}
          - \frac{152}{9} \zeta_2
          - \frac{160}{27} \zeta_3
          \bigg\}

      + C_A C_F n_f   \bigg\{
          - \frac{2653}{27}
          + \frac{40}{3} \zeta_2
          + \frac{32}{5} \zeta_2^2
          + \frac{616}{9} \zeta_3
          \bigg\}

      + C_A^2 n_f   \bigg\{
          - \frac{117778}{729}
          + \frac{26780}{81} \zeta_2
          - \frac{232}{15} \zeta_2^2
          + \frac{1888}{9} \zeta_3
          \bigg\}

      + C_A^3   \bigg\{
          \frac{563231}{729}
          - \frac{113600}{81} \zeta_2
          + \frac{3488}{15} \zeta_2^2
          + 192 \zeta_5
          - \frac{34292}{27} \zeta_3
          + \frac{176}{3} \zeta_3 \zeta_2
          \bigg\}\,,
\end{autobreak} 
\\ 
\begin{autobreak} 
\varphi_{g,3}^{(1)} =

         C_A n_f^2   \bigg\{
          \frac{56}{27}
          \bigg\}

      + C_A C_F n_f   \bigg\{
          4
          - \frac{8}{3} \zeta_2
          \bigg\}

      + C_A^2 n_f   \bigg\{
          \frac{1528}{81}
          - \frac{248}{9} \zeta_2
          - 8 \zeta_3
          \bigg\}

      + C_A^3   \bigg\{
          - \frac{18988}{81}
          + \frac{1280}{9} \zeta_2
          + \frac{448}{3} \zeta_3
          \bigg\}\,,
\end{autobreak} 
\\ 
\begin{autobreak} 
\varphi_{g,3}^{(2)} =

       C_A n_f^2   \bigg\{
          \frac{8}{27}
          \bigg\}

      + C_A^2 n_f   \bigg\{
          \frac{164}{27}
          + \frac{2}{3} \zeta_2
          \bigg\}

      + C_A^3   \bigg\{
          - \frac{1432}{27}
          + \frac{40}{3} \zeta_2
          \bigg\}\,,
\end{autobreak} 
\\ 
\begin{autobreak} 
\varphi_{g,3}^{(3)} =

       C_A^2 n_f   \bigg\{
          \frac{32}{27}
          \bigg\}

      + C_A^3   \bigg\{
          - \frac{176}{27}
          \bigg\}\,.
\end{autobreak} 
\end{align} 
For SU$(n_c)$ gauge group, the color factor $C_F=(n_c^2-1)/2 n_c$ and $C_A=n_c$.  Here $n_f$ is the number of active flavors.
%---------------------------------------------------------------

\subsection{Resumming next-to SV terms}  \label{resNSV}
The resummation of threshold logarithms can be conveniently done in Mellin space.  
The threshold limit $z \rightarrow 1$ in $z$ space translates to $N\rightarrow \infty$ in the Mellin space. 
To include both SV and  NSV terms we need to keep $\ln N$ as well as ${\cal O}(1/N)$
terms in the large $N$ limit.
In the large $N$ limit one encounters order one terms from $\omega=2 \beta_0 a_s(\mu_R^2) \ln N$ at every order in $a_s$ in
the exponent which spoils the reliability of fixed order truncation of the
perturbative series.  This can be solved by reorganising the series by using
the resummed strong coupling constant.  
Up to ${\cal O}(1/N)$ in $N$ space, the $N$ space resumed  will reproduce the
fixed order result.
The Mellin moment of $\Delta_g$ was computed in \cite{Ajjath:2020ulr}
and is given by 
\begin{eqnarray}
\label{DeltaN}
\Delta_{g,N}(\mu_R^2,\mu_F^2) = C_0^g(\mu_R^2,\mu_F^2) \exp\left(
\Psi_{N}^g (\mu_F^2,\textcolor{black}{\mu_R^2})
\right)\,,
\end{eqnarray}
where
\begin{eqnarray}\label{eq:Mellin}
\Psi_{N}^g(\mu_F^2,\textcolor{black}{\mu_R^2}) = 2 \int_0^1 dz z^{N-1}\Psi_{\cal D}^g (\mu_F^2,\textcolor{black}{\mu_R^2},z) .
\end{eqnarray}
We split all the $1/N^0$ and $1/N$ terms in  $\Psi_{N}^g$ as
\begin{eqnarray}
\label{eq:Psi}
\Psi_{N}^g = \Psi_{\rm{sv},N}^g + \Psi_{\rm{nsv},N}^g\,.
\end{eqnarray}
In other words, we split $\Psi_N^g$ in such a way that $\Psi_{\rm{sv},N}^g$ contains  $\ln^j N, j=0,1,\cdots$ terms
while $\Psi_{{\rm nsv},N}^g$ contains terms of the form $(1/N) \ln^j N, j=0,1,\cdots$.
It is well known that $\Psi_{\rm{sv},N}^g$ takes the following form:
\begin{eqnarray}
\label{PsiSVN}
        \Psi_{\rm{sv},N}^g = \ln(g_0^g(a_s(\mu_R^2))) + g_1^g(\omega)\ln N + \sum_{i=0}^\infty a_s^i(\mu_R^2) g_{i+2}^g(\omega) \,,
\end{eqnarray}
with
\begin{align}
\label{lng0}
\ln g_0^g(a_s(\mu_R^2)) = \sum_{i=1}^\infty a_s^i(\mu_R^2)~ g_{0,i}^g \,.
\end{align}
The exponents $g_{0,i}^g$ and $g_i^g$ are listed in the Appendices \ref{app:g0g} and \ref{app:gN}. . The $g^g_i$ coefficients are universal whereas
the constant $g_0^g$ contains only $N$ independent pieces 
resulting  from the Mellin moment of $\Psi_{\cal D}^g (\mu_F^2,z)$. 
The constant $g_0^g$ is ambiguous and is defined by the condition that $\Psi_{\rm{sv},N}^g - \ln(g_0^g) = 0$ when $N=1$.
This ambiguity can be exploited to define various resummation schemes.
The $N$-independent coefficients $C^g_0$ and $g_0^g$ are related to the coefficients $\tilde g_0^g$ given in the paper \cite{H:2019dcl,Ajjath:2020rci} using the following relation,
\begin{equation}
\tilde g_0^g(\mu_R^2, \mu_F^2) = C_0^g(\mu_R^2,\mu_F^2) \  g_0^g(a_s(\mu_R^2))
\end{equation}
which is expanded in terms of $a_s(\mu_R^2)$ as,
\begin{equation} \label{g0tg}
  \tilde g_0^g(a_s(\mu_R^2)) = \sum_{i=0}^\infty a_s^i(\mu_R^2) \tilde g^g_{0,i}\quad \,,
\end{equation}
the exponents $\tilde g^g_{0,i}$ are listed in the Appendix \ref{app:g0t}. 

The NSV function $\Psi_{\rm{nsv},N}^g$ is found to be
\begin{align}
\label{PsiNSVN}
 \Psi_{\rm{nsv},N}^g = {1 \over N} 
\sum_{i=0}^\infty a_s^i(\mu_R^2) \bigg ( \bar g_{i+1}^g(\omega)
%\nonumber\\
%&& 
%+ {1 \over N} \Bigg(h^g_{00}(\omega) + h^g_{01}(\omega) \ln(N)
+ h^g_{i}(\omega,N) \bigg)\,,
\end{align}
with 
%  \Bigg)\,,
%\nonumber\\
\begin{align}\label{hNSV}
h^g_i(\omega,N) = \sum_{k=0}^{i} h^g_{ik}(\omega)~ \ln^k N.
\end{align}
where $\bar g^g_i(\omega)$ and $h^g_{ik}(\omega)$ are presented in the Appendices \ref{app:gbN} and \ref{app:hN}, respectively. 

We find that in each of the exponents 
$g_i^g(\omega)$, $\overline g_i^g(\omega)$ and $ h_{ik}^g(\omega)$ 
from SV as well as NSV, we are resumming in the Mellin space ``order one" $\omega$ terms to all orders in perturbation theory. In the large $N$ expansion, one finds that the $\ln N$ terms are always associated
with {\it Euler-Mascheroni} constant, $\gamma_E$.  As these $\gamma_E$ terms can spoil the reliability of
perturbative expansion, they are often combined with $\ln N$.   This results in a change in the definition of 
order one term $\omega$ as \textcolor{black}{$\overline{\omega} = 2 a_s(\mu_R^2) \beta_0 \ln \overline N$} with $\overline
N = \exp(\gamma_E) N$.  This alters the numerical predictions considerably, see \cite{Das:2019btv}. 
We call the former one as standard {\it N-exponentiation} and the latter by 
{\it $\overline N$-exponentiation}.
In addition to these $\gamma_E$ terms,  one encounters $N$ independent pieces that are mostly proportional to irrational
terms namely Euler zeta terms $\zeta_i$.  Unlike $\gamma_E$, there exists no systematic way to sum them up. In other
words, there exists ambiguity in whether one should keep them in the exponent or not.    
While there are many ways we can treat these terms, we adopt two distinct ways namely
{\it soft exponentiation} and {\it all exponentiation}.  
In {\it soft exponentiation},
the Mellin moment of entire soft-collinear function \cite{Ravindran:2005vv,Ravindran:2006cg}
which contains all the $+$ distributions as well as the $\delta(1-z)$ terms is exponentiated.
In the latter scheme , we exponentiate the form factor along with the soft-collinear function in the Mellin space. 
In \cite{Bonvini:2014joa,Bonvini:2016frm}, this approach was used  to study the inclusive Higgs boson production in gluon fusion.
They found that the numerical predictions are less sensitive to the choice of factorisation and renormalisation sales 
compared to the standard threshold approach.
This approach is theoretically supported by the fact that the Sudakov K+G differential equation \cite{Sudakov:1954sw, Mueller:1979ih, Collins:1980ih, Sen:1981sd, Ravindran:2005vv,Ravindran:2006cg} that the form factors satisfy give solutions of the form of   
exponential of $N$ independent terms.

In \cite{Ajjath:2020rci,Ajjath:2021lvg}, we have presented various schemes that can deal with $N$ independent terms for the numerical study of Drell-Yan process.
We briefly describe them below for completeness:
\begin{itemize}
% \item {\bf {\tt Standard} $N$ {\tt exponentiation}:} In this scheme
% the $N$ independent term $g_0^g$ is defined by the condition that $\Psi^g_{\rm{sv},N} - \ln(g_0^g)=0$ when $N=1$.

\item {\textbf{\textit{Standard} $N$ {\textit{ exponentiation}}}:}  In this scheme, we exponentiate only the large-$N$ pieces that contribute to the CFs while keeping all the $N$-independent pieces outside the exponential. This will only alter the $\Psi_{{\rm sv},N}^g$ whereas the $\Psi_{{\rm nsv},N}^g$ will remain same as it contains only $N$-dependent terms. Therefore  we write:
\begin{eqnarray}
\Delta_{g,N}(m_H^2,\omega) = \tilde g_0^g(m_H^2) \exp\left(
G_{{\rm sv}, N}^g (m_H^2,\omega) + \Psi_{{\rm nsv}, N}^g (m_H^2,\omega) 
\right)\,,
\end{eqnarray}
where NSV part is defined in \eqref{PsiNSVN}. And the SV part is given as,
\begin{equation}
    G_{{\rm sv}, N}^g (m_H^2,\omega) = \Psi_{{\rm sv}, N}^g (m_H^2,\omega) - \ln(g_0^g) 
\end{equation}
 which can be obtained from the Mellin moment of $\Psi_{\cal D}^g$ and keeping only those terms that vanish when $N=1$. 
% This is the standard approach adopted all along this paper and we compare other prescriptions with the $N$-exponentiation scheme in later sections.
%The coefficient $g_0^q$ is obtained by setting $\Psi^q_{{\rm SV},N} - \ln g_0^q=0$ when $N=1$.

\item {\textbf{\textit{Standard} $\Nb${\textit{ exponentiation}}}:} 
Here,  we compute Mellin moment in the large 
$\Nb  \big(= N\exp(\gamma_{E}) \big)$  limit that means the entire $\gamma_{E}^{}$ dependent terms 
are exponentiated through $\overline{N}$. In this scheme, the CFs $\Delta_{g,\overline N}$ can be expressed as
\begin{align}
\Delta_{g,\overline N}(m_H^2,\mu_R^2,\mu_F^2) =  \bar{ \tilde{g}}_0^g(m_H^2,\mu_R^2,\mu_F^2) \exp\left(
G_{{\rm sv},\overline N}^g (m_H^2,\overline{\omega}) + \Psi_{{\rm nsv}, \overline N}^g (m_H^2,\overline{\omega}) 
\right) \,.
\end{align}
Here the $\bar{ \tilde{g}}_0^g(m_H^2,\mu_R^2,\mu_F^2)$ are given by 
\begin{equation}
\label{g0nb}
\bar{ \tilde{g}}_0^g(m_H^2,\mu_R^2,\mu_F^2) = C_0^g(m_H^2,\mu_R^2,\mu_F^2) \  \bar{{ g}}_0^g(a_s(\mu_R^2)) \,,
\end{equation}

where $\overline g_0^g$ results from the $N$-independent part of the Mellin transformation of the distributions and  the  $\gamma_E$'s are all absorbed into the $\overline N$-dependent functions. Here the $\overline N$-dependent part of SV, i.e., $G_{{\rm sv},\overline N}^g$, is obtained by setting $\Psi^g_{{\rm sv},\overline N} - \ln \overline g_0^g=0$ 
when $\overline N=1$.
Hence, $G^g_{{\rm sv}, \overline{N}} $ and $\Psi_{{\rm nsv}, \overline N}^g$ are given by
\begin{align}\label{eq:gnb}
G^g_{{\rm sv}, \overline{N}} &= g^g_1(\overline \omega) \ln \overline{N} + \sum_{i=0}^\infty a_s^i(\mu_R^2)~ g^g_{i+2}(\overline{\omega})\,,
% \end{align}
% \begin{align}
\\
  \Psi_{\rm{nsv},\overline N}^g &= {1 \over N} 
\sum_{i=0}^\infty a_s^i(\mu_R^2) \bigg ( \bar g_{i+1}^g(\overline \omega)
+ h^g_{i}(\overline \omega,\overline N) \bigg)\,,
\end{align}
with 
\begin{align}\label{hnb}
h^g_i(\overline \omega,\overline N) = \sum_{k=0}^{i} h^g_{ik}(\overline \omega)~ \ln^k \overline N.
\end{align}
Numerically, this can make difference at every logarithmic accuracy. In \cite{Das:2019btv}, it was found that the $\overline{N}$-exponentiation shows a faster convergence compared to 
the $N$-exponentiation for the charged and neutral DIS processes. This is the standard approach adopted all along this paper and we compare other prescriptions with the $\Nb$-exponentiation scheme in later sections.

\item {\textbf{\textit{Soft}{\textit{ exponentiation}}}:}  
Here, we exponentiate complete finite part of soft-collinear function $\mathrm \Phi_g$ to obtain   
\begin{align}\label{eq:resN}
\Delta_{g,\Nb}(m_H^2,\overline{\omega}) = \tilde{g}^{g,\rm {Soft}}_0(m_H^2) \exp \Big( \Psi^{g,\rm {Soft}}_{{\rm sv}, \Nb}(m_H^2,\overline{\omega}) + \Psi^{g}_{{\rm nsv}, \Nb}(m_H^2,\overline{\omega})\Big) \,.
\end{align}
with
\begin{align}\label{eq:gnbsoft}
\Psi_{{\rm sv},\Nb}^{g,\rm {Soft}}(m_H^2,\overline{\omega}) = \ln \overline N ~g^{g,\rm Soft}_1(\overline{\omega}) + \sum_{i=0}^{\infty}  a_s^i(\mu_R^2)~ g^{g,\rm Soft}_{i+2}(m_H^2,\overline{\omega}) \,.
\end{align}
Here the exponents $g_i^{g,\rm Soft}$ contain the complete finite part coming from the soft-collinear function.
The remaining $N$-independent terms coming from  finite part of form factor and AP kernels contribute to $\tilde{g}^{g,\rm{Soft}}_0$, whose expansion in powers of $a_s$ is given as:
\begin{align}\label{eq:g0bsoft}
\textcolor{black}{
\tilde{g}^{g,\rm{Soft}}_{0}(m_H^2,\mu_R^2,\mu_F^2) = 1+\sum_{i=1}^{\infty} a_s^i ~ \tilde{g}^{g,\rm{Soft}}_{0i}(m_H^2,\mu_R^2,\mu_F^2) \,.}
\end{align}
The $N$-independent constants $\tilde{g}^{g, \rm Soft}_{0i}(m_H^2)$ and the resummed exponents $g_i^{g, \rm Soft}(\overline{\omega})$ are 
listed in Appendix \ref{app:softexp}.

\item {\textbf{\textit{All}{\textit{ exponentiation}}}:}   
This scheme arises in light of \eqref{masterq}  which is a consequence of the differential equations satisfied by each of the building blocks leading to an all-order exponential structure for  $\Delta_{g,N}$. Here we study the numerical impact of the entire
result taken to the exponent as given below.
\begin{align}\label{eq:resall}
\Delta_{g,\Nb}(m_H^2,\overline{\omega}) = \exp \Big( \Psi_{{\rm sv},\Nb}^{g,\rm All}(m_H^2,\overline{\omega}) +  \Psi_{{\rm nsv},\Nb}^{g}(m_H^2,\overline{\omega})  \Big) \,,
\end{align}
where \begin{align} \label{gnall}
\Psi_{{\rm sv},\Nb}^{g,\rm All}(m_H^2,\overline{\omega}) = \ln \overline N ~g_1^{q,\rm All}(m_H^2) + \sum_{i=0}^{\infty} a_s^i ~g_{i+2}^{g,\rm All}(m_H^2,\overline{\omega})  \,.
\end{align}
\end{itemize}
The exponent in \eqref{eq:resall} contains both $N$ dependent and independent terms which are listed in Appendix \ref{app:Allexp}. In \cite{Bonvini:2014joa,Bonvini:2016frm}
this scheme was used to study the  inclusive cross section for the production of Higgs boson in gluon fusion at the LHC. 
For similar studies on DY and DIS processes in $\overline{\rm MS}$ schemes, see \cite{Eynck:2003fn}. \\
In \cite{Ajjath:2020rci,Ajjath:2021lvg}, we had studied how various schemes discussed so far can affect the predictions
of invariant mass distribution of lepton pairs in the DY process. 
% We extended similar study taking into account next to SV terms in \cite{H:2019dcl} for the production
% of lepton pairs in DY process.   
In the present work, we perform a similar analysis with the resummed NSV exponent for the Higgs boson production and report their
numerical impact. In order to distinguish between SV and SV+NSV resummed results, we denote the NSV included resummed results by $\overline{\rm N^nLL}$. We use fixed order results up to NNLO level and the resummed results up to $\overline{\rm NNLL}$ accuracy. The resummed results are matched to the fixed order result 
in order to avoid any double counting of threshold logarithms.    
% We use the publicly available code TROLL, \cite{Bonvini:2014joa} to obtain fixed order prediction up to NNLO and in house routines to incorporate resummed NSV predictions.
The resummed result at a given accuracy, say $\overline{\rm N^nLL}$, is
obtained by taking the difference between the resummed result and the same truncated 
up to order $a_s^n$. Hence, it contains contributions from the SV and NSV 
terms to all orders in perturbation theory starting from $a_s^{n+1}$ on wards:   
%%%
\textcolor{black}{
\begin{align}\label{eq:matched}
\sigma_N^{\rm {H},\rm {N^nLO+\overline {\rm N^nLL}}} &= 
\sigma_N^{\rm{H},\rm {N^nLO}} +
%\sigma^{(0)} 
%\sum_{ab\in\{g\}}
 {\pi G_B^2(\mu_R^2) \over 4 (N^2-1)}
  \int_{c-i\infty}^{c+i\infty} \frac{dN}{2\pi i} (\tau)^{-N} %\delta_{a\overline b}
  f_{g,N}(\mu_F^2) f_{g,N}(\mu_F^2) \nn\\
&\times \bigg( \Delta_{g,N} \bigg|_{\overline {\rm {N^nLL}}} - {\Delta_{g,N}}\bigg|_{tr ~\rm {N^nLO}}     \bigg)  \,.
\end{align}}
We use PDFs known to NNLO accuracy and their Mellin space counterparts ($f_{i,N}$) can be obtained using QCD-PEGASUS \cite{Vogt:2004ns}. 
However, we follow \cite{Catani:1989ne,Catani:2003zt} to directly
deal with PDFs in the $z$ space. 
%While inverting  Mellin space back to $\tau$ space  
%we use the {\it Minimal prescription} \cite{Catani:1996yz} procedure. 
In the above eq.\ (\ref{eq:matched}) the second term represents the resummed result truncated 
to N$^n$LO order.  \textcolor{black}{  In Table \ref{tab:res}, we list the resummation coefficients which are required to obtain the resummed $\Delta_{g,N}$ at a given logarithmic accuracy.} In the next section, we present the numerical results for the production Higgs boson in gluon fusion
at the LHC. 

\begin{table}[h!]
 \centering
 \begin{center}
 \begin{small}
 {\renewcommand{\arraystretch}{1.6}
 \begin{tabular}{|P{3.5cm}||P{3.5cm}|}
  \rowcolor{lightgray}
  Logarithmic Accuracy & Resummed Exponents    
    \\
 %   \rowcolor{lightgray}
 %                   Accuracy &   Exponents  & \\
  \hline
 $\overline{\rm LL}$&	$\tilde g^g_{0,0},g^g_1,\overline g^g_1,h^g_0$     \\
  \hline
 $\overline{\rm NLL}$ &$\tilde g^g_{0,1},g^g_2,\overline g^g_2, h^g_1$  \\
  \hline
 $\overline{\rm NNLL}$&$\tilde g^g_{0,2},g^g_3,\overline g^g_3, h^g_2$   \\
   \hline
  %\hline
 \end{tabular}}
 \end{small}
 \end{center}
 	\caption{\label{tab:res}  The set of resummation coefficients \Big\{$\tilde g^g_{0,i},g^g_i(\omega),\overline{g}^g_i(\omega), h^g_{i}(\omega)$\Big\} which is required to compute $\Delta_{g,N}$ at a given logarithmic accuracy.  }
 \end{table}

%I need the following expressions  (in terms of color factors for gg channel) in tex format:
%\begin{itemize}
%\item  $C_0^g$ in 2.6
%\item $\overline G^g_{SV}$ in eqn 2.8
%\item $A,B,C,D,f$ to third/fourth order for $g$
%\item $\varphi_{g,i}^k$ in 2.13
%\item
%all $g_0, g_i$ in 2.17
%\item
%all $\overline g_i$ , $h_{ik}^g$ in 2.20 and 2.21
%\item 
%expressions for various resummation schemes
%\end{itemize}

%\documentclass[a4paper,11pt]{article}
%\pdfoutput=1
%\usepackage{jheppub}
%\usepackage{graphicx,color}
%\usepackage{amsmath,autobreak}
%\usepackage{array}
%\usepackage{ulem}
%\newcolumntype{P}[1]{>{\centering\arraybackslash}p{#1}}
%\allowdisplaybreaks
% ************** Used Packages **************
%\RequirePackage{ifpdf}
%\usepackage{amsmath}
%\usepackage{mathtools}

%\usepackage{autobreak}

%\usepackage[final]{pdfpages}
%\usepackage{ifpdf}
%\usepackage{slashed}

%\usepackage{hyperref}

%\usepackage{color}
%\usepackage{graphics}

%\usepackage{etoolbox}
%\usepackage{fixmath}

%\usepackage{caption}
%\usepackage{subcaption}
%\usepackage{amsfonts}
%\usepackage{multirow}
%\usepackage{epstopdf}

% For making math symbol large: \mathlarger
%\usepackage{relsize}

%\usepackage{float}

% For tables
%\usepackage[table]{xcolor}
%\usepackage{tabularx}

%\allowdisplaybreaks

%\newcommand{\fig}[1]{fig.\ (\ref{#1})}

%\begin{document}

\section{Numerical Analysis} \label{pheno}

In this section, we discuss the numerical impact of resummed soft virtual plus next-to-soft virtual (SV+NSV) corrections for the Higgs boson production in the gluon fusion channel at the LHC. The leading order process is initiated via gluon-gluon fusion through top-quark loop, which is integrated out in the large $m_t$-limit. Throughout the computation we use, the top quark mass $m_t=173.2$ GeV, the Higgs mass $m_H=125$ GeV and the Fermi constant $G_F = 4541.63$ pb. The fixed order numerical predictions up to $\rm NNLO$ have been obtained using an in-house FORTRAN code and we have validated our  results with the publicly available fixed order code \href{http://www.ge.infn.it/~bonvini/higgs/}{ggHiggs} \cite{Bonvini:2014joa}. 
We have performed the inverse Mellin transformation of the resummed $N$-space result using an in-house FORTRAN code. Minimal prescription \cite{Catani:1996yz} has been used to deal with the Landau pole in the Mellin inversion routines. We use the \textbf{MMHT2014(68cl)} \cite{Harland-Lang:2014zoa} parton distribution set and \textbf{LHAPDF} \cite{Buckley:2014ana} interface to provide the strong coupling constant $a_s$ with $n_f = 5$ active massless quark flavours throughout.  The detailed analysis is done for 13 TeV LHC, however in the later sections we have extended our analysis to other collider energies as well. %\textcolor{red}{\sout{For better clarity, $\mu_R$ and $\mu_F$ in the plots lie in the range $(0.2,3) m_H$.} }

We begin by investigating the importance of NSV logarithms in comparison to the SV distributions, present in the fixed order predictions for the gluon induced Higgs boson productions. Hence, we restrict ourselves to the partonic coefficient function $\Delta_g$ given in Eq. \ref{masterq} which consists of $\rm SV$ and  $\rm NSV$ terms. The $\rm SV$ part consists of $\delta(1-z)$ and ${\cal D}_k(z)$ while the $\rm NSV$ part consists of collinear logarithms $\ln^k (1-z)$.  
%coming only from diagonal channel i.e. $gg$ channel. 
 We compute the percentage contribution of the $\rm NSV$ logarithms to the hadronic cross section to analyse their relevance. At $\rm NLO$, we find the $\rm SV$ comprises $73.16\%$ of the Born contribution while the $\rm NSV$ contribute to only $45.81\%$. However, from $\rm NNLO$ onwards the contributions coming from $\rm NSV$ logarithms become more significant than that of $\rm SV$. For instance, at $\rm NNLO$ the $\rm NSV$ logarithms give $58.91\%$,  whereas $\rm SV$ $15.81\%$ of the Born cross-section. Similar trend follows at $\rm N^3LO$ level where $\rm NSV$ contribution is $25.83\%$ and $\rm SV$ has a negative contribution of $-2.29\%$ of the total Born cross section. In Table \ref{tab:N2LO}, we provide the individual contributions coming from different powers of $\rm SV$ distributions and $\rm NSV$ logarithms at $\rm NNLO$ accuracy. The corresponding percentages for $\rm SV$ and $\rm NSV$ contributions at $\rm N^3LO$ accuracy are given in \cite{Anastasiou:2014lda}. From this analysis, one can conclude that with increasing order of perturbative corrections, NSV logarithms become phenomenologically more relevant than the SV distributions.

%The significant increase in the percentage contribution of $\rm NSV$ logarithms over $\rm SV$ from $\rm NNLO$ accuracy onwards establishes its importance. 
\begin{table}[h] 
 \begin{center}
 \begin{small}
% %\newcolumntype{P}[1]{>{\centering\arraybackslash}p{#1}}
 {\renewcommand{\arraystretch}{1.5}
 \begin{tabular}{|P{1.3cm}||P{1.5cm} |P{1.8cm}||P{1.5cm}|P{1.5cm}|}
 \hline
 \rowcolor{lightgray}
 \multicolumn{1}{|c||}{$\mu_R$ = $\mu_F$ (GeV)}
     & \multicolumn{2}{c|}{ SV}   
     &\multicolumn{2}{c|}{ NSV}  
     \\ 
  \hline
  \hline
 %  \hline
 \multirow{5}{5cm} { \hspace{1cm} 125} &  $\mathcal{D}_3$ &  45.3\% & $\ln^3(1-z)$&  52.64\% \\
 & $\mathcal{D}_2$ &  4.87\% & $\ln^2(1-z)$&   37.34\% \\
  & $\mathcal{D}_1$ &  -10.60\% & $\ln^1(1-z)$&   -7.45\% \\
  & $\mathcal{D}_0$ & -25.51\% & $\ln^0(1-z)$&   -23.62\% \\
  & $\delta(1-z)$ & 1.75\% & & \\
  \hline
  %\hline
 %  \cline{2-5}
 %   \cline{2-5}
 %\rowcolor{lightgray} 
 \multicolumn{1}{|c||}{\hspace{0.2cm} TOTAL}
     & \multicolumn{2}{c|}{ \hspace{2cm} 15.81\%}   
     &\multicolumn{2}{c|}{\hspace{2cm}  58.91\%}  
     \\ 
 \hline
 %Drell-Yan &8.59\%&5.44\% & 9.82\% & 2.62\% & 1.49\%&-1.00\%\\
 %\hline
 \end{tabular}}
 \caption{\% contribution of SV distributions and NSV logarithms to the Born cross section at NNLO at $\mu_R$ = $\mu_F=125$ GeV.}
 \label{tab:N2LO}
 \end{small}
 \end{center}
 \end{table}
Moreover, in case of Higgs production through gluon fusion, the diagonal $gg$-channel constitutes the dominant contribution to the hadronic cross section. For example, the contribution from the $gg$ channel is $48.35 \%$ of the full NLO cross section whereas the correction from the $qg$-channel is only $-0.79\%$. Similarly the $a_s^2$ correction from the $gg$ channel is $19.71 \%$  and from the $qg$ is around $2\%$ of the NNLO cross section. The other sub-dominant channels are more suppressed in comparison to the $qg$-channel. Since there is a significant increase in the percentage contribution of $\rm NSV$ logarithms over $\rm SV$ from $\rm NNLO$ accuracy on wards from the diagonal channel, which is also the main contributor to the hadronic cross section, it would be interesting to see how phenomenologically important the resummed $\rm NSV$ logarithms are, in addition to the well established $\rm SV$ resummation.

In the context of threshold resummation in \cite{Bonvini:2014joa}, it was found that the resummation of the SV distributions to N$^3$LL accuracy change the NNLO results by 20\% at $\mu_R=\mu_F=m_H$ and 5.5\% at $\mu_R=\mu_F=m_H/2$. Compared to
fixed order predictions, the SV resummed results were found to depend moderately  on the renormalization scale
%, which drives the theoretical uncertainty of the fixed order computation. Also,
while there was increase in the dependence on the factorization scale.  Recently in \cite{Ajjath:2021lvg}, the SV resummation was extended to include the resummed effects from the NSV logarithms for the lepton pair production in Drell-Yan process. It was found that owing to the large coefficients of the NSV logarithms, the resummation of such logarithms lead to a considerable amount of enhancement of the fixed order cross section. The renormalization scale dependency were also seen to reduce compared to the fixed order results.  However, the uncertainty from factorization scale increases as we did not include the resummed contributions of NSV terms, namely the
collinear logarithms  from the off-diagonal channels. 

%{\color{red} In this paper, we perform similar such analysis for the case of gluon induced Higgs boson productions.} Since the 
%
\textcolor{black}{
In summary, we find that in the dominant $gg$ channel, as the order of perturbation increases,  the numerical coefficients of collinear logarithms increase leading to contributions that are larger than those from  SV distributions.   In addition,
there exists theoretical results for the resummation of NSV terms to all orders up to NNLL accuracy \cite{Ajjath:2020ulr} and hence using the latter, it is desirable to study their numerical importance.  In this context, we ask the following questions that can shed light on the importance of NSV terms and also unravel the role of beyond NSV terms in the context of Higgs boson production in gluon fusion at the LHC.}
%Now in case of Higgs production through gluon fusion we raise the following questions concerning SV+NSV resummation, which we aim to unravel in the subsequent discussion:
\begin{itemize}
    \item How much is the enhancement of the resummed cross section in comparison to the fixed order ones, owing to the large coefficients of the NSV logarithms ?
    \item Since $gg$-channel is the dominant contributor to the hadronic cross section, what is the impact of resummation of NSV terms from the  collinear logarithms on the unphysical scales, namely $\mu_F$ and $\mu_F$ ?
    \item How different is the impact of SV+NSV resummation from the well established threshold resummation ?
\end{itemize}
%\\
We address the above questions in the following sections. 

\subsection{Fixed-order vs Resummed results} 

In this section, we study the numerical relevance of resuming the $\rm NSV$ logarithms along with the $\rm SV$ distributions at $\overline {\rm LL}$, $\overline{\rm NLL}$ and $\overline{\rm NNLL}$ accuracy matched to the corresponding fixed order results using Eq. \ref{eq:matched}. We investigate the dependence of $\rm SV + \rm NSV$ resummed cross section on the choices of renormalisation $\mu_R$ and factorisation $\mu_F$ scales to study their perturbative uncertainties. We start by plotting the inclusive cross section as a function of both $\mu_R$ and $\mu_F$ simultaneously in a 3D graph for centre-of-mass energy $13$ TeV. 
%This is done to get an overall picture of how the cross section depends on these unphysical scales. 

\begin{figure}[ht]
%\begin{center}
\hspace*{-1.2cm}
\includegraphics[width=.48\textwidth]{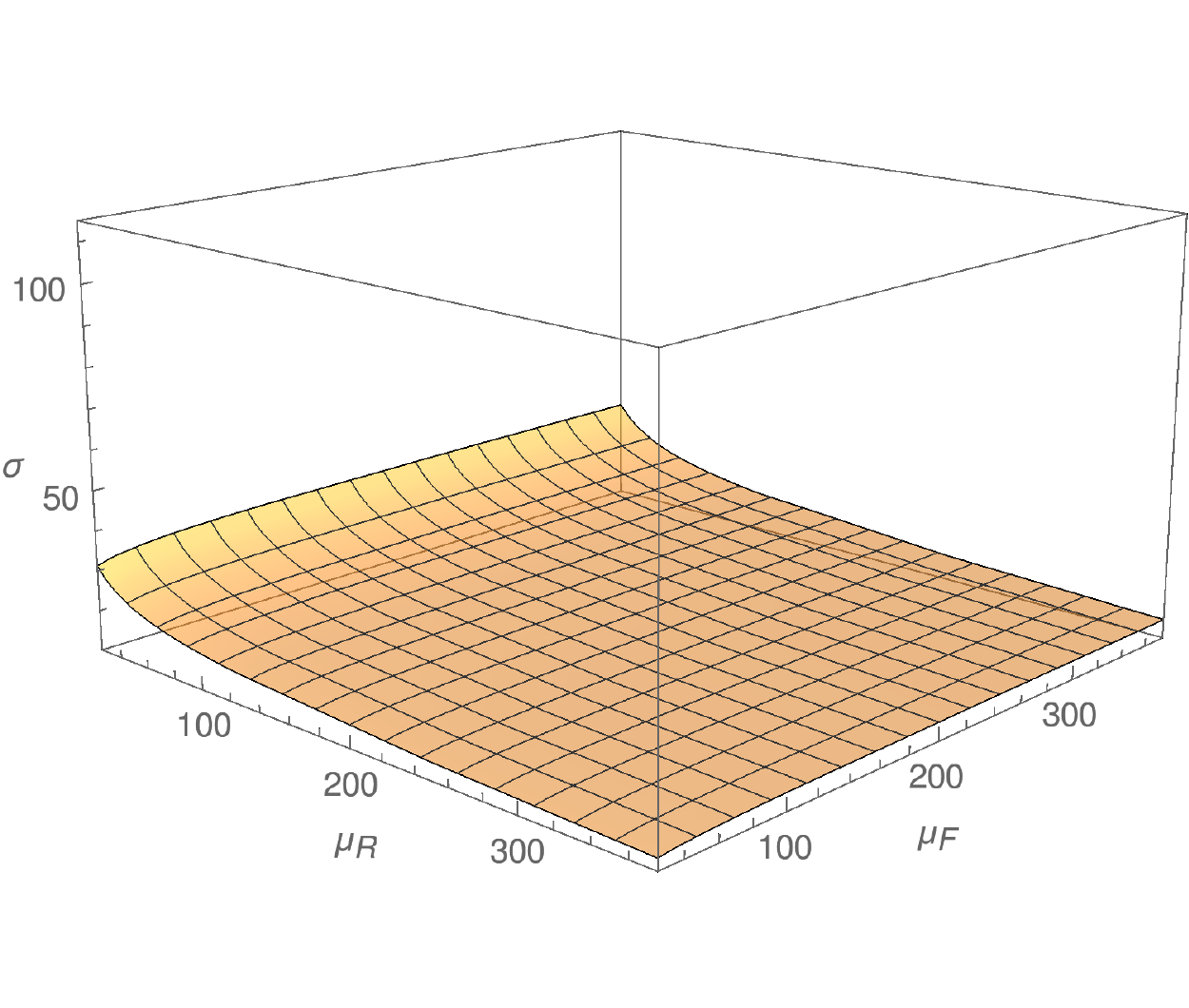}
\includegraphics[width=.48\textwidth]{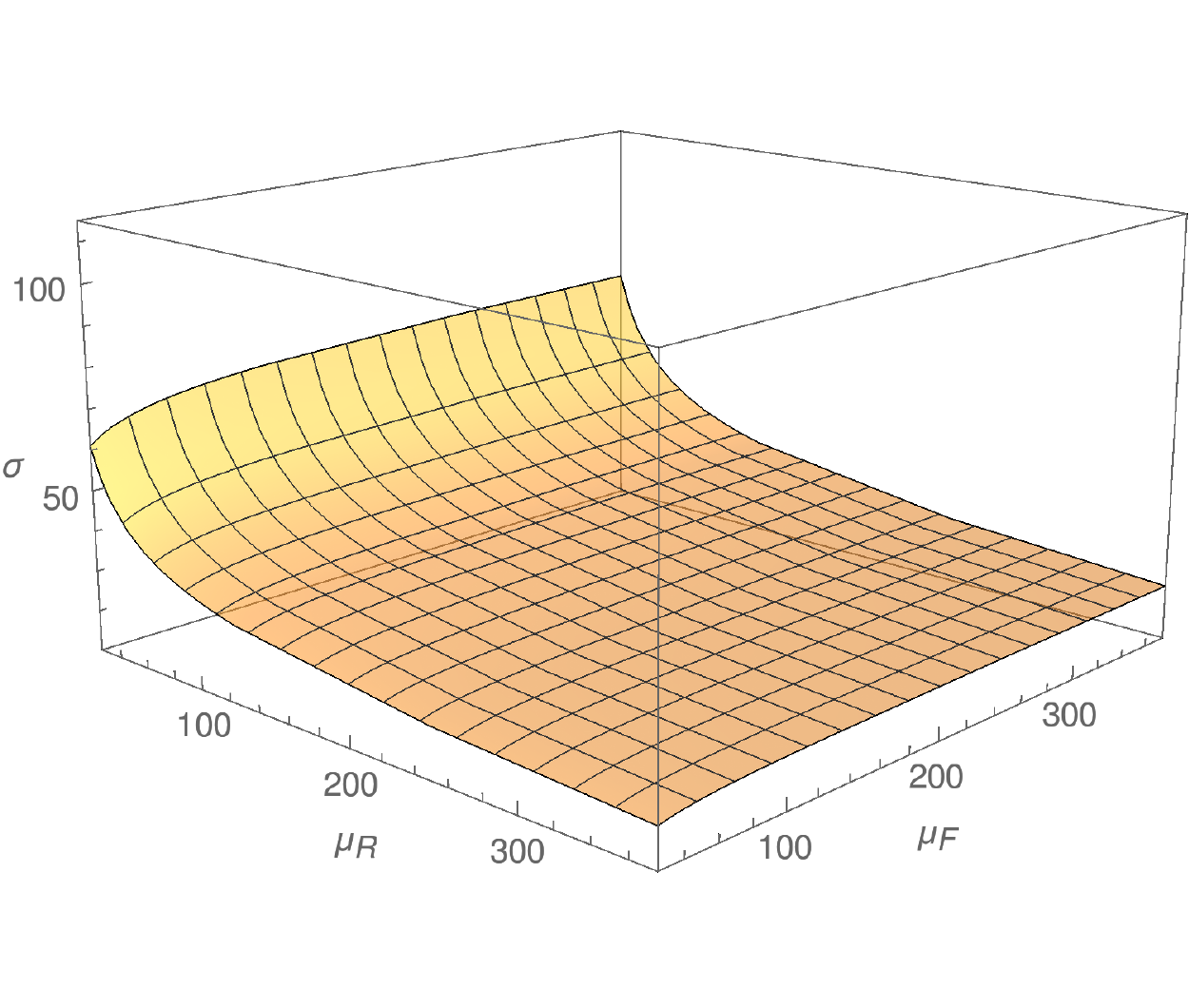}
\\(a)
%\end{center}
%\caption{\small{$\mu_{R}$ variation for $\mu_F = m_H/2$ for \textcolor{red}{N-Exp}(left panel) and for \textcolor{red}{$\overline{N}$-Exp}(right panel) }}
\label{fig5}
\end{figure}

\begin{figure}[ht]
%\begin{center}
\hspace*{-1.2cm}
\includegraphics[width=.48\textwidth]{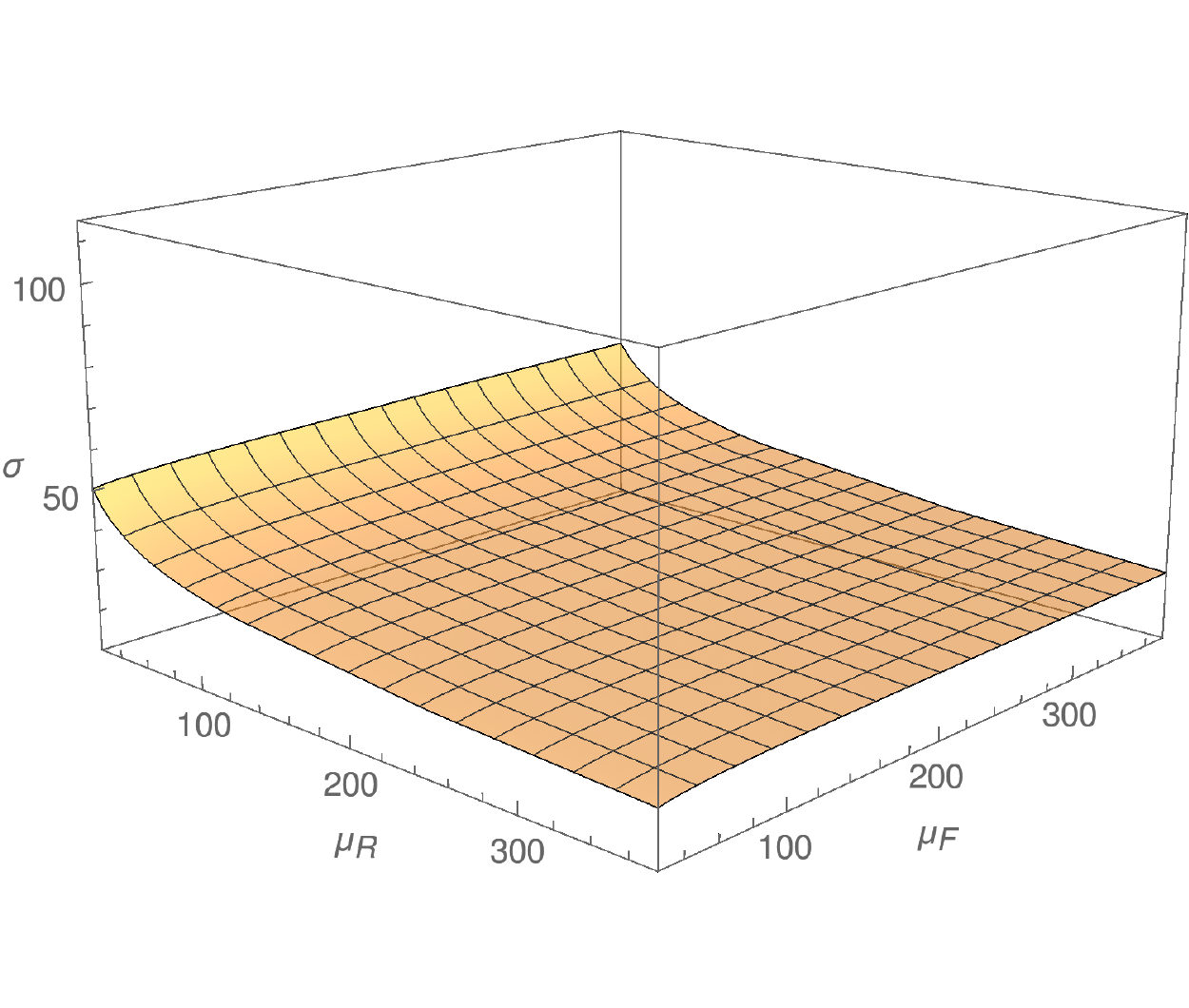}
\includegraphics[width=.48\textwidth]{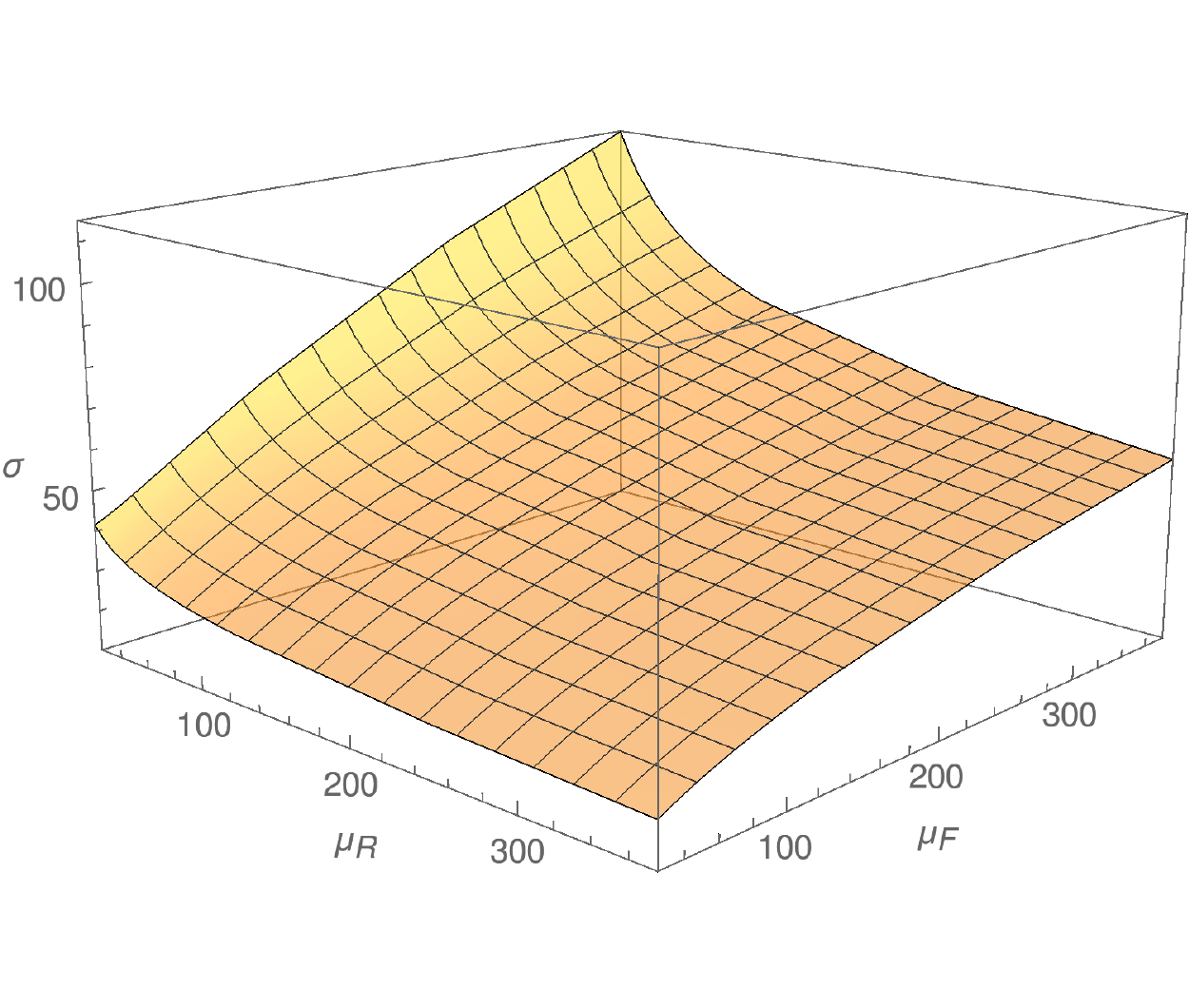}
\\(b)\\
%\end{center}
%\caption{\small{$\mu_{R}$ variation for $\mu_F = m_H/2$ for \textcolor{red}{N-Exp}(left panel) and for \textcolor{red}{$\overline{N}$-Exp}(right panel) }}
\label{fig5}
\end{figure}

\begin{figure}[ht]
%\begin{center}
\hspace*{-1.2cm}
\includegraphics[width=.48\textwidth]{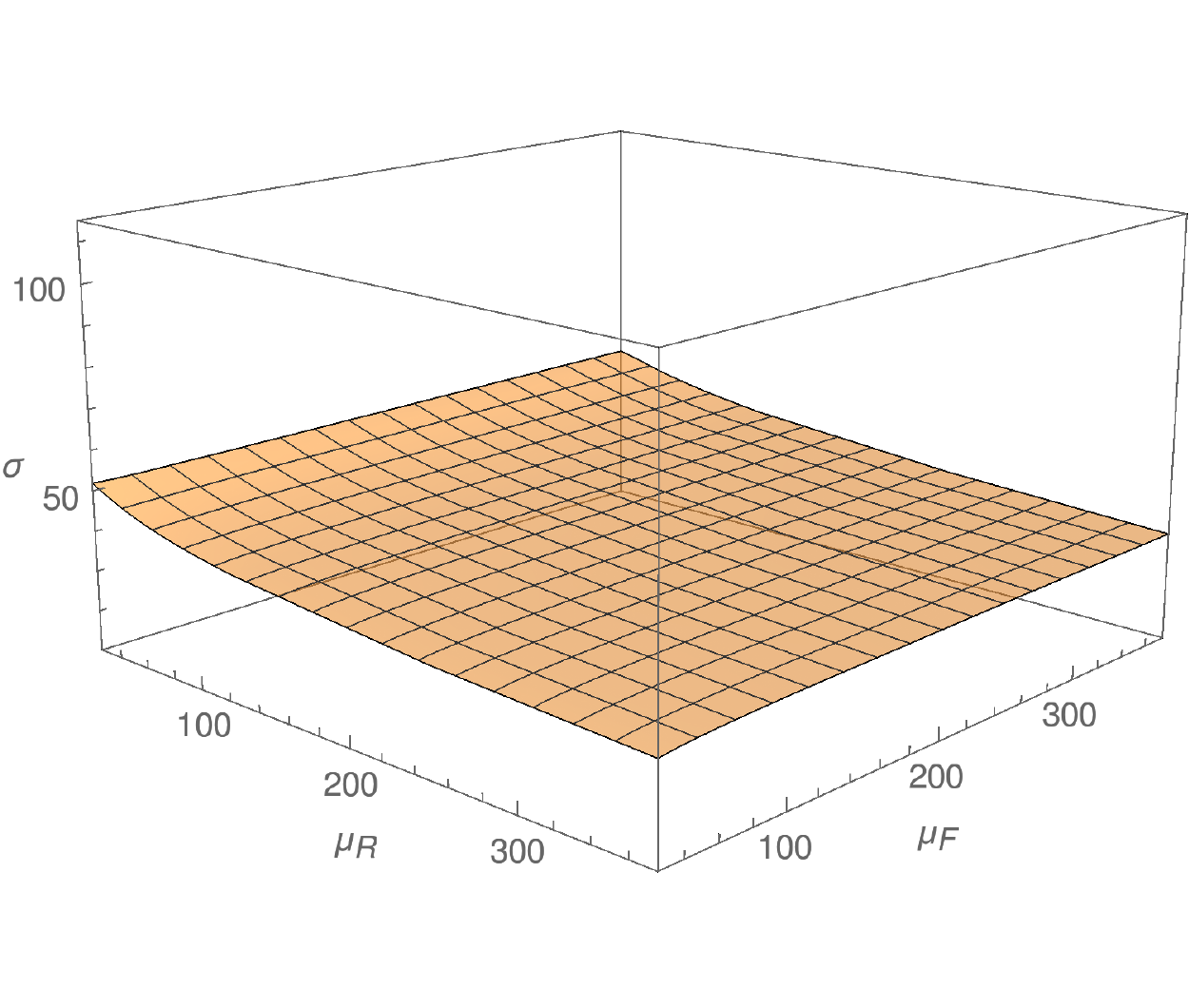}
\includegraphics[width=.48\textwidth]{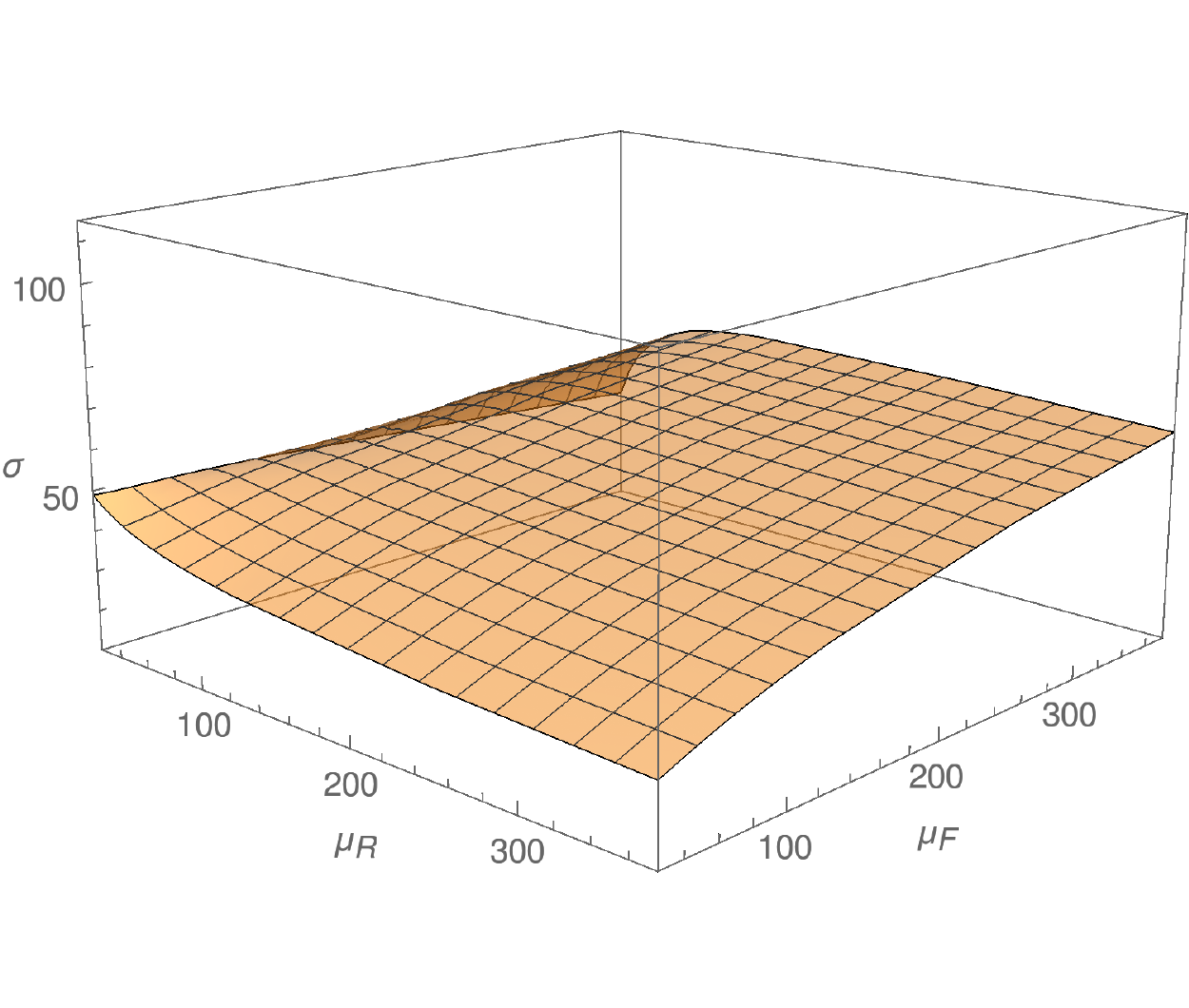}
\\(c)\\
%\end{center}
\caption{\small{The cross section $\sigma$ in picobarns[pb] w.r.t $\mu_R$ and $\mu_F$ for fixed order(left panel) vs resumed prediction(right panel) at (a)LO vs $\rm LO+\overline{LL}$, (b)NLO vs $\rm NLO+\overline{NLL}$ and (c) NNLO vs $\rm NNLO+\overline{NNLL}$ accuracy for $\sqrt{S} = 13$ TeV.}}
\label{fig1}
\end{figure}
On the left panel of subpanels (a), (b) and (c) of Fig.\ref{fig1}, the fixed order predictions for the cross section have been plotted at the accuracy $\rm LO$, $\rm NLO$ and $\rm NNLO$ respectively. The right panels show the $\rm SV+NSV$ resumed cross section matched to the fixed order at $\rm LO + \overline{\rm LL}$, $\rm NLO + \overline{\rm NLL}$ and $\rm NNLO + \overline{\rm NNLL}$ accuracy at (a), (b) and (c) subpanels respectively. The scales $\mu_R$ and $\mu_F$ are varied in the range $0.2 \le \{ \frac{\mu_R}{m_H},\frac{\mu_F}{m_H} \} \le 3$. The fixed order predictions contain contribution from all the channels while the resummed result consists of logarithms coming only from the diagonal $gg$ channel. There is a comprehensible reduction in the scale dependency of the cross section when we go from $\rm LO$ to $\rm NNLO$. Thus as expected, the scale dependence is reduced when the higher-order corrections are included in fixed order result. Whereas, from the resummed predictions on the right panel of Fig.\ref{fig1}, we find that the scale dependency increases in comparison to the corresponding fixed order predictions. However, if we consider the region in the vicinity of $\{ \mu_R, \mu_F \} = \{ \frac{m_H}{2},\frac{m_H}{2} \} $ and vary these scales in such a way that the ratio $\frac{\mu_R}{\mu_F}$ is not larger than 2 or smaller than 1/2, then we find that the scale dependency of the resummed result becomes comparable to the corresponding fixed order result at $\rm NNLO$ accuracy. The overall scale uncertainty using the canonical 7-point variation method, ranges between $(+8.92\%,-10.12\%)$ for $\rm NNLO$ whereas it is $(+ 11.90\%, - 8.32\%)$ for $\rm NNLO + \overline{NNLL}$ which validates our former statement. In addition to this, there is a significant enhancement of 4.76\% to the NLO cross section due to the addition of resummed result at $ \overline{\rm NLL}$ accuracy.
%\sout{in the cross section for the resummed result compared to its fixed order counterpart till $\rm NLO$ accuracy in the above specified region.
% \textcolor{red}{\sout{At $\rm NNLO$, however the perturbative series tends to converge with a decrease of $3.15 \%$ in cross section by the inclusion of $\rm NSV$ resummed logarithms. Hence it gives a stable prediction, justifying the truncation of the series at this order.}}
Similarly, the inclusion of the SV+NSV resummed result decreases the NNLO cross section by 3.15\% at $\{ \mu_R, \mu_F \} = \{ \frac{m_H}{2},\frac{m_H}{2} \} $ with an overall scale uncertainty comparable to the fixed order ones. There is also a better perturbative convergence of the resummed result at this particular choice of central scale in comparison to the corresponding fixed order results. Henceforth, in rest of the paper, because of the above arguments, we keep the central scales at  $\{ \mu_R, \mu_F \} = \{ \frac{m_H}{2},\frac{m_H}{2} \} $. We would also like to mention that the resummation scheme chosen for the rest of the paper is $\overline{N}$ exponentiation. The reasons for this particular choice of scheme will be discussed in the last section of this paper.

We now proceed to study the quantitative impact of the $\rm SV+NSV$ resummed corrections against the fixed order results using the K-factor, defined by
\begin{equation}
    \mathrm{K} = \frac{\sigma \big( \mu_R = \mu_F = m_H/2  \big)}{\sigma^{\rm{LO}} \big( \mu_R = \mu_F = m_H/2  \big)}\,.
\end{equation}
Below, in Table \ref{Tab:Table2}, we present K-factors of fixed order and resummed results. We have set renormalisation  and factorisation scales at $m_H/2 = 62.5$ GeV. We find that the cross section is enhanced by $78.96\%$ when the resummed $\rm SV+NSV$ logarithms are added at $\rm \overline{LL}$ accuracy to $\rm LO$ prediction and by $4.76 \%$ when $\rm \overline{NLL}$ is added to $\rm NLO$ prediction. The inclusion of $\rm \overline{NNLL}$ to $\rm NNLO$ however, decreases the cross section by $3.15 \%$. As discussed before, this hints to better perturbative convergence at $\rm NNLO + \overline{NNLL}$ level improving the reliability of perturbative predictions at this accuracy.

\begin{table}[htp] 
\begin{center}
\begin{small}
%\newcolumntype{P}[1]{>{\centering\arraybackslash}p{#1}}
{\renewcommand{\arraystretch}{1.7}
\begin{tabular}{|P{1.5cm} ||P{1.5cm}|p{1.2cm}||P{1.5cm}|P{1.5cm}|}
\rowcolor{lightgray}
     \multicolumn{1}{c|}{ $\rm LO+\overline{\rm LL}$}   
    &\multicolumn{1}{c||}{ $\rm NLO$ }  
    &\multicolumn{1}{c|}{$\rm NLO+\overline{\rm NLL}$}  
    &\multicolumn{1}{c||}{$\rm NNLO$ } 
    & \multicolumn{1}{c|}{$\rm NNLO+\overline{\rm NNLL}$ }\\
%    & \multicolumn{1}{c|}{NNLO$_{q\bar q}$+NNLL(sv+nsv)}\\
 \hline
%  \hline
 $1.7896$ & $1.6392$  & $1.7173$ & $1.9432$ & $1.882$ \\
\hline

\end{tabular}}
\caption{The $\mathrm{K}$-factor values for resummed result in comparison to the fixed order ones at central scale value $\mu_R = \mu_F = \frac{m_H}{2}$ at 13 TeV LHC.}
\label{Tab:Table2}
\end{small}
\end{center}
\end{table}

Few comments are in order from the above table. We notice that the percentage enhancement in the cross section at $\rm NNLO + \overline{NNLL}$ is $9.59 \%$ over $\rm NLO + \overline{NLL}$ whereas it is $18.55 \%$ when we go from $\rm NLO$ to $\rm NNLO$ accuracy. Therefore, inclusion of resummed $\rm SV+NSV $ logarithms improves the reliability of perturbative predictions. Also, the K-factor values are closer for $\rm NNLO$ and $\rm NNLO + \overline{NNLL}$ as compared to $\rm NLO$ and $\rm NLO + \overline{NLL}$. This suggests that the resummed contributions to the fixed order cross section decrease as we go to higher orders in perturbation theory.

From the above discussion on the K-factors, we have seen that the $\rm SV + NSV $ resummed contribution enhances the cross section as compared to the corresponding fixed order predictions till $\rm NLO$ accuracy. The 3D graph in Fig.\ref{fig1} shows the overall increase in the $\mu_R$ and $\mu_F$ scale variations when we add the resumed predictions till $\rm \overline{NLL}$ accuracy. However, at $\rm NNLO + \overline{NNLL}$ accuracy, the cross section decreases and the overall scale uncertainty gets closer to the corresponding fixed order prediction in the vicinity of $\{\mu_R,\mu_F\} = \{m_H/2,m_H/2\}$. This suggests that the truncation at NNLO improved by resummed contributions at $\overline{\rm NNLL}$ accuracy provides a reliable perturbative predictions.  In the following, we study the impact of  $\mu_R$ and $\mu_F$ scales separately by keeping one of them fixed. \\
\begin{figure}[ht]
%\begin{center}
\hspace*{-2cm}
\includegraphics[scale =0.4]{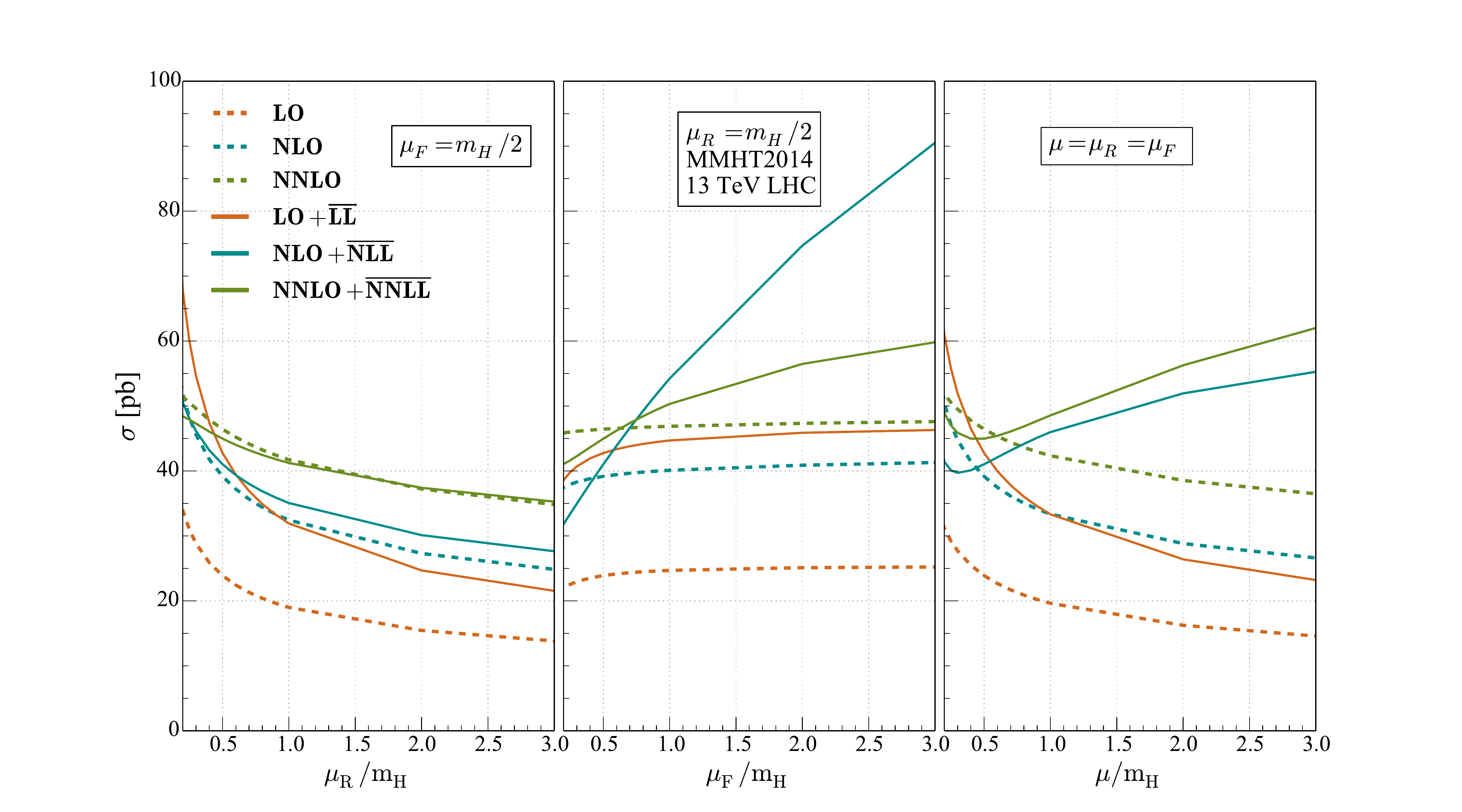}
%\end{center}
\caption{\small{Scale variation of the resummed results against the fixed order w.r.t the renormalisation scale($\mu_R$) in the first panel, factorisation scale($\mu_F$) in the second panel and $\mu_R = \mu_F = \mu$ in the third panel at 13 TeV LHC.}}
\label{fig2}
\end{figure}

We start by plotting the cross sections at various orders of fixed order as well as resumed result as a function of $\mu_R$ and $\mu_F$ in the range $(0.2, 3)m_H$ in Fig.\ref{fig2}. The quantitative values of uncertainty is calculated in the range $(1/4,1)m_H$ such that the ratio of $\frac{\mu_R}{\mu_F}$ is not greater than 2 or smaller than 1/2. In the first panel, the cross section is varied w.r.t $\mu_R$, keeping $\mu_F$ held fixed at $\frac{m_H}{2}$. The uncertainty ranges between $(+17.21\% , -14.55 \% )$ for $\rm NLO + \overline{NLL}$ whereas it is between $(+23.21\%,-17.18\% )$ for $\rm NLO$. Similarly, it ranges between $(+6.54\% ,-8.32 \%)$ for $\rm NNLO + \overline{NNLL}$ whereas for $\rm NNLO$, it is between $(+8.85\%,-10.12\% )$. Thus, we observe that $\mu_R$ uncertainty decreases when the $\rm SV+NSV $ resummed contributions are added to the fixed order predictions. Also, it decreases significantly when we go from $\rm NLO + \overline{NLL}$ level to $\rm NNLO + \overline{NNLL}$ level. 
%This suggests that $\mu_R$ uncertainty decreases on including higher order corrections and also, higher order resummed logarithms to the corresponding fixed order results. 
The maximum increments and minimum decrements in the cross section from the value at the central scale $\mu_R=\mu_F=\frac{m_H}{2}$ in the range $\mu_R = \{1/4,1\}m_H$ are given in Table \ref{Tab:Table3} for both fixed order as well as resummed results. From  Fig.\ref{fig1}, we observed that the scale uncertainty arising by varying both the scales simultaneously was larger for the resummed cross section compared to the corresponding fixed order results. However, here we notice that $\mu_R$ dependency is less for resummed cross sections indicating the role of $\mu_F$ in the scale uncertainty of Fig.\ref{fig1}.  Next, we study the variation of $\mu_F$ in fixed order as well as resummed results.

 \begin{table}[htp] 
\begin{center}
\begin{small}
%\newcolumntype{P}[1]{>{\centering\arraybackslash}p{#1}}
{\renewcommand{\arraystretch}{1.7}
\begin{tabular}{|P{1.8cm} |P{1.9cm}||p{1.8cm}|P{1.8cm}||P{1.8cm}|P{1.8cm}|}
\rowcolor{lightgray}
    \multicolumn{1}{c|}{ $\rm LO$}   
    &\multicolumn{1}{c||}{$\rm LO+\overline{\rm LL}$ }  
    &\multicolumn{1}{c|}{$\rm NLO$}  
    &\multicolumn{1}{c||}{$\rm NLO+\overline{\rm NLL}$} 
    
    & \multicolumn{1}{c|}{$\rm NNLO$}
    & \multicolumn{1}{c|}{$\rm NNLO+\overline{\rm {NNLL}}$}
    \\ 
 \hline
%  \hline
\hline
 $ 23.8940^{+7.08}_{-4.90}$ & $ 42.7612^{+17.27}_{-10.85} $
& $ 39.1681^{+9.09}_{-6.73} $ 
& $ 41.0325^{+7.06}_{-5.97} $
& $ 46.4304^{+4.11}_{-4.70} $ 
& $ 44.9685^{+2.94}_{-3.74} $    \\
\hline

\end{tabular}}
\caption{Values of resummed cross section (\textcolor{black}{in pb)} at 13 TeV LHC at various orders in comparison to the fixed order results at central scale value $\mu_R = \mu_F = \frac{m_H}{2}$. The maximum increment and decrements around central scale value are calculated by varying $\mu_R$ at fixed $\mu_F = \frac{m_H}{2}$.}
\label{Tab:Table3}
\end{small}
\end{center}
\end{table} 

The second panel of Fig.\ref{fig2} shows the dependence of cross section on factorisation scale $\mu_F$ keeping $\mu_R$ held fixed at $m_H/2$. We notice that the fixed order predictions till $\rm NNLO$ order have negligible dependence on $\mu_F$. This is also reflected from the maximum increments and minimum decrements from the central value (for $\mu_R=\mu_F=m_H/2$) for fixed order predictions given in Table \ref{Tab:Table4}. However, in the case of resummed cross section, Table \ref{Tab:Table4} shows significant increments and decrements around the central value. For instance, for the $\rm LO + \overline{LL}$, the uncertainty lies between $(+4.51\%,-6.92\%)$ while for $\rm NLO + \overline{NLL}$ it changes to $(+ 32.17\%,- 18.62\%)$. At $\rm NNLO + \overline{NNLL}$ accuracy, the $\mu_F$ uncertainty comes down to $(+11.90\%,-7.56\%)$.
We have seen previously in Fig.\ref{fig1} that the overall scale uncertainty due to renormalisation and factorisation scales was maximum for $\rm NLO + \overline{NLL}$ case. This suggests that the scale uncertainties for the resummed result is driven by $\mu_F$ variation. But we also know that $\mu_F$ variation entangles contribution arising from different channels. Hence,
the naive expectation is that the resummed predictions for the off-diagonal collinear logarithms may be accounted as the compensating factor for the $\mu_F$ dependency coming from diagonal channels.
%\textcolor{red}{\sout{This suggests that the uncertainty arising from the $\mu_F$ variations is the dominant factor in the case of resummed predictions. Since $\mu_F$ variation entangles the contributions coming from the off-diagonal channels, it is attributed to the fact that the resummed predictions lack off-diagonal channels which are important for compensating the $\mu_F$ dependence coming from diagonal channels \cite{Ajjath:2021lvg}.}} 
However, this can not account for $\mu_F$ uncertainty in the case of Higgs production through gluon fusion because  the off-diagonal channels have minuscule contributions at each order in perturbation theory.   We will interpret  this in the next section  by studying the behaviour of $\rm SV$ resummed results which contain only distributions and delta functions. \\

\begin{table}[H] 
\begin{center}
\begin{small}
%\newcolumntype{P}[1]{>{\centering\arraybackslash}p{#1}}
{\renewcommand{\arraystretch}{1.7}
\begin{tabular}{|P{1.8cm} |P{1.8cm}||p{1.8cm}|P{1.8cm}||P{1.8cm}|P{1.5cm}|}
\rowcolor{lightgray}
     \multicolumn{1}{c|}{ $\rm LO$}   
    &\multicolumn{1}{c||}{$\rm LO+\overline{\rm LL}$ }  
    &\multicolumn{1}{c|}{$\rm NLO$}  
    &\multicolumn{1}{c||}{$\rm NLO+\overline{\rm NLL}$} 
    
    & \multicolumn{1}{c|}{$\rm NNLO$}
    & \multicolumn{1}{c|}{$\rm NNLO+\overline{\rm {NNLL}}$}
    \\ 
 \hline
  $ 23.8940^{+0.80}_{-1.35} $ 
& $ 42.7612^{+1.93}_{-2.96} $
& $ 39.1681^{+0.93}_{-1.23} $ 
& $ 41.0325^{+13.20}_{-7.64} $ 
& $ 46.4304^{+0.44}_{-0.43} $ 
& $ 44.9685^{+5.35}_{-3.40} $    \\
\hline
%Drell-Yan &8.59\%&5.44\% & 9.82\% & 2.62\% & 1.49\%&-1.00\%\\
%\hline
\end{tabular}}
\caption{Values of resummed cross section (in pb) at various orders in comparison to the fixed order results at central scale value $\mu_R = \mu_F = \frac{m_H}{2}$. The maximum increment and decrements around central scale value are calculated by varying $\mu_F$ at fixed $\mu_R = \frac{m_H}{2}$}
\label{Tab:Table4}
\end{small}
\end{center}
\end{table}

In the third panel of Fig.\ref{fig2}, we vary $\mu_R$ and $\mu_F$ simultaneously by setting $\mu_R = \mu_F = \mu$. This plot basically is the diagonal line from the axis in the 3D graph of Fig.\ref{fig1} where $\mu_R = \mu_F$. We observe that for fixed order predictions, the overall uncertainty is dictated by the behavior of $\mu_R$ uncertainty. On the other hand, in the case of resummed predictions, it is the $\mu_F$ dependency which plays the major role. In Table \ref{Tab:Table5} given below, we have presented the values of cross section for fixed order as well as resummed predictions for central scale value $\mu_R = \mu_F = m_H/2$. \\

\begin{table}[H] 
\begin{center}
\begin{small}
%\newcolumntype{P}[1]{>{\centering\arraybackslash}p{#1}}
{\renewcommand{\arraystretch}{1.7}
\begin{tabular}{|P{1.8cm} |P{1.8cm}||p{1.8cm}|P{1.8cm}||P{1.8cm}|P{1.5cm}|}
\rowcolor{lightgray}
     \multicolumn{1}{c|}{ LO}   
    &\multicolumn{1}{c||}{LO+$\overline{\rm LL}$ }  
    &\multicolumn{1}{c|}{NLO}  
    &\multicolumn{1}{c||}{NLO+$\overline{\rm NLL}$} 
    
    & \multicolumn{1}{c|}{NNLO}
    & \multicolumn{1}{c|}{NNLO+$\overline{\rm {NNLL}}$}
    \\ 
 \hline
  $ 23.8940^{+5.33}_{-4.27}$ 
& $ 42.7612^{+13.03}_{-9.44} $ 
& $ 39.1681^{+7.99}_{-5.82}$ 
& $ 41.0325^{+4.95}_{-1.32} $ 
& $ 46.4304^{+4.13}_{-4.09}$ 
& $ 44.9685^{+3.58}_{-0.004} $    \\
\hline
%Drell-Yan &8.59\%&5.44\% & 9.82\% & 2.62\% & 1.49\%&-1.00\%\\
%\hline
\end{tabular}}
\caption{Values of resummed cross section (in pb) at various orders in comparison to the fixed order results at central scale value $\mu_R = \mu_F = \frac{m_H}{2}$. The maximum increment and decrements around central scale value are calculated by setting $\mu_R = \mu_F = \mu$ and varying $\mu$.}
\label{Tab:Table5}
\end{small}
\end{center}
\end{table}

In summary, we  studied the behaviour of resummed $\rm SV+ NSV$ logarithms w.r.t $\mu_R$ and $\mu_F$ scales and compared it against fixed order results. We found that in comparison to the fixed order the $\mu_R$ uncertainty decreases for the resummed results. But for the $\mu_F$ variation we found that the uncertainty is more at $\rm NLO + \overline{NLL}$ than at $\rm NNLO + \overline{NNLL}$ with larger factorization scale dependence compared to their fixed order counterparts. In addition, we compared the K-factors  of resummed results against the fixed order predictions to understand the perturbative convergence. In the next section we focus on the role of resummed $\rm NSV$ terms against $\rm SV$ resummed contribution. \\

\subsection{$\rm SV$ resummation vs $\rm SV+NSV$ resummation}

In the previous section, we compared the $\rm SV + NSV$ resummed results with the fixed order predictions. The K-factors in Table \ref{Tab:Table2} show that there is a significant enhancement till $\rm NLO+\overline{NLL}$ order. %\textcolor{red}{\sout{Although, at $\rm NNLO+\overline{NNLL}$ level accuracy, the perturbative corrections decrease slightly, which combined with the observation that around central scale value the overall $\mu_R$ and $\mu_F$ scale dependency of this level becomes more or less similar to the corresponding fixed order result, hints towards stability of the perturbative series. This further suggests that at $\rm NNLO+\overline{NNLL}$ level, the perturbative series becomes more reliable.}}
 We find that 
the contribution from resummed SV+NSV terms decreases the fixed order $\rm NNLO$ predictions when matched at 
$\rm NNLO+\overline{NNLL}$ level
leading to better perturbative convergence. We also observed that $\mu_R$ scale uncertainty gets improved at each order upon including the resummed corrections. On the other hand, uncertainty from $\mu_F$ scale  increases significantly by the inclusion of resummed cross section especially at $\rm NLO+\overline{NLL}$ accuracy. Now, in this section we will perform a detailed analysis on the inclusion of resummed NSV logarithms by comparing it with the SV resummed results. We will also try to reason out the behaviour of $\rm SV + NSV$ resummed results w.r.t $\mu_F$ and $\mu_R$ uncertainty variations.

\begin{table}[H] 
\begin{center}
\begin{small}
%\newcolumntype{P}[1]{>{\centering\arraybackslash}p{#1}}
{\renewcommand{\arraystretch}{1.7}
\begin{tabular}{|P{1.5cm}||P{1.5cm}|P{1.2cm}||P{1.5cm}|P{1.5cm}|}
\rowcolor{lightgray}
     \multicolumn{1}{c||}{ $\rm NLO+\rm NLL$ }  
    &\multicolumn{1}{c|}{$\rm NLO+\overline{\rm NLL}$}  
    &\multicolumn{1}{c||}{$\rm NNLO + \rm NNLL$} 
    & \multicolumn{1}{c|}{$\rm NNLO+\overline{\rm NNLL}$ }\\

 \hline

  $1.59$
& $1.7173$ 
& $1.89$
& $1.882$  \\
\hline

\end{tabular}}
\caption{The K-factor values for $\rm NSV$ resummed result in comparison to the $\rm SV$ resummed predictions at central scale value $\mu_R = \mu_F = \frac{m_H}{2}$. }
\label{Tab:Table6}
\end{small}
\end{center}
\end{table}

We begin with the K-factors for $\rm SV$ resummed results as well as for $\rm SV + NSV$ resumed results at $\rm NLO$ and $\rm NNLO$ accuracy given in Table \ref{Tab:Table6}. We find that the inclusion of resummed NSV logarithms enhances the cross section by $8.01 \%$ when we go from $\rm NLL$ to $\rm \overline{NLL}$ accuracy. Whereas, at $\rm NNLO + \overline{NNLL}$ the cross section decreases slightly by $0.42 \%$. This shows the reduction of resummed corrections while taking into account the higher logarithmic accuracy. We also observe that the K-factors  are closer for $\rm NLO + \overline{NLL}$ and $\rm NNLO + \overline{NNLL}$ as compared to $\rm NLO + NLL$ and $\rm NNLO + NNLL$. This suggests better perturbative convergence when $\rm NSV$ logarithms are taken into consideration. 

\begin{figure}[ht]
\begin{center}
\hspace*{-1cm}
\includegraphics[scale =0.4]{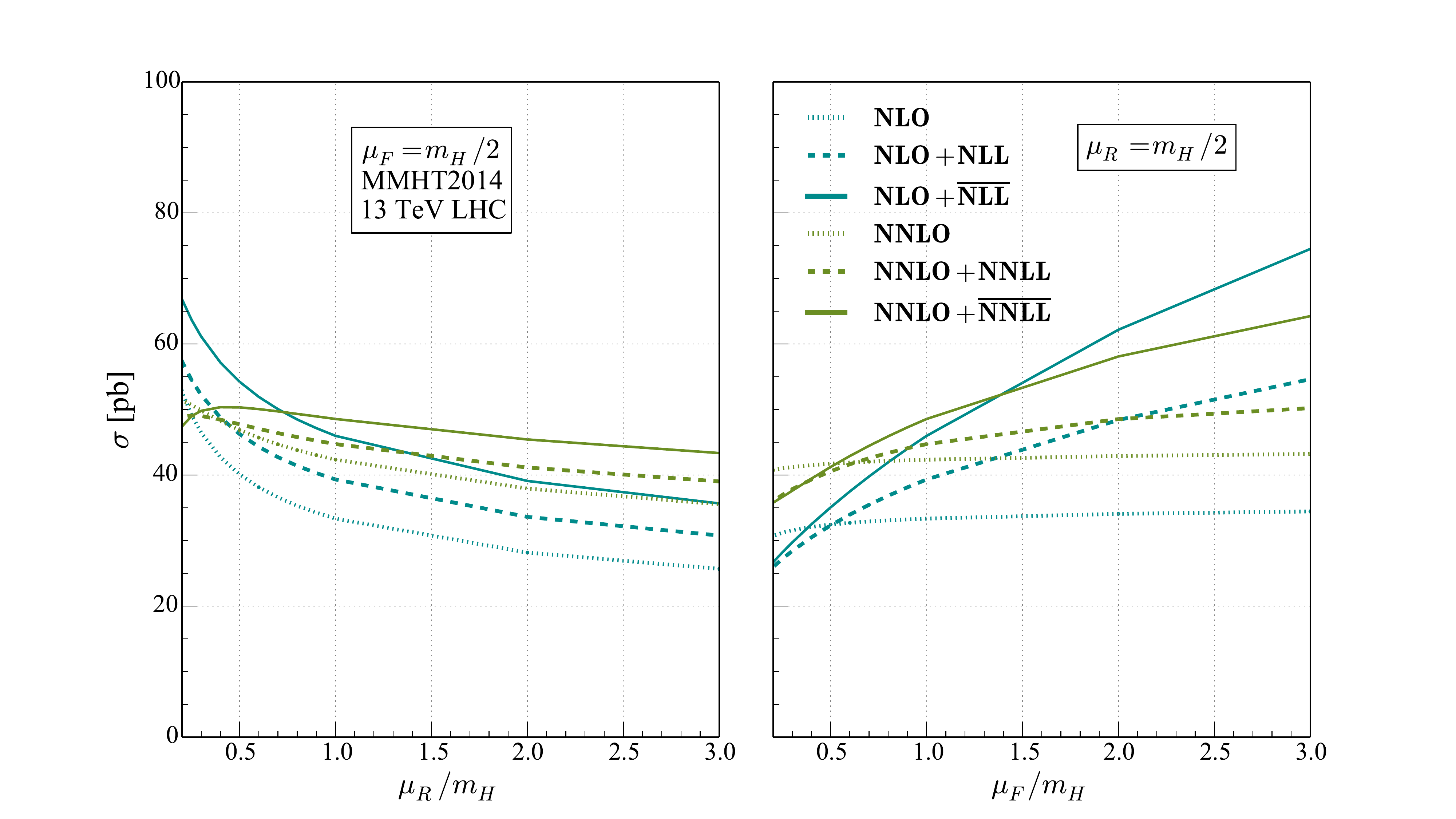}
\end{center}
\caption{\small{Scale variation of the $\rm SV$ and $\rm NSV$ resummed results against the fixed order w.r.t the renormalisation scale($\mu_R$) in the first panel and factorisation scale($\mu_F$) in the second panel at 13 TeV LHC.}}
\label{fig3}
\end{figure}

We now move on to study the scale uncertainties arising from the $\rm NSV$ logarithms. For this, we do a comparative study of $\mu_R$ and $\mu_F$ variations separately between fixed order, resummed $\rm SV$  and resummed $\rm SV + NSV$ results. In the first panel of Fig. \ref{fig3}, the cross section is plotted as a function of $\mu_R$ keeping $\mu_F = m_H/2$. The $\mu_R$ uncertainty for $\rm NLO$ lies in the range $(+23.21\% ,-17.18\%)$, for $\rm NLO + NLL$ between $(+18.57 \% ,-14.99 \%)$ and for $\rm NLO + \overline{NLL}$, it lies between $(+17.21\% ,-14.55\%)$. Similarly, at second order, the uncertainty lies between $(+8.85\% ,-10.12\%)$ for $\rm NNLO$, $(+9.58\% ,-10.02\%)$ for $\rm NNLO + NNLL$ and for $\rm NNLO + \overline{NNLL}$, it ranges between 
$(+6.54\% ,-8.32\%)$. We notice that at $\rm NLO$ accuracy, inclusion of resummed $\rm SV$ as well as resummed $\rm SV+NSV$ contributions improve the renormalisation scale uncertainty. %\textcolor{red}{ \sout{This is expected because the $\mu_R$ uncertainty gets cancelled when we add higher order terms through resummation of $\rm SV$ and $\rm NSV$ logarithms.}} 
Thus as expected, the inclusion of higher order logarithmic corrections within a particular channel  reduces the $\mu_R$ uncertainty. We also observe that the uncertainties at $\rm NLO + NLL$ and $\rm NLO + \overline{NLL}$ are comparable. For instance, the fixed order result at $\rm NLO$ consists of $73.16\%$ of $\rm SV$ contribution and $45.81\%$ of $\rm NSV$ contribution. Thus, $\rm NSV$ being the sub-dominant contributor does not allow much improvement in the uncertainty when resummed $\rm NSV$ logarithms are added to the resummed $\rm SV$ distributions. 
% {\color{red}\sout{That is why uncertainties at $\rm NLO + NLL$ and $\rm NLO + \overline{NLL}$ are more or less similar.}}
Now, at $\rm NNLO$ accuracy, the inclusion of resummed $\rm SV$ contribution does not improve the $\mu_R$ scale dependency significantly over the fixed order result.   
% \textcolor{red}{\sout {This is because the $\rm SV$ contribution to the $\rm NNLO$ cross section, arising from the distributions present at the two-loop is only $15.81\%$. Therefore, including higher order resummed $\rm SV$ terms won't make any substantial change to the uncertainty of fixed order result.}} 
{\color{black}However,  at the $\rm NNLO + \overline{NNLL}$, there is a comprehensible reduction in $\mu_R$ scale variation. We find that the SV distributions at two-loop contribute to $15.81\%$ of the Born cross section, while the NSV logarithms at the same order contribute to overall $58.91\%$ to the fixed order result at NNLO. This suggests a significant improvement in the $\mu_R$ scale variation while including the resummation effects of NSV logarithms}.
% {\color{red}\sout{This suggests that inclusion of resummed $\rm NSV$ logarithms improves the uncertainty at this level. \textcolor{red}{\sout{One of the reason for this is}} \textcolor{red}{This is also because of the fact} that $\rm NSV$ logarithms are the dominant contributor to the fixed order result at $\rm NNLO$ with an overall $58.91\%$ contribution  \textcolor{red}{coming from the two-loop NSV logarithms}. Thus, the higher order terms in the resummed $\rm NSV$ result improves the $\mu_R$ uncertainty arising from the fixed order NSV terms. }}

% \textcolor{red}{\sout{Also, we know that the resummed $\rm NSV$ logarithms give rise to spurious beyond $\rm NSV$ logarithms because of Mellin inversion from N-space to z-space.}}
% \textcolor{red}{In addition, we also know that owing to the ``inexact" Mellin inversion, any resummed result carries certain spurious terms.}
\textcolor{black}{From the above analysis we find that the resummed result leads to a reduction of the renormalization scale uncertainty. Let us now take a deeper look into this reduction. In general any resummed result carries informations of the all-order corrections arising from the quantity we are resumming. Further, owing to the ``inexact" Mellin inversion, it also carries certain spurious terms  which are beyond the precision of the resummed quantity.} For instance, the resummed NSV logarithms at $\rm NNLO+\overline{NNLL}$ accuracy incorporates to the fixed order results the all-order corrections arising from the summation of the next-to-next-to leading towers of NSV logarithms 
%{\color{red} \sout{to all orders }} 
and certain spurious terms which are beyond NSV. Now looking at the reduction of the scale uncertainty from NNLO to $\rm NNLO+NNLL$ to $\rm NNLO+\overline{NNLL}$, it is clear that the all-order corrections present in the resummed results leads to the improvement {\color{black} in} renormlization scale uncertainty. Now this reduction is more at $\rm NNLO+\overline{NNLL}$ than at $\rm NLO+\overline{NLL}$ owing to the significant contribution of the NSV logarithms at NNLO. 
%\textcolor{red}{\sout{In order to understand how these spurious contributions effect the uncertainty, we compare the $z$-space $\rm NSV$ contribution to the fixed order at $\rm NNLO$ with the Mellin inverted $\rm NSV$ contribution which will also contain beyond $\rm NSV$ logarithms. We find that the Mellin inverted z-space $\rm NSV$ contribution is less dependent on $\mu_R$ scale. Thus, the inclusion of beyond $\rm NSV$ terms(even though we have considered the spurious ones) improve the scale uncertainty. This suggests that if we could resum the beyond $\rm NSV$ terms present at higher orders then the $\mu_R$ scale uncertainty will show further betterment.}} 
In Table \ref{Tab:Table7} we list the values of resumed $\rm SV$ and $\rm SV+NSV$ cross sections as well as fixed order result till $\rm NNLO$  accuracy for $\mu_R = \mu_F = m_H/2$. The maximum increments and decrements are calculated from the central scale value in the range $\frac{1}{4} \le \frac{\mu_R}{m_H} \le 1$ keeping $\mu_F = m_H/2$.

\begin{table}[H] 
\begin{center}
\begin{small}
%\newcolumntype{P}[1]{>{\centering\arraybackslash}p{#1}}
{\renewcommand{\arraystretch}{1.7}
\begin{tabular}{|P{1.8cm}|P{1.8cm}||p{1.8cm}|P{1.8cm}||P{1.8cm}|P{1.5cm}|}
\rowcolor{lightgray}
     \multicolumn{1}{c|}{ NLO}   
    &\multicolumn{1}{c||}{NLO+$\rm NLL$ }  
    &\multicolumn{1}{c|}{NLO+$\overline{\rm NLL}$}  
    &\multicolumn{1}{c||}{NNLO} 
    
    & \multicolumn{1}{c|}{NNLO+$\rm NNLL$}
    & \multicolumn{1}{c|}{NNLO+$\overline{\rm {NNLL}}$}
    \\ 
 \hline

  $ 39.1681^{+9.09}_{-6.73} $
& $ 38.0142^{+7.06}_{-5.70} $
& $ 41.0325^{+7.06}_{-5.97} $ 
& $ 46.4304^{+4.11}_{-4.70} $  
& $ 45.0904^{+4.32}_{-4.52} $ 
& $ 44.9685^{+2.94}_{-3.74} $    \\
\hline

\end{tabular}}
\caption{Values of resummed $\rm SV + NSV$ cross section (in pb) at various orders in comparison to the fixed order results and resummed $\rm SV$ predictions at central scale value $\mu_R = \mu_F = \frac{m_H}{2}$.}
\label{Tab:Table7}
\end{small}
\end{center}
\end{table}

\begin{figure}[ht]
%\begin{center}
%\hspace*{-1cm}
\includegraphics[scale =0.55]{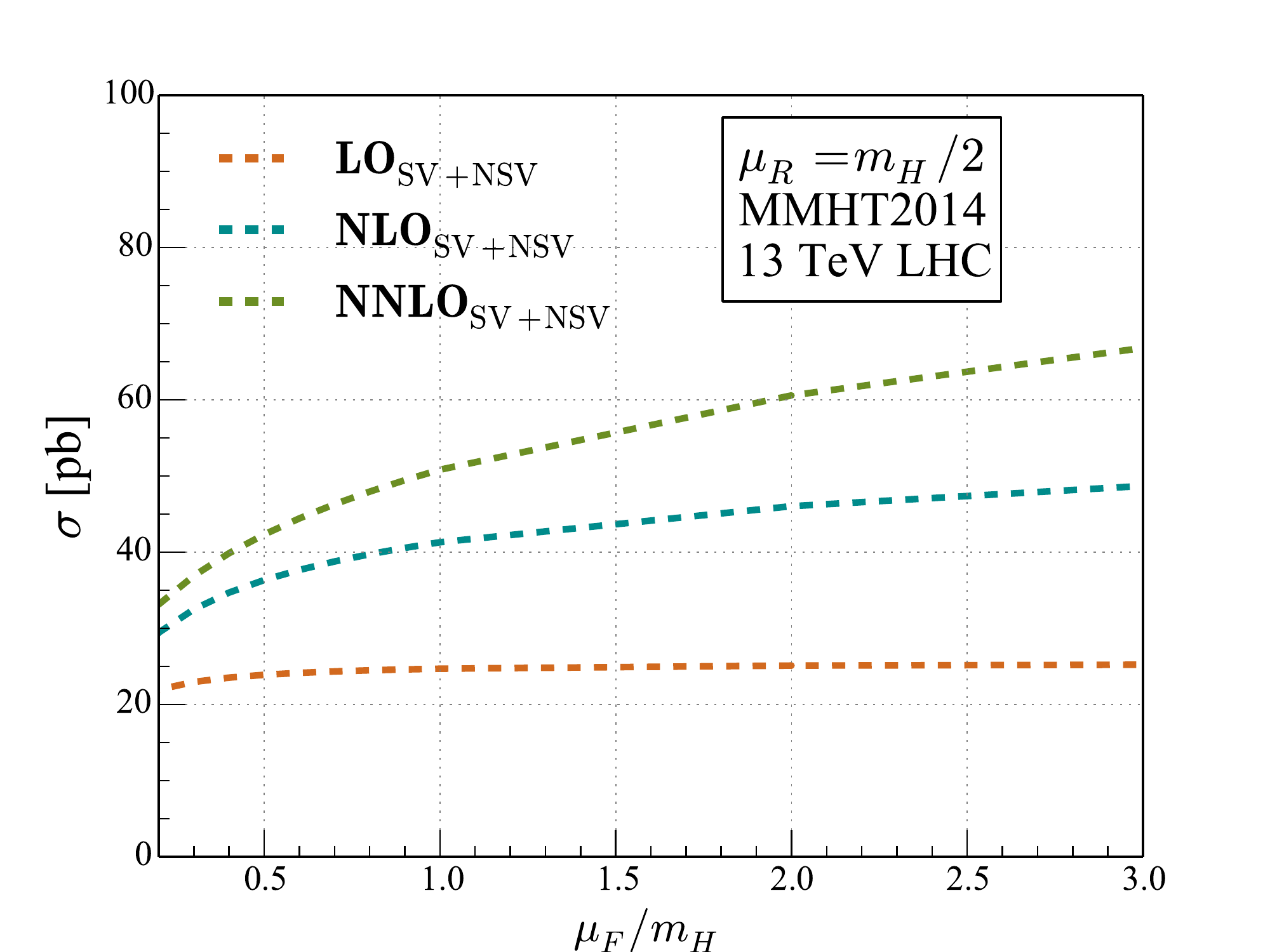}
%\end{center}
\caption{\small{$\mu_F$ scale dependence of $\rm SV + NSV$ part of the fixed order result till $\rm NNLO$ accuracy for Higgs boson production via gluon fusion at $\mu_R = m_H/2$ and $13$ TeV LHC.}}
\label{fig4}
\end{figure}

Having established the reasons for the behaviour of resummed $\rm NSV$ logarithms w.r.t $\mu_R$ variation, we proceed  to study the scale uncertainty w.r.t the factorisation scale $\mu_F$. The second panel in Fig.\ref{fig3}, shows the variation in cross section as a function of $\mu_F$ keeping $\mu_R = m_H/2$. 
%As seen earlier, the fixed order predictions are almost insensitive to $\mu_F$ variation. Let us try to understand the reason behind this.  In order to do that, 
Let us first consider the plot in Fig.\ref{fig4} where we depicted the fixed order result truncated to SV+NSV accuracy against $\mu_F$ variation. We find that there is a significant increase in scale dependency  when we go from $\rm NLO$ to $\rm NNLO$ accuracy.
Recall that the $\mu_F$ dependence in the fixed order predictions given in Fig. \ref{fig3} is extremely mild.
This suggests that the large $\mu_F$ dependence that we see in in Fig.\ref{fig4} is expected to get compensation from beyond NSV terms in the threshold expansion. This is because the fixed order analysis confirms the significant contributions arising from beyond SV terms with increase in the order of perturbative expansion \cite{Anastasiou:2014lda}.
%\textcolor{red}{\sout{In addition, the increasing trend of $\mu_F$  dependence as we increase the order from $\rm NLO$ to $\rm NNLO$  implies that the percentage contribution of beyond $\rm NSV$ terms increases with the increase in the order of perturbation theory.}} 

Our next observation is on the behaviour of resummed SV results against $\mu_F$ variation. We find that the uncertainties at $\rm NLO + NLL$ and $\rm NNLO + NNLL$ lie in the range $(+21.68\% ,-14.84\% )$ and $(+5.90\% ,-6.18\% )$ respectively. Interestingly, we observe  that the resummed $\rm SV$ cross section at $\rm NLO + NLL$  is more sensitive to $\mu_F$
compared to fixed order $\rm NLO$ results.   It is mainly due to the {\color{black}spurious terms that contributes beyond $\rm SV$,} arising from the ``inexact" Mellin inversion \textcolor{black}{of the $N$-space} resummed result.   Unlike $\rm NLO+NLL$,  at $\rm NNLO + NNLL$, the $\mu_F$ dependence  drops down  significantly but is still more than the fixed order counterpart. Hence the resummation of next-to-next-to leading SV distributions,  compensates the spurious terms of $\rm NLO+NLL$ with higher order logarithmic corrections. But again the reduction in the uncertainty will not be as much as compared to the fixed order due to the presence of the residual spurious terms arising from the Mellin inversion at $\rm NNLO + NNLL$ accuracy. %\textcolor{red}{\sout{This is because of relatively smaller contribution  from $\rm SV$ terms at NNLO level compared to NLO level, for example it goes down from $73.16\%$ at $\rm NLO$ to $15.81\%$ at $\rm NNLO$.  Hence, at second order, Mellin inversion of small resummed SV  will give  less spurious beyond $\rm SV$ terms that sensitive to $\mu_F$.}} 

Now let  us try to understand the behaviour of $\rm NSV$ logarithms based on our findings in fixed order and resummed $\rm SV$ predictions.
Recall that the resummed $\rm SV$ results suggest that spurious beyond $\rm SV$ terms  
spoil the $\mu_F$ scale dependence. Also, the behaviour of fixed order results w.r.t $\mu_F$ and the Fig.\ref{fig4} suggest that  $\mu_F$ dependence of both NSV and beyond $\rm NSV$ terms  increase with order of perturbative expansion and hence any truncated result will have more uncertainties than the fixed order corrections. 
%These indicate that the $\rm NSV$ terms are very much sensitive to factorisation scale.   
This explains the $\mu_F$ dependence of $\rm SV + NSV$ resummed prediction given in the second panel of Fig.\ref{fig3}. 
  
In addition, we also observe in Fig.\ref{fig3} that the $\mu_F$ dependency goes down from (+32.17\%, -18.61\%) at $\rm NLO + \overline{NLL}$ to (+11.89\%, -7.56\%) at $\rm NNLO + \overline{NNLL}$. If we compare the uncertainty of the SV resummed result, which was in the range of $(+21.68\% ,-14.84\% )$ at $\rm NLO + NLL$ we can conclude that the bulk of the uncertainty in SV+NSV resummation at this particular accuracy arises from the SV resummation. Now at the next logarithmic accuracy, the uncertainty coming from the SV resummation, which is $(+5.90\% ,-6.18\% )$ at $\rm NNLO + NNLL$, is comparable to the uncertainties arising from the resummation of the NSV logarithms. It has be noted that resummation of NSV logarithms was never supposed to compensate the uncertainties arising from the resummation of SV distributions. But owing to the ``inexact" Mellin Inversion, the spurious terms which are developed in the SV resummation at the NSV accuracy gets corrected through NSV resummation while beyond NSV spurious terms from both the distributions and logarithms will remain uncompensated and aid to the $\mu_F$ uncertainty. Also similar to the SV resummation, higher order logarithmic corrections from the NSV terms will compensate for the lower spurious ones thereby reducing the uncertainties even more.

In Table \ref{Tab:Table8}, we have provided the fixed order as well as resummed cross section values at the central scale for various perturbative orders. We have also given the maximum increment and decrements at each order around the central value by varying $\mu_F$ in the range $\{ \frac{1}{4}, 1 \}m_H$ keeping $\mu_R = m_H/2$ fixed. These are used to calculate the corresponding percentage uncertainties given above. \\

\begin{table}[ht] 
\begin{center}
\begin{small}
%\newcolumntype{P}[1]{>{\centering\arraybackslash}p{#1}}
{\renewcommand{\arraystretch}{1.7}
\begin{tabular}{|P{1.8cm} |P{1.8cm}||p{1.8cm}|P{1.8cm}||P{1.8cm}|P{1.5cm}|}
\rowcolor{lightgray}
     \multicolumn{1}{c|}{ $\rm NLO$}   
    &\multicolumn{1}{c||}{$\rm NLO+\rm NLL$ }  
    &\multicolumn{1}{c|}{$\rm NLO+\overline{\rm NLL}$}  
    &\multicolumn{1}{c||}{$\rm NNLO$} 
    
    & \multicolumn{1}{c|}{$\rm NNLO+\rm NNLL$}
    & \multicolumn{1}{c|}{$\rm NNLO+\overline{\rm {NNLL}}$}
    \\ 
 \hline
%  \hline
  $ 39.1681^{+0.93}_{-1.23} $ 
& $ 38.0142^{+8.24}_{-5.64} $
& $ 41.0325^{+13.20}_{-7.64} $  
& $ 46.4304^{+0.44}_{-0.43} $ 
& $ 45.0904^{+2.66}_{-2.79} $ 
& $ 44.9685^{+5.35}_{-3.40} $    \\
\hline

\end{tabular}}
\caption{Values of resummed $\rm SV + NSV$ cross section (in pb) at various orders in comparison to the fixed order results and resummed $\rm SV$ predictions at 13 TeV LHC at central scale value $\mu_R = \mu_F = \frac{m_H}{2}$. The uncertainties around the central scale value is calculated by varying $\mu_F$ keeping $\mu_R = m_H/2$ fixed.}
\label{Tab:Table8}
\end{small}
\end{center}
\end{table}
\textcolor{black}{In summary, we find that the resummation of NSV terms reduces the uncertainty resulting from $\mu_R$ when the logarithmic accuracy is improved from next-to-leading to next-to-next-to-leading order.  
% {\color{red}\sout{ In the case of $\mu_R$ we know that any change resulting from variation of $\mu_R$ at a given perturbative order will be compensated by the remaining higher order terms.
% In the resummation, we sum up certain SV as well as NSV logarithms to all orders along with their $\mu_R$ dependency and hence the presence of these $\mu_R$ dependent terms to all orders leads to the reduction in the $\mu_R$ uncertainty .}}
 {\color{black} As we know, any change resulting from the $\mu_R$ variation at a given perturbative order will be compensated by the remaining higher order terms. Hence the resummation naturally reduces the dependence on this unphysical scale, as  we sum up certain SV and NSV logarithms to all orders, which include higher order $\mu_R$ dependent terms as well}.
However, this is not case for the
%{\color{red} \sout{scale $\mu_F$ variation} }
$\mu_F$ sensitivity}. Unlike $\mu_R$ dependent terms, the $\mu_F$ dependent terms do not cancel between different perturbative orders.   On the contrary, we find that at each perturbative order SV, NSV and beyond NSV terms have different $\mu_F$ dependence. Hence truncation of higher order threshold terms, or equivalently approximation based on threshold expansion can lead to residual $\mu_F$ scale dependency at a particular order. This gets amplified further due to resummation where only certain SV and NSV logarithms are resummed to all orders due to approximation in the threshold expansion. This  also hints towards the importance of beyond $\rm NSV$ terms  for better accuracy of predictions at each level.    In the next section, we study the importance of $\rm SV + NSV $ resummed contributions to inclusive cross section at various collider energies.

%%%%%%%Surabhi's write-up
% In summary, we find that along with $\rm NSV$, beyond $\rm NSV$ contributions also increase with the increase in the order of perturbation theory. Similar to $\rm NSV$ logarithms, the inclusion of beyond $\rm NSV$ logarithms can improve the renormalisation scale uncertainty at each order. The $\rm NSV$ logarithms however, spoils the scale uncertainty w.r.t the factorisation scale. At fixed order, we saw that the significant variation brought in by $\rm NSV$ logarithms is well compensated by beyond $\rm NSV$ logarithms. Similar effect is observed in resummed case as well. This  hints towards the importance of beyond $\rm NSV$ terms as well for better accuracy of predictions at each level.    In the next section, we study the importance of $\rm SV + NSV $ resummed contributions to inclusive cross section at various collider energies. \\
%%%%%%%%%%%%%

\subsection{Numerical results for different collider energies}

From  Fig.\ref{fig1}, we deduced that the appropriate central scale for the analysis would be $\mu_R = \mu_F = \frac{m_H}{2}$. In literature, one finds the choice $\mu_R = \mu_F = m_H$  as well to be the central scale. We will do a short analysis in this section to justify our choice of central scale. For this, we will vary the cross section as a function of hadronic centre-of-mass energy. It will also help us to understand the behaviour of $\rm SV+NSV$ resummed cross section at different collider energies. \\

\begin{figure}[ht]
%\begin{center}
\hspace*{-1.4 cm}
\includegraphics[scale =0.35]{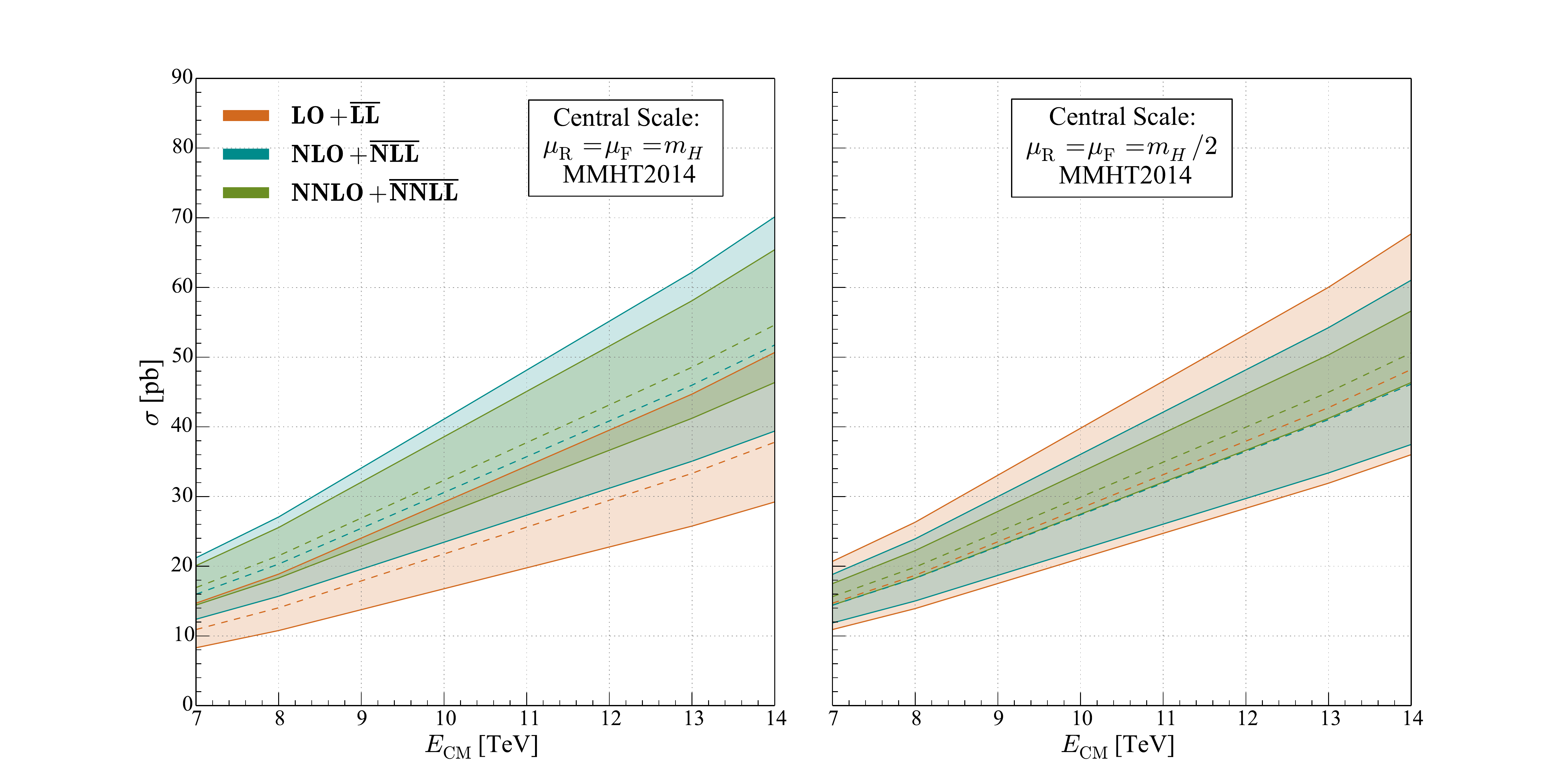}
%\end{center}
\caption{\small{$\rm SV + NSV$ resummed cross sections plotted against the hadronic centre-of-mass energy($\rm E_{CM}$) till $\rm NNLO + \overline{NNLL}$ accuracy. The band corresponds to the scale variation around the central scale choices $\mu_R = \mu_F = m_H$ and $\mu_R = \mu_F = m_H/2$ for the left and the right panel respectively. The dotted line represents the value of the cross section at the central scale.}}
\label{fig5}
\end{figure}

In Fig.\ref{fig5}, we plot the $\rm SV +NSV$ resummed cross section against the hadronic centre-of-mass energy from $7$ TeV to $14$ TeV. The bands in the plot correspond to the scale variations obtained by using the canonical 7-point variation where $\{ \mu_R,\mu_F \}$ is varied in the range $\{ \frac{m_x}{2} , 2 m_x \}$, keeping the ratio $\frac{\mu_R}{\mu_F}$ not larger than 2 and smaller than 1/2. For the left panel, $m_x = m_H$ and for the right panel, $m_x = m_H/2$. At 13 TeV LHC, the enhancement in the cross section from the central scale $\mu_R = \mu_F = m_H$ values while going from $\rm LO+\overline{LL}$ to $\rm NLO+\overline{NLL}$ is $37.99 \%$ and it is $5.60 \%$ from $\rm NLO+\overline{NLL}$ to $\rm NNLO+\overline{NNLL}$. The large increment in the cross section for $\rm NLO+\overline{NLL}$ over $\rm LO+\overline{LL}$ hints towards inclusion of more and more higher order terms and therefore results in loss of credibility of the perturbative prediction. If we look at these enhancements for central scale $\mu_R = \mu_F = m_H/2$ values, it is $-4.04 \%$ while going from $\rm LO+\overline{LL}$ to $\rm NLO+\overline{NLL}$ and $9.59 \%$ for $\rm NLO+\overline{NLL}$ to $\rm NNLO+\overline{NNLL}$. Also, we observe that for the plot corresponding to the central scale $\mu_R = \mu_F = m_H/2$,  the $\rm NLO+\overline{NLL}$ scale uncertainty band is included within the $\rm LO+\overline{LL}$ band and the $\rm NNLO+\overline{NNLL}$ scale uncertainty band is included within that of $\rm NLO+\overline{NLL}$. However, this is not the case with the plot corresponding to central scale $\mu_R = \mu_F = m_H$. Thus, the above two observations regarding the enhancement and the uncertainty band plot indicate that the perturbative expansion of the hadronic cross section is more convergent and therefore reliable for the right panel. 

From the above plots, we also observe that the uncertainty due to renormalisation and factorisation scales increases with the energy of the collider for both the choices of central scales. The reason for this large uncertainty at high $\rm E_{CM}$ could be the lack of knowledge of the $\rm PDF$ sets at these energies. 

Let us now look at this theoretical uncertainty related to the choice of central scale for $\rm SV+NSV$ resummed cross section by comparing it with the fixed order and $\rm SV$ resummed results. Tables \ref{Tab:Table9} and \ref{Tab:Table10} present the cross sections and related scale uncertainty for fixed order, $\rm SV$ and $\rm SV+NSV$ resummed results at $\rm NNLO$ accuracy for different collider energies for the central scales $\mu_R=\mu_F=m_H$ and $\mu_R=\mu_F=m_H/2$ respectively. The scale uncertainty in the tables has been calculated by varying the renormalisation and factorisation scales and using the canonical 7-point variation approach.
%\begin{center}
%The table \ref{table:1}
\begin{table}[h!]
\begin{small}
\centering
{\renewcommand{\arraystretch}{1.6}
 \begin{tabular}{||P{1.3cm}||P{3.3cm}|P{3.3cm}|P{3.7cm} ||} 
 \hline
 $\sqrt{S}$ & NNLO (pb) & NNLO+NNLL (pb)  & $\rm NNLO+\overline{NNLL}$ (pb)  \\ [1.2ex] 
 \hline\hline 
 
 7 TeV   & ${14.5570}^{+1.60}_{-1.54}$   
 &  ${15.4870}^{+1.20}_{-1.36}$  
 &  ${16.9331}^{+3.15}_{-2.51}$  \\ [1.2ex]
 
 8 TeV   &  ${18.5517}^{+2.03}_{-1.95}$  
 &  ${19.7026}^{+1.56}_{-1.76}$ 
 &  ${21.5068}^{+4.05}_{-3.20}$  \\   [1.2ex] 
 
 13 TeV   & ${42.3392}^{+4.53}_{-4.41}$   
 &  ${44.7214}^{+3.81}_{-4.15}$  
 &  ${48.5533}^{+9.54}_{-7.33}$  \\ [1.2ex] 
 
 14 TeV   &  ${47.6973}^{+5.09}_{-4.96}$  
 &  ${50.3438}^{+4.33}_{-4.70}$ 
 &  ${54.6158}^{+10.80}_{-8.26}$  \\   [1.2ex]  
 
% 100 TeV  &  ${714.3105}^{+74.13}_{-74.05}$  
% &  ${743.3359}^{+76.75}_{-79.88}$ 
% &  ${793.4355}^{+175.16}_{-126.01}$  \\   [1.2ex]
 
 \hline
\end{tabular}}
\caption{Fixed-order + Resummed result at different centre of mass energies at the LHC. The scale uncertainty has been estimated using 7-point scale variation around the central scale  $\mu_R = \mu_F = m_H (125 ~\rm GeV)$.}
\label{Tab:Table9}
\end{small}
\end{table}
%\end{center}

We start by calculating the quantitative effect of including the resummed $\rm SV$ and $\rm SV+ NSV$ logarithms to the fixed order prediction at the above mentioned central scales. We find that at 13 TeV LHC, the inclusion of resummed $\rm SV + NSV$ result increases the $\rm NNLO$ prediction by $14.68 \%$ for the central scale $\mu_R = \mu_F = m_H$ whereas for central scale $\mu_R = \mu_F = m_H/2$, the $\rm NNLO$ prediction decreases by $3.15 \%$. Similarly, inclusion of only resumed $\rm SV$ logarithms enhances the $\rm NNLO$ cross section by $5.6 \%$ for central scale $\mu_R = \mu_F = m_H$ while it decreases the cross section by $2.89 \%$ for central scale $\mu_R = \mu_F = m_H/2$. Similar trend follows at all other values of collider energies. The comparatively smaller difference between resumed result and fixed order result at $\rm NNLO$ accuracy makes the prediction at the central scale $\mu_R = \mu_F =m_H/2$ much more reliable. 

%\newpage
%\begin{center}
%The table \ref{table:1}
\begin{table}[h!]
\begin{small}
\centering
{\renewcommand{\arraystretch}{1.6}
 \begin{tabular}{||P{1.3cm}||P{3.3cm}|P{3.3cm}|P{3.7cm} ||} 
 \hline
 $\sqrt{S}$ & NNLO (pb) & NNLO+NNLL (pb)  & $\rm NNLO+\overline{NNLL}$ (pb)  \\ [1.2ex] 
 \hline\hline 
 
 7 TeV   & ${16.1185}^{+1.43}_{-1.60}$   
 &  ${15.6267}^{+1.43}_{-1.50}$  
 &  ${15.6390}^{+1.87}_{-1.21}$  \\   [1.2ex] 
 
 8 TeV   & ${20.4934}^{+1.82}_{-2.04}$  
 &  ${19.8753}^{+1.84}_{-1.93}$ 
 &  ${19.8735}^{+2.37}_{-1.57}$  \\   [1.2ex] 
 
 13 TeV   & ${46.4304}^{+4.14}_{-4.70}$  
 &  ${45.0904}^{+4.32}_{-4.52}$  
 &  ${44.9685}^{+5.35}_{-3.74}$  \\   [1.2ex] 
 
 14 TeV   & ${52.2523}^{+4.66}_{-5.31}$  
 &  ${50.7547}^{+4.89}_{-5.11}$ 
 &  ${50.6006}^{+6.02}_{-4.24}$  \\   [1.2ex] 
 
% 100 TeV  & ${767.0335}^{+80.02}_{-86.37}$  
% &  ${749.0435}^{+85.20}_{-85.59}$ 
% &  ${743.1405}^{+86.62}_{-75.72}$  \\   [1.2ex]
 
 \hline
\end{tabular}}
\caption{Fixed-order+ Resummed gluon fusion cross section at different centre of mass energies at the LHC. The scale uncertainty has been estimated using 7-point scale variation around the central scale  $\mu_R = \mu_F = m_H/2$.}
\label{Tab:Table10}
\end{small}
\end{table}
%\end{center}

Next, we observe that the scale dependence due to $\mu_R$ and $\mu_F$ is less in the case of fixed order as well as $\rm SV$ and $\rm SV+NSV$ resumed results if the central scale is $\mu_R = \mu_F = m_H/2$. Though the uncertainties don't differ comprehensibly for the fixed order and $\rm SV$ resumed results but the difference is significant in the case of $\rm SV + NSV$ resumed result between the two different choices of central scales. At 13 TeV LHC, for the central scale $\mu_R = \mu_F = m_H$, the uncertainty for $\rm NNLO + \overline{NNLL}$ accuracy lies between $+ 19.65\% , - 15.10\%$ whereas this goes down significantly to $+ 11.90\%, - 8.32\%$ for the central scale $\mu_R = \mu_F = m_H/2$. 

\begin{figure}[ht!]
%\begin{center}
\hspace*{-1.4 cm}
\includegraphics[scale =0.35]{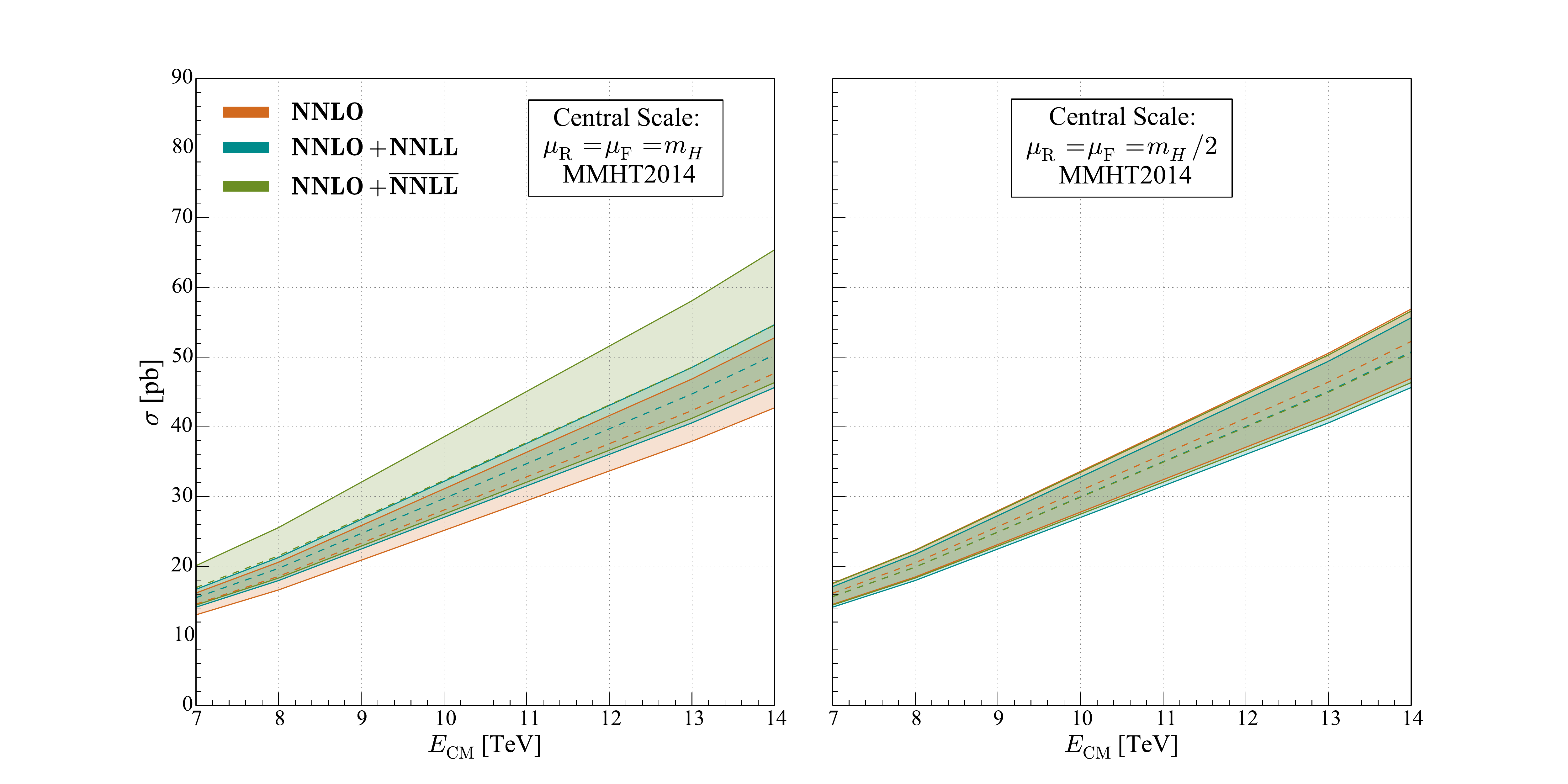}
%\end{center}
\caption{\small{Fixed order, $\rm SV$ resummed and $\rm SV + NSV$ resummed cross sections plotted against the hadronic centre-of-mass energy($\rm E_{CM}$) at second order in perturbation theory. The band corresponds to the scale variation around the central scale choices $\mu_R = \mu_F = m_H$ and $\mu_R = \mu_F = m_H/2$ for the left and the right panel respectively. The dotted line represents the value of the cross section at the central scale.}}
\label{fig6}
\end{figure}

From the tables(\ref{Tab:Table9},\ref{Tab:Table10}) given above, we can also see that the relations between fixed-order, $\rm SV$ and 
$\rm SV + NSV$ resummed results and their uncertainty bands are similar for all the collider energies considered in the table. For better understanding, in Fig.\ref{fig6}, we present the plots for fixed-order, $\rm SV$ and $\rm SV+NSV$ resumed cross sections at $\rm NNLO$ accuracy, at various collider energies for both the central scale choices.

Having established the reasons for the choice of central scale $\mu_R = \mu_F = m_H/2$, we now proceed to the last part of our numerical analysis where we will see the effect of $\rm SV + NSV$ resummation in different schemes namely, $N$ exponentiation, $\overline{N}$ exponentiation, \textit{all exponentiation} and \textit{soft exponentiation}. 

\subsection{$\rm SV+NSV$ resummation in different schemes}
In this section, we will explore why we chose $\overline{N}$ exponentiation to do the numerical analysis in the previous sections. We will also do the comparative study between $N$ exponentiation, $\overline{N}$ exponentiation, \textit{all exponentiation} and \textit{soft exponentiation} to see how the enhancements in the cross section at different orders and scale uncertainty gets modified due to $\rm SV + NSV$ resummation across these schemes. 

We start by studying the numerical impact of $N$ and $\overline{N}$ exponentiation. As we know, for the case of $\overline{N}$ exponentiation, we resum $\gamma_E$ terms along with $\ln N$ whereas for $N$ exponentiation we only resum the $\ln N$ terms. We have already done detailed analysis of $\rm SV+NSV$ resummation effects in $\overline{N}$ exponentiation in the previous sections. Let us now see how the behaviour changes by the inclusion of resummed $\rm SV+NSV$ results in the $N$ exponentiation. 

\begin{table}[H] 
\begin{center}
\begin{small}
%\newcolumntype{P}[1]{>{\centering\arraybackslash}p{#1}}
{\renewcommand{\arraystretch}{1.7}
\begin{tabular}{|p{2.5cm}||P{2.5cm} ||P{2.5cm}|P{2.5cm}|}
\rowcolor{lightgray}
\multicolumn{1}{c||}{Scheme}
    & \multicolumn{1}{c|}{ $\rm LO+\overline{\rm LL}$}  
    &\multicolumn{1}{c|}{$\rm NLO+\overline{\rm NLL}$} 
    & \multicolumn{1}{c|}{$\rm NNLO+\overline{\rm NNLL}$ }\\
%    & \multicolumn{1}{c|}{NNLO$_{q\bar q}$+NNLL(sv+nsv)}\\
 \hline

$N$-Exp & 1.2067 & 1.7629 &  1.9483  \\
\hline

$\overline{N}$-Exp & 1.7896 &  1.7173 &  1.882 \\
\hline

%Drell-Yan &8.59\%&5.44\% & 9.82\% & 2.62\% & 1.49\%&-1.00\%\\
%\hline
\end{tabular}}
\caption{The K-factor values for resummed result for $N$ and $\overline{N}$ exponentiation at central scale $\mu_R = \mu_F = m_H/2$ for $13$ TeV LHC.}
\label{Tab:Table11}
\end{small}
\end{center}
\end{table}

For this, we first compare the K-factor values for these two schemes at various orders given in Table \ref{Tab:Table11}. We observe that for $N$ exponentiation, there is an increment of $46.09\%$ for $\rm NLO + \overline{NLL}$ cross section over $\rm LO + \overline{LL}$ and $10.51\%$ while going from $\rm NLO + \overline{NLL}$ accuracy to $\rm NNLO + \overline{NNLL}$. On the other hand, for $\overline{N}$ exponentiation as we have seen earlier, there is a reduction of $4.04\%$ in the cross section at $\rm NLO + \overline{NLL}$ compared to $\rm LO + \overline{LL}$ and an enhancement of $9.59\%$ while going from $\rm NLO + \overline{NLL}$ accuracy to $\rm NNLO + \overline{NNLL}$. These percentages show that the perturbative convergence for $\rm SV + NSV$ resummed result in $\overline{N}$ exponentiation is better than $N$ exponentiation. 

Next, we look at the uncertainty due to renormalisation($\mu_R$) and factorisation($\mu_F$) scales. From Table \ref{Tab:Table12}, we find that the uncertainty calculated using 7-point canonical approach by varying $\mu_R$ and $\mu_F$ is less for $N$ exponentiation compared to $\overline{N}$ exponentiation till $\rm NLO + \overline{NLL}$ accuracy. However, at $\rm NNLO + \overline{NNLL}$ accuracy, the uncertainty of $\rm SV + NSV$ resummed result for $N$ exponentiation(lies between $\{+9.19\%,-8.68\%\}$) becomes comparable to $\overline{N}$ exponentiation(lies between $\{+11.90\%,-8.32\%\}$). Hence, the good perturbative convergence is achieved by the resummed result in  $\overline{N}$ exponentiation which improves the reliability of the perturbative predictions and it becomes the deciding factor in choosing $\overline{N}$ exponentiation over $N$ exponentiation for numerical analysis in the previous sections. \\

\begin{table}[H] 
\begin{center}
\begin{small}
%\newcolumntype{P}[1]{>{\centering\arraybackslash}p{#1}}
{\renewcommand{\arraystretch}{1.7}
\begin{tabular}{|p{2.5cm}||P{2.5cm} ||P{2.5cm}||P{2.5cm}||P{2.5cm}|}
\rowcolor{lightgray}
\multicolumn{1}{c||}{Order}
    & \multicolumn{1}{c|}{ $N-Exp$}   
    &\multicolumn{1}{c|}{$\overline{N}-Exp$}  
    &\multicolumn{1}{c||}{$All-Exp$} 
    & \multicolumn{1}{c|}{$Soft-Exp$ }\\
%    & \multicolumn{1}{c|}{NNLO$_{q\bar q}$+NNLL(sv+nsv)}\\
 \hline

$\rm LO+\overline{LL}$ & $ 28.8333^{+9.44}_{-6.34}$ & $42.7612^{+17.27}_{-10.85}$ &  $42.7612^{+17.27}_{-10.85}$ & $ 42.7612^{+17.27}_{-10.85}$ \\
\hline

$\rm NLO+\overline{NLL}$ & $ 42.1234^{+9.24}_{-7.05}$ & $ 41.0325^{+13.20}_{-7.64}$ & $53.4675^{+12.60}_{-8.62}$ & $ 44.9558^{+14.43}_{-8.82} $ \\
\hline

$\rm NNLO+\overline{NNLL}$ & $ 46.5527^{+4.28}_{-4.04}$ & $ 44.9685^{+5.35}_{-3.74}$ & $47.4358^{+5.41}_{-2.02} $ & $ 46.5944^{+5.10}_{-4.67} $ \\
\hline

%Drell-Yan &8.59\%&5.44\% & 9.82\% & 2.62\% & 1.49\%&-1.00\%\\
%\hline
\end{tabular}}
\caption{A comparison of $\rm SV + NSV$ resummed cross sections (in pb) between different resummation schemes at central scale $\mu_R = \mu_F = m_H/2$ for $13$ TeV LHC.}
\label{Tab:Table12}
\end{small}
\end{center}
\end{table}

In the end we discuss on numerical impact of $\rm SV + NSV$ resummed effects by giving a short analysis on the other two schemes namely, \textit{All exponentiation} and \textit{Soft exponentiation}. In \textit{all exponentiation} scheme, we exponentiate the Mellin moment of complete soft-collinear function along with the form factor contributions whereas in \textit{soft exponentiation} scheme, we only exponentiate the soft-collinear part. We find that the $\rm SV + NSV$ resummed result in \textit{soft exponentiation} is closest to our best prediction given in $\overline{N}$ exponentiation. From Table \ref{Tab:Table12}, we see that the maximum increments and decrements around the central scale value is comparable for \textit{soft exponentiation} and $\overline{N}$ exponentiation schemes till $\rm NNLO + \overline{NNLL}$ accuracy. Also, the enhancement while going from $\rm NLO + \overline{NLL}$ to $\rm NNLO + \overline{NNLL}$ accuracy is $3.64 \%$ in soft exponentiation scheme which is close to $9.59\%$ increase in the case of $\overline{N}$ exponentiation scheme as compared to other schemes. We observe a striking feature in the case of \textit{all exponentiation} scheme. There is an increase in cross section at $\rm NNLO + \overline{NNLL}$ compared to $\rm NLO + \overline{NLL}$ accuracy in all the schemes except \textit{all exponentiation}. In this case, the cross section decreases by $11.28\%$ at $\rm NNLO + \overline{NNLL}$ accuracy. Also, the scale uncertainties coming from $\mu_R$ and $\mu_F$ variations are found to be least in \textit{all exponentiation} scheme at $\rm NLO + \overline{NLL}$ and $\rm NNLO + \overline{NNLL}$ level with latter lying between ($+11.40\%,-4.26\%$).

\section{Conclusions} \label{concl}
Extensive study on the Higgs bosons produced at the LHC has been going on both in experimental
as well as in the theoretical fronts.  They provide not only stringent tests of the SM but also
constraints on models beyond SM.  Variety of observables involving Higgs bosons
measured with unprecedented accuracy have been compared against state-of-the art
predictions.  One such observable is the inclusive cross section of single Higgs boson production
and the predictions are known in perturbative QCD to N$^3$LO accuracy in fixed order and to
N$^3$LL in resummed framework.  The electroweak corrections at NNLO further improves the
predictions.  These estimates computed using effective theory approach
are mildly improved when exact mass dependence is included.    

Further, partial results beyond third order for SV part of the cross section are known and they are found to improve the
predictions.  In addition, there are remarkable developments to include
terms from next to SV contributions beyond third order and also to systematically
sum them up to all orders, through resummation framework.  Recent numerical studies
with resummed NSV contributions at leading logarithmic approximation strongly justify their
phenomenological importance.  The present paper incorporates NSV terms beyond
leading logarithmic approximation, in particular for the dominant diagonal channel beyond NLO
level and reports their numerical impact at the 13 TeV LHC.  Note that the theoretical set up for the
resummation of NSV terms from the non-diagonal channels is still a open problem
beyond LL approximation.  We expect that their numerical impact will be subdominant
compared to the diagonal channel.  Fixed order predictions already hints the importance
of not only NSV terms but also beyond NSV terms and our study confirms this
in the resummed framework. Our numerical predictions on the K factors
taking into account the resummed contributions from
NSV terms beyond leading logarithmic approximation imply better perturbative
convergence compared to fixed order or SV resummed contributions. For instance, the cross-section at $\rm NNLO + \overline{NNLL}$ accuracy reduces by $3.15\%$ as compared to the $\rm NNLO$ result for the central scale $\mu_R = \mu_F = m_H/2$ at 13 TeV LHC. In addition, the dependence on the renormalisation scale gets reduced upon the inclusion of
NSV resummed making the prediction more reliable.  However, we find that the sensitivity
to factorisation scale increases in the presence of resummed NSV terms implying the
importance of beyond NSV terms within the resummed framework.
It is well known that the SV resummed predictions depend on how we treat $N$ independent
terms leading to different schemes.  We have compared numerical predictions from these schemes
in the presence of NSV terms and found that $\overline N$ exponentiation scheme shows better perturbative convergence and
less sensitive to unphysical scales.  In summary, we have presented a detailed study on the
numerical importance of  resummed NSV terms for the production of Higgs bosons at the LHC.
We find that the inclusion of resummed NSV terms improves perturbative convergence and
reduces the uncertainty from the choice of renormalisation scale.

%\end{document}

\section{Acknowledgements}
We thank Claude Duhr, Moch and Bonvini for useful discussions throughout this project. We also thank Claude Duhr and Bernhard Mistlberger for providing third order results for the inclusive reactions.  
We would like to thank L. Magnea and E. Laenen for their encouragement to work on this area. In addition we would also like to thank the computer administrative unit of IMSc  for their help and support.
%% Appendix
\appendix
\section{Anomalous dimensions} \label{app:anodim}

Here we present all the anomalous dimensions used in performing the resummation.
\subsection*{Cusp anomalous dimensions $A_i^g$} 
In the following we list the cusp anomalous dimensions $A_i^g$ till four-loop level:  
 \begin{align} 
\begin{autobreak} 
 A^g_1 =
        4 C_A  \,,
      
\end{autobreak} 
\\ 
\begin{autobreak} 
A^g_2 =
  C_A n_f   \bigg\{
          - \frac{40}{9}
          \bigg\}

       + C_A^2   \bigg\{
           \frac{268}{9}
          - 8 \zeta_2
          \bigg\}\,,
      
\end{autobreak} 
\\ 
\begin{autobreak} 
 A^g_3 =
  C_A n_f^2   \bigg\{
          - \frac{16}{27}
          \bigg\}

       + C_A C_F n_f   \bigg\{
          - \frac{110}{3}
          + 32 \zeta_3
          \bigg\}

       + C_A^2 n_f   \bigg\{
          - \frac{836}{27}
          + \frac{160}{9} \zeta_2
          - \frac{112}{3} \zeta_3
          \bigg\}

       + C_A^3   \bigg\{
           \frac{490}{3}
          - \frac{1072}{9} \zeta_2
          + \frac{176}{5} \zeta_2^2
          + \frac{88}{3} \zeta_3
          \bigg\}\,,
      
\end{autobreak} 
\\
\begin{autobreak} 
A^g_4 =
          \frac{d_A^{abcd} d_A^{abcd}}{N_A}   \bigg\{
          - 128 \zeta_2
          - \frac{7936}{35} \zeta_2^3
          + \frac{3520}{3} \zeta_5
          + \frac{128}{3} \zeta_3
          - 384 \zeta_3^2
          \bigg\}

       +  n_f \frac{d_A^{abcd} d_F^{abcd}}{N_A}  \bigg\{
            256 \zeta_2
          - \frac{1280}{3} \zeta_5
          - \frac{256}{3} \zeta_3
          \bigg\}

       + C_A n_f^3   \bigg\{
          - \frac{32}{81}
          + \frac{64}{27} \zeta_3
          \bigg\}

       + C_A C_F n_f^2   \bigg\{
           \frac{2392}{81}
          + \frac{64}{5} \zeta_2^2
          - \frac{640}{9} \zeta_3
          \bigg\}

       + C_A C_F^2 n_f   \bigg\{
           \frac{572}{9}
          - 320 \zeta_5
          + \frac{592}{3} \zeta_3
          \bigg\}

       + C_A^2 n_f^2   \bigg\{
           \frac{923}{81}
          - \frac{608}{81} \zeta_2
          - \frac{224}{15} \zeta_2^2
          + \frac{2240}{27} \zeta_3
          \bigg\}

       + C_A^2 C_F n_f   \bigg\{
          - \frac{34066}{81}
          + \frac{440}{3} \zeta_2
          - \frac{352}{5} \zeta_2^2
          + 160 \zeta_5
          + \frac{3712}{9} \zeta_3
          - 128 \zeta_3 \zeta_2
          \bigg\}

       + C_A^3 n_f   \bigg\{
          - \frac{24137}{81}
          + \frac{20320}{81} \zeta_2
          - \frac{352}{15} \zeta_2^2
          + \frac{2096}{9} \zeta_5
          - \frac{23104}{27} \zeta_3
          + \frac{448}{3} \zeta_3 \zeta_2
          \bigg\}

       + C_A^4   \bigg\{
           \frac{84278}{81}
          - \frac{88400}{81} \zeta_2
          + \frac{3608}{5} \zeta_2^2
          - \frac{20032}{105} \zeta_2^3
          - \frac{3608}{9} \zeta_5
          + \frac{20944}{27} \zeta_3
          - \frac{352}{3} \zeta_3 \zeta_2
          - 16 \zeta_3^2
          \bigg\}\,,

\end{autobreak} 
\end{align} 

where $n_f$ is the number of active quark flavours in the theory. The quadratic Casimirs $C_F$ and $C_A$ are given by  
$\frac{n_c^2-1}{2~ nc}$ and $n_c$ respectively.
The quartic Casimirs are given by
\begin{align}
 \frac{d_A^{abcd}d_A^{abcd}}{N_A} &= \frac{n_c^2 (n_c^2 + 36)}{24}, \qquad
 \frac{d_A^{abcd}d_F^{abcd}}{N_A} = \frac{n_c (n_c^2 + 6)}{48},
% \frac{d_F^{abcd}d_A^{abcd}}{N_F} &=  \frac{(n_c^2-1)(n_c^2+6)}{48}, 
% \frac{d_F^{abcd}d_F^{abcd}}{N_F} = \frac{(n_c^2-1)(n_c^4 - 6 n_c^2 + 18)}{96n_c^3},
%\frac{d_F^{abcd}d_A^{abcd}}{N_A} &=  \frac{n_c (n_c^2+6)}{48}, \qquad
%\frac{d_F^{abcd}d_F^{abcd}}{N_A} = \frac{(n_c^4 - 6 n_c^2 + 18)}{96n_c^2},
\end{align} 
with $N_A = n_c^2 -1$ where $n_c = 3$ for QCD.
\subsection*{Collinear anomalous dimensions $B_i^q$}

The collinear anomalous dimensions $B_i^g$ are given till three-loop as, 
\begin{align} 
\begin{autobreak} 
B_1^g =
         n_f   \bigg\{
          - \frac{2}{3}
          \bigg\}

       + C_A   \bigg\{
           \frac{11}{3}
          \bigg\}\,,
\end{autobreak} 
\\ 
\begin{autobreak} 
B_2^g =
        C_F n_f   \bigg\{
          - 2
          \bigg\}

       + C_A n_f   \bigg\{
          - \frac{8}{3}
          \bigg\}

       + C_A^2   \bigg\{
           \frac{32}{3}
          + 12 \zeta_3
          \bigg\}\,,
\end{autobreak} 
\\ 
\begin{autobreak} 
B_3^g =
          C_F n_f^2   \bigg\{
           \frac{11}{9}
          \bigg\}

       + C_F^2 n_f  

       + C_A n_f^2   \bigg\{
           \frac{29}{18}
          \bigg\}

       + C_A C_F n_f   \bigg\{
          - \frac{241}{18}
          \bigg\}

       + C_A^2 n_f   \bigg\{
          - \frac{233}{18}
          - \frac{8}{3} \zeta_2
          - \frac{4}{3} \zeta_2^2
          - \frac{80}{3} \zeta_3
          \bigg\}

       + C_A^3   \bigg\{
           \frac{79}{2}
          + \frac{8}{3} \zeta_2
          + \frac{22}{3} \zeta_2^2
          - 80 \zeta_5
          + \frac{536}{3} \zeta_3
          - 16 \zeta_3 \zeta_2
          \bigg\}\,.
\end{autobreak} 
\end{align}
\subsection*{Soft anomalous dimensions $f_i^g$}
The soft anomalous dimensions $f_i^g$ till three-loop are given as
\begin{align}
\begin{autobreak} 
f_1^g =  
0
,   
\end{autobreak} 
\\ 
\begin{autobreak} 
f_2^g =
          C_A n_f   \bigg\{
          - \frac{112}{27}
          + \frac{4}{3} \zeta_2
          \bigg\}

         + C_A^2   \bigg\{
            \frac{808}{27}
          - \frac{22}{3} \zeta_2
          - 28 \zeta_3
          \bigg\}\,.
\end{autobreak} 
\\ 
\begin{autobreak} 
f_3^g =
         C_A n_f^2   \bigg\{
          - \frac{2080}{729}
          - \frac{40}{27} \zeta_2
          + \frac{112}{27} \zeta_3
          \bigg\}

       + C_A C_F n_f   \bigg\{
          - \frac{1711}{27}
          + 4 \zeta_2
          + \frac{32}{5} \zeta_2^2
          + \frac{304}{9} \zeta_3
          \bigg\}

       + C_A^2 n_f   \bigg\{
          - \frac{11842}{729}
          + \frac{2828}{81} \zeta_2
          - \frac{96}{5} \zeta_2^2
          + \frac{728}{27} \zeta_3
          \bigg\}

       + C_A^3   \bigg\{
           \frac{136781}{729}
          - \frac{12650}{81} \zeta_2
          + \frac{352}{5} \zeta_2^2
          + 192 \zeta_5
          - \frac{1316}{3} \zeta_3
          + \frac{176}{3} \zeta_3 \zeta_2
          \bigg\}\,.
\end{autobreak}
\end{align} 
\subsection*{NSV anomalous dimensions $C_i^g$ \& $D_i^g$ }
The NSV anomalous dimensions $C_i^g$ and  $D_i^g$ till three-loop are given as

\begin{align}
\begin{autobreak}
C_1^g = 
0,
\end{autobreak}\\
\begin{autobreak}
C_2^g = 
        16  C_A^2 \,,

\end{autobreak}\\
\begin{autobreak}
C_3^g =  
        C_A^2 n_f   \bigg\{
          - \frac{320}{9}
          \bigg\}

       + C_A^3   \bigg\{
           \frac{2144}{9}
          - 64 \zeta_2
          \bigg\}\,.
\end{autobreak}\\
%\end{align}
%\subsection*{NSV anomalous dimensions $D_i^q$}
%
%The NSV anomalous dimensions $D_i^q$ till three-loop are given as
%\begin{align}
\begin{autobreak}
D_1^g = 
-4 C_A, 
\end{autobreak} \\
\begin{autobreak}
D_2^g =
         C_A n_f   \bigg\{
           \frac{40}{9}
          \bigg\}

       + C_A^2   \bigg\{
          - \frac{268}{9}
          + 8 \zeta_2
          \bigg\}\,,
\end{autobreak} \\
\begin{autobreak}
D_3^g = 
         C_A n_f^2   \bigg\{
           \frac{16}{27}
          \bigg\}

       + C_A C_F n_f   \bigg\{
           \frac{110}{3}
          - 32 \zeta_3
          \bigg\}

       + C_A^2 n_f   \bigg\{
           \frac{908}{27}
          - \frac{160}{9} \zeta_2
          + \frac{112}{3} \zeta_3
          \bigg\}

       + C_A^3   \bigg\{
          - 166
          + \frac{1072}{9} \zeta_2
          - \frac{176}{5} \zeta_2^2
          + \frac{56}{3} \zeta_3
          \bigg\}\,.
\end{autobreak}
\end{align}
%\\
%--------------------------- SV--------------------------------------%
\subsection*{SV Threshold Exponents }
The function $\overline{G}^g_{SV}\big(a_s(q^2(1-z)^2),\epsilon\big)$ given in \ref{phicint} is related to the threshold exponent $\textbf{D}^g\big(a_s(q^2(1-z)^2),\epsilon\big)$ via Eq.(46) of \cite{Ravindran:2006cg} where the universal $\mathbf{D}_i^g$ coefficients, till three-loop, are given as,

\begin{align} 
\begin{autobreak} 
\mathbf{D}^g_1 =
             0,   
\end{autobreak} 
\\ 
\begin{autobreak} 
\mathbf{D}^g_2 =
           C_A n_f   \bigg\{
           \frac{224}{27}
          - \frac{32}{3} \zeta_2
          \bigg\}

         + C_A^2   \bigg\{
          - \frac{1616}{27}
          + \frac{176}{3} \zeta_2
          + 56 \zeta_3
          \bigg\}\,, 
\end{autobreak} 
\\ 
\begin{autobreak} 
\mathbf{D}^g_3 =
          C_A n_f^2   \bigg\{
          - \frac{3712}{729}
          + \frac{640}{27} \zeta_2
          + \frac{320}{27} \zeta_3
          \bigg\}

       + C_A C_F n_f   \bigg\{
           \frac{3422}{27}
          - 32 \zeta_2
          - \frac{64}{5} \zeta_2^2
          - \frac{608}{9} \zeta_3
          \bigg\}

       + C_A^2 n_f   \bigg\{
           \frac{125252}{729}
          - \frac{29392}{81} \zeta_2
          + \frac{736}{15} \zeta_2^2
          - \frac{2480}{9} \zeta_3
          \bigg\}

       + C_A^3   \bigg\{
          - \frac{594058}{729}
          + \frac{98224}{81} \zeta_2
          - \frac{2992}{15} \zeta_2^2
          - 384 \zeta_5
          + \frac{40144}{27} \zeta_3
          - \frac{352}{3} \zeta_3 \zeta_2
          \bigg\}\,.
\end{autobreak} 
\end{align}

 \section{Resummation Coefficients for the $N$ exponentiation} \label{app:Nexp}
 \subsection{The $N$-independent coefficients $g_{0,i}^g $} \label{app:g0g}
 The $N$-independent coefficients $g_{0,i}^g $ in Eq. \ref{lng0}, till three-loop,
are given by
\begin{align}
\begin{autobreak}
g_{0,0}^g = 
     0\,,
\end{autobreak}\\
\begin{autobreak}
 g_{0,1}^g=

          C_A   \bigg\{
           8 \zeta_2
          - 8 \gamma_E L_{qr}
          + 8 \gamma_E L_{fr}
          + 8 \gamma_E^2
          \bigg\}\,,
\end{autobreak}\\
\begin{autobreak}
    g_{0,2}^g=

        C_A n_f   \bigg\{
          - \frac{64}{9} \zeta_3
          - \frac{80}{9} \zeta_2
          + \frac{16}{3} L_{qr} \zeta_2
          - \frac{224}{27} \gamma_E
          + \frac{80}{9} \gamma_E L_{qr}
          - \frac{8}{3} \gamma_E L_{qr}^2
          - \frac{80}{9} \gamma_E L_{fr}
          + \frac{8}{3} \gamma_E L_{fr}^2
          - \frac{80}{9} \gamma_E^2
          + \frac{16}{3} \gamma_E^2 L_{qr}
          - \frac{32}{9} \gamma_E^3
          \bigg\}

       + C_A^2   \bigg\{
           \frac{352}{9} \zeta_3
          + \frac{536}{9} \zeta_2
          - 16 \zeta_2^2
          - \frac{88}{3} L_{qr} \zeta_2
          + \frac{1616}{27} \gamma_E
          - 56 \gamma_E \zeta_3
          - \frac{536}{9} \gamma_E L_{qr}
          + 16 \gamma_E L_{qr} \zeta_2
          + \frac{44}{3} \gamma_E L_{qr}^2
          + \frac{536}{9} \gamma_E L_{fr}
          - 16 \gamma_E L_{fr} \zeta_2
          - \frac{44}{3} \gamma_E L_{fr}^2
          + \frac{536}{9} \gamma_E^2
          - 16 \gamma_E^2 \zeta_2
          - \frac{88}{3} \gamma_E^2 L_{qr}
          + \frac{176}{9} \gamma_E^3
          \bigg\}\,,
\end{autobreak}\\
\begin{autobreak}
    g_{0,3}^g=

        C_A n_f^2   \bigg\{
           \frac{1280}{81} \zeta_3
          + \frac{800}{81} \zeta_2
          - \frac{64}{45} \zeta_2^2
          - \frac{256}{27} L_{qr} \zeta_3
          - \frac{320}{27} L_{qr} \zeta_2
          + \frac{32}{9} L_{qr}^2 \zeta_2
          + \frac{3712}{729} \gamma_E
          + \frac{64}{9} \gamma_E \zeta_3
          - \frac{800}{81} \gamma_E L_{qr}
          + \frac{160}{27} \gamma_E L_{qr}^2
          - \frac{32}{27} \gamma_E L_{qr}^3
          - \frac{32}{27} \gamma_E L_{fr}
          - \frac{160}{27} \gamma_E L_{fr}^2
          + \frac{32}{27} \gamma_E L_{fr}^3
          + \frac{800}{81} \gamma_E^2
          - \frac{320}{27} \gamma_E^2 L_{qr}
          + \frac{32}{9} \gamma_E^2 L_{qr}^2
          + \frac{640}{81} \gamma_E^3
          - \frac{128}{27} \gamma_E^3 L_{qr}
          + \frac{64}{27} \gamma_E^4
          \bigg\}

       + C_A C_F n_f   \bigg\{
          - \frac{64}{3} \zeta_3
          - \frac{220}{3} \zeta_2
          + 64 \zeta_2 \zeta_3
          + 16 L_{qr} \zeta_2
          - \frac{3422}{27} \gamma_E
          + \frac{608}{9} \gamma_E \zeta_3
          + \frac{64}{5} \gamma_E \zeta_2^2
          + \frac{220}{3} \gamma_E L_{qr}
          - 64 \gamma_E L_{qr} \zeta_3
          - 8 \gamma_E L_{qr}^2
          - \frac{220}{3} \gamma_E L_{fr}
          + 64 \gamma_E L_{fr} \zeta_3
          + 8 \gamma_E L_{fr}^2
          - \frac{220}{3} \gamma_E^2
          + 64 \gamma_E^2 \zeta_3
          + 16 \gamma_E^2 L_{qr}
          - \frac{32}{3} \gamma_E^3
          \bigg\}

       + C_A^2 n_f   \bigg\{
          - \frac{18496}{81} \zeta_3
          - \frac{16408}{81} \zeta_2
          + \frac{256}{9} \zeta_2 \zeta_3
          + \frac{256}{5} \zeta_2^2
          + \frac{2816}{27} L_{qr} \zeta_3
          + \frac{4624}{27} L_{qr} \zeta_2
          - \frac{64}{3} L_{qr} \zeta_2^2
          - \frac{352}{9} L_{qr}^2 \zeta_2
          - \frac{125252}{729} \gamma_E
          + \frac{1808}{27} \gamma_E \zeta_3
          + \frac{1648}{81} \gamma_E \zeta_2
          - \frac{32}{5} \gamma_E \zeta_2^2
          + \frac{16408}{81} \gamma_E L_{qr}
          - \frac{320}{9} \gamma_E L_{qr} \zeta_2
          - \frac{2312}{27} \gamma_E L_{qr}^2
          + \frac{32}{3} \gamma_E L_{qr}^2 \zeta_2
          + \frac{352}{27} \gamma_E L_{qr}^3
          - \frac{1672}{27} \gamma_E L_{fr}
          - \frac{224}{3} \gamma_E L_{fr} \zeta_3
          + \frac{320}{9} \gamma_E L_{fr} \zeta_2
          + \frac{2312}{27} \gamma_E L_{fr}^2
          - \frac{32}{3} \gamma_E L_{fr}^2 \zeta_2
          - \frac{352}{27} \gamma_E L_{fr}^3
          - \frac{16408}{81} \gamma_E^2
          + \frac{320}{9} \gamma_E^2 \zeta_2
          + \frac{4624}{27} \gamma_E^2 L_{qr}
          - \frac{64}{3} \gamma_E^2 L_{qr} \zeta_2
          - \frac{352}{9} \gamma_E^2 L_{qr}^2
          - \frac{9248}{81} \gamma_E^3
          + \frac{128}{9} \gamma_E^3 \zeta_2
          + \frac{1408}{27} \gamma_E^3 L_{qr}
          - \frac{704}{27} \gamma_E^4
          \bigg\}

       + C_A^3   \bigg\{
           \frac{56960}{81} \zeta_3
          + \frac{62012}{81} \zeta_2
          - \frac{4576}{9} \zeta_2 \zeta_3
          - \frac{12656}{45} \zeta_2^2
          + \frac{352}{5} \zeta_2^3
          - \frac{7744}{27} L_{qr} \zeta_3
          - \frac{14240}{27} L_{qr} \zeta_2
          + \frac{352}{3} L_{qr} \zeta_2^2
          + \frac{968}{9} L_{qr}^2 \zeta_2
          + \frac{594058}{729} \gamma_E
          + 384 \gamma_E \zeta_5
          - \frac{24656}{27} \gamma_E \zeta_3
          - \frac{12784}{81} \gamma_E \zeta_2
          + \frac{352}{3} \gamma_E \zeta_2 \zeta_3
          - \frac{176}{5} \gamma_E \zeta_2^2
          - \frac{62012}{81} \gamma_E L_{qr}
          + 352 \gamma_E L_{qr} \zeta_3
          + \frac{2144}{9} \gamma_E L_{qr} \zeta_2
          - \frac{352}{5} \gamma_E L_{qr} \zeta_2^2
          + \frac{7120}{27} \gamma_E L_{qr}^2
          - \frac{176}{3} \gamma_E L_{qr}^2 \zeta_2
          - \frac{968}{27} \gamma_E L_{qr}^3
          + \frac{980}{3} \gamma_E L_{fr}
          + \frac{176}{3} \gamma_E L_{fr} \zeta_3
          - \frac{2144}{9} \gamma_E L_{fr} \zeta_2
          + \frac{352}{5} \gamma_E L_{fr} \zeta_2^2
          - \frac{7120}{27} \gamma_E L_{fr}^2
          + \frac{176}{3} \gamma_E L_{fr}^2 \zeta_2
          + \frac{968}{27} \gamma_E L_{fr}^3
          + \frac{62012}{81} \gamma_E^2
          - 352 \gamma_E^2 \zeta_3
          - \frac{2144}{9} \gamma_E^2 \zeta_2
          + \frac{352}{5} \gamma_E^2 \zeta_2^2
          - \frac{14240}{27} \gamma_E^2 L_{qr}
          + \frac{352}{3} \gamma_E^2 L_{qr} \zeta_2
          + \frac{968}{9} \gamma_E^2 L_{qr}^2
          + \frac{28480}{81} \gamma_E^3
          - \frac{704}{9} \gamma_E^3 \zeta_2
          - \frac{3872}{27} \gamma_E^3 L_{qr}
          + \frac{1936}{27} \gamma_E^4
          \bigg\}\,.
\end{autobreak}
\end{align}
Here, $\gamma_E$ is the Euler-Mascheroni constant and $\zeta_i$'s are the Riemann zeta functions. In the aforementioned equations, $L_{qr} = \ln \big(\frac{q^2}{\mu_R^2}\big)$ and $L_{fr} = \ln \big(\frac{\mu_F^2}{\mu_R^2}\big)$.

\subsection{The $N$-independent coefficients $\Tilde{g}^g_0$} \label{app:g0t}
The $N$-independent coefficients $\Tilde{g}^g_{0,i}$ in Eq. \ref{g0tg}, till three-loop, are given by

\begin{align}
\begin{autobreak}

   \Tilde{g}^g_{0,0} =
          1
         \,,
\end{autobreak}\\
\begin{autobreak}
  \Tilde{g}^g_{0,1} =

          n_f   \bigg\{
           \frac{4}{3} L_{fr}
          \bigg\}

       + C_A   \bigg\{
          16 \zeta_2
          - \frac{22}{3} L_{fr}
          - 8 \gamma_E L_{qr}
          + 8 \gamma_E L_{fr}
          + 8 \gamma_E^2
          \bigg\}

         \,,
\end{autobreak}\\
\begin{autobreak}
   \Tilde{g}^g_{0,2} =

          n_f^2   \bigg\{
           \frac{4}{3} L_{fr}^2
          \bigg\}

       + C_F n_f   \bigg\{
          - \frac{67}{3}
          + 16 \zeta_3
          + 4 L_{qr}
          + 4 L_{fr}
          \bigg\}

       + C_A n_f   \bigg\{
          - \frac{80}{3}
          - \frac{88}{9} \zeta_3
          - \frac{160}{9} \zeta_2
          + 8 L_{qr}
          + \frac{32}{3} L_{qr} \zeta_2
          + \frac{16}{3} L_{fr}
          + \frac{64}{3} L_{fr} \zeta_2
          - \frac{44}{3} L_{fr}^2
          - \frac{224}{27} \gamma_E
          + \frac{80}{9} \gamma_E L_{qr}
          - \frac{8}{3} \gamma_E L_{qr}^2
          - \frac{80}{9} \gamma_E L_{fr}
          - \frac{32}{3} \gamma_E L_{fr} L_{qr}
          + \frac{40}{3} \gamma_E L_{fr}^2
          - \frac{80}{9} \gamma_E^2
          + \frac{16}{3} \gamma_E^2 L_{qr}
          + \frac{32}{3} \gamma_E^2 L_{fr}
          - \frac{32}{9} \gamma_E^3
          \bigg\}

       + C_A^2   \bigg\{
           93
          - \frac{308}{9} \zeta_3
          + \frac{1072}{9} \zeta_2
          + 92 \zeta_2^2
          - 24 L_{qr}
          + 24 L_{qr} \zeta_3
          - \frac{176}{3} L_{qr} \zeta_2
          - \frac{64}{3} L_{fr}
          - 24 L_{fr} \zeta_3
          - \frac{352}{3} L_{fr} \zeta_2
          + \frac{121}{3} L_{fr}^2
          + \frac{1616}{27} \gamma_E
          - 56 \gamma_E \zeta_3
          - \frac{536}{9} \gamma_E L_{qr}
          - 112 \gamma_E L_{qr} \zeta_2
          + \frac{44}{3} \gamma_E L_{qr}^2
          + \frac{536}{9} \gamma_E L_{fr}
          + 112 \gamma_E L_{fr} \zeta_2
          + \frac{176}{3} \gamma_E L_{fr} L_{qr}
          - \frac{220}{3} \gamma_E L_{fr}^2
          + \frac{536}{9} \gamma_E^2
          + 112 \gamma_E^2 \zeta_2
          - \frac{88}{3} \gamma_E^2 L_{qr}
          + 32 \gamma_E^2 L_{qr}^2
          - \frac{176}{3} \gamma_E^2 L_{fr}
          - 64 \gamma_E^2 L_{fr} L_{qr}
          + 32 \gamma_E^2 L_{fr}^2
          + \frac{176}{9} \gamma_E^3
          - 64 \gamma_E^3 L_{qr}
          + 64 \gamma_E^3 L_{fr}
          + 32 \gamma_E^4
          \bigg\}
         \,,
\end{autobreak}\\
\begin{autobreak}
   \Tilde{g}^g_{0,3} = 

         n_f^3   \bigg\{
          \frac{32}{27} L_{fr}^3
          \bigg\}

       + C_F n_f^2   \bigg\{
           \frac{8962}{81}
          - \frac{224}{3} \zeta_3
          - \frac{184}{9} \zeta_2
          - \frac{32}{45} \zeta_2^2
          - \frac{104}{3} L_{qr}
          + \frac{64}{3} L_{qr} \zeta_3
          + \frac{8}{3} L_{qr}^2
          - \frac{290}{9} L_{fr}
          + \frac{64}{3} L_{fr} \zeta_3
          + \frac{16}{3} L_{fr} L_{qr}
          + \frac{28}{3} L_{fr}^2
          \bigg\}

       + C_F^2 n_f   \bigg\{
           \frac{608}{9}
          - 320 \zeta_5
          + \frac{592}{3} \zeta_3
          - 4 L_{qr}
          - 2 L_{fr}
          \bigg\}

       + C_A n_f^2   \bigg\{
           \frac{2515}{27}
          + \frac{3344}{81} \zeta_3
          - \frac{1328}{81} \zeta_2
          - \frac{224}{15} \zeta_2^2
          - \frac{370}{9} L_{qr}
          - \frac{352}{27} L_{qr} \zeta_3
          - \frac{640}{27} L_{qr} \zeta_2
          + \frac{16}{3} L_{qr}^2
          + \frac{64}{9} L_{qr}^2 \zeta_2
          - \frac{349}{9} L_{fr}
          - \frac{352}{27} L_{fr} \zeta_3
          - \frac{640}{27} L_{fr} \zeta_2
          + \frac{32}{3} L_{fr} L_{qr}
          + \frac{128}{9} L_{fr} L_{qr} \zeta_2
          + \frac{116}{9} L_{fr}^2
          + \frac{64}{3} L_{fr}^2 \zeta_2
          - \frac{176}{9} L_{fr}^3
          + \frac{3712}{729} \gamma_E
          + \frac{64}{9} \gamma_E \zeta_3
          - \frac{800}{81} \gamma_E L_{qr}
          + \frac{160}{27} \gamma_E L_{qr}^2
          - \frac{32}{27} \gamma_E L_{qr}^3
          - \frac{992}{81} \gamma_E L_{fr}
          + \frac{320}{27} \gamma_E L_{fr} L_{qr}
          - \frac{32}{9} \gamma_E L_{fr} L_{qr}^2
          - \frac{160}{9} \gamma_E L_{fr}^2
          - \frac{32}{3} \gamma_E L_{fr}^2 L_{qr}
          + \frac{416}{27} \gamma_E L_{fr}^3
          + \frac{800}{81} \gamma_E^2
          - \frac{320}{27} \gamma_E^2 L_{qr}
          + \frac{32}{9} \gamma_E^2 L_{qr}^2
          - \frac{320}{27} \gamma_E^2 L_{fr}
          + \frac{64}{9} \gamma_E^2 L_{fr} L_{qr}
          + \frac{32}{3} \gamma_E^2 L_{fr}^2
          + \frac{640}{81} \gamma_E^3
          - \frac{128}{27} \gamma_E^3 L_{qr}
          - \frac{128}{27} \gamma_E^3 L_{fr}
          + \frac{64}{27} \gamma_E^4
          \bigg\}

       + C_A C_F n_f   \bigg\{
          - \frac{63991}{81}
          + 160 \zeta_5
          + \frac{1184}{3} \zeta_3
          - \frac{3404}{9} \zeta_2
          + 384 \zeta_2 \zeta_3
          + \frac{176}{45} \zeta_2^2
          + \frac{616}{3} L_{qr}
          - \frac{352}{3} L_{qr} \zeta_3
          + 96 L_{qr} \zeta_2
          - \frac{44}{3} L_{qr}^2
          + \frac{1715}{9} L_{fr}
          - \frac{352}{3} L_{fr} \zeta_3
          + 64 L_{fr} \zeta_2
          - \frac{88}{3} L_{fr} L_{qr}
          - \frac{154}{3} L_{fr}^2
          - \frac{3422}{27} \gamma_E
          + \frac{608}{9} \gamma_E \zeta_3
          + \frac{64}{5} \gamma_E \zeta_2^2
          + 252 \gamma_E L_{qr}
          - 192 \gamma_E L_{qr} \zeta_3
          - 40 \gamma_E L_{qr}^2
          - 252 \gamma_E L_{fr}
          + 192 \gamma_E L_{fr} \zeta_3
          + 40 \gamma_E L_{fr}^2
          - 252 \gamma_E^2
          + 192 \gamma_E^2 \zeta_3
          + 48 \gamma_E^2 L_{qr}
          + 32 \gamma_E^2 L_{fr}
          - \frac{32}{3} \gamma_E^3
          \bigg\}

       + C_A^2 n_f   \bigg\{
          - \frac{98059}{81}
          + \frac{808}{9} \zeta_5
          + \frac{296}{81} \zeta_3
          - \frac{38168}{81} \zeta_2
          - \frac{784}{3} \zeta_2 \zeta_3
          - \frac{4696}{135} \zeta_2^2
          + \frac{4058}{9} L_{qr}
          - \frac{736}{27} L_{qr} \zeta_3
          + \frac{12560}{27} L_{qr} \zeta_2
          + 120 L_{qr} \zeta_2^2
          - \frac{136}{3} L_{qr}^2
          + 16 L_{qr}^2 \zeta_3
          - \frac{704}{9} L_{qr}^2 \zeta_2
          + \frac{3109}{9} L_{fr}
          + \frac{2144}{27} L_{fr} \zeta_3
          + \frac{10256}{27} L_{fr} \zeta_2
          + \frac{376}{3} L_{fr} \zeta_2^2
          - \frac{272}{3} L_{fr} L_{qr}
          + 32 L_{fr} L_{qr} \zeta_3
          - \frac{1408}{9} L_{fr} L_{qr} \zeta_2
          - \frac{1090}{9} L_{fr}^2
          - 48 L_{fr}^2 \zeta_3
          - \frac{704}{3} L_{fr}^2 \zeta_2
          + \frac{968}{9} L_{fr}^3
          - \frac{125252}{729} \gamma_E
          + \frac{1808}{27} \gamma_E \zeta_3
          - \frac{9104}{81} \gamma_E \zeta_2
          - \frac{32}{5} \gamma_E \zeta_2^2
          + \frac{33688}{81} \gamma_E L_{qr}
          + \frac{704}{9} \gamma_E L_{qr} \zeta_3
          + \frac{2240}{9} \gamma_E L_{qr} \zeta_2
          - \frac{4040}{27} \gamma_E L_{qr}^2
          - \frac{352}{3} \gamma_E L_{qr}^2 \zeta_2
          + \frac{352}{27} \gamma_E L_{qr}^3
          - \frac{10904}{81} \gamma_E L_{fr}
          - \frac{2048}{9} \gamma_E L_{fr} \zeta_3
          - \frac{2240}{9} \gamma_E L_{fr} \zeta_2
          - \frac{3328}{27} \gamma_E L_{fr} L_{qr}
          - 64 \gamma_E L_{fr} L_{qr} \zeta_2
          + \frac{352}{9} \gamma_E L_{fr} L_{qr}^2
          + \frac{2456}{9} \gamma_E L_{fr}^2
          + \frac{544}{3} \gamma_E L_{fr}^2 \zeta_2
          + \frac{352}{3} \gamma_E L_{fr}^2 L_{qr}
          - \frac{4576}{27} \gamma_E L_{fr}^3
          - \frac{33688}{81} \gamma_E^2
          - \frac{704}{9} \gamma_E^2 \zeta_3
          - \frac{2240}{9} \gamma_E^2 \zeta_2
          + \frac{8144}{27} \gamma_E^2 L_{qr}
          + \frac{448}{3} \gamma_E^2 L_{qr} \zeta_2
          - \frac{992}{9} \gamma_E^2 L_{qr}^2
          + \frac{64}{3} \gamma_E^2 L_{qr}^3
          + \frac{1088}{9} \gamma_E^2 L_{fr}
          + \frac{448}{3} \gamma_E^2 L_{fr} \zeta_2
          + 64 \gamma_E^2 L_{fr} L_{qr}
          + \frac{64}{3} \gamma_E^2 L_{fr} L_{qr}^2
          - \frac{1696}{9} \gamma_E^2 L_{fr}^2
          - \frac{320}{3} \gamma_E^2 L_{fr}^2 L_{qr}
          + 64 \gamma_E^2 L_{fr}^3
          - \frac{14624}{81} \gamma_E^3
          - \frac{128}{3} \gamma_E^3 \zeta_2
          + \frac{5248}{27} \gamma_E^3 L_{qr}
          - 64 \gamma_E^3 L_{qr}^2
          - \frac{2432}{27} \gamma_E^3 L_{fr}
          - \frac{128}{3} \gamma_E^3 L_{fr} L_{qr}
          + \frac{320}{3} \gamma_E^3 L_{fr}^2
          - \frac{2624}{27} \gamma_E^4
          + \frac{640}{9} \gamma_E^4 L_{qr}
          + \frac{128}{9} \gamma_E^4 L_{fr}
          - \frac{256}{9} \gamma_E^5
          \bigg\}

       + C_A^3   \bigg\{
           \frac{215131}{81}
          + \frac{3476}{9} \zeta_5
          - \frac{130828}{81} \zeta_3
          + 96 \zeta_3^2
          + \frac{186880}{81} \zeta_2
          - \frac{2024}{3} \zeta_2 \zeta_3
          + \frac{119692}{135} \zeta_2^2
          + \frac{3872}{15} \zeta_2^3
          - \frac{8284}{9} L_{qr}
          - 160 L_{qr} \zeta_5
          + \frac{16424}{27} L_{qr} \zeta_3
          - \frac{38704}{27} L_{qr} \zeta_2
          + 352 L_{qr} \zeta_2 \zeta_3
          - 660 L_{qr} \zeta_2^2
          + 88 L_{qr}^2
          - 88 L_{qr}^2 \zeta_3
          + \frac{1936}{9} L_{qr}^2 \zeta_2
          - 761 L_{fr}
          + 160 L_{fr} \zeta_5
          - \frac{2872}{27} L_{fr} \zeta_3
          - \frac{32944}{27} L_{fr} \zeta_2
          - 352 L_{fr} \zeta_2 \zeta_3
          - \frac{2068}{3} L_{fr} \zeta_2^2
          + 176 L_{fr} L_{qr}
          - 176 L_{fr} L_{qr} \zeta_3
          + \frac{3872}{9} L_{fr} L_{qr} \zeta_2
          + \frac{2486}{9} L_{fr}^2
          + 264 L_{fr}^2 \zeta_3
          + \frac{1936}{3} L_{fr}^2 \zeta_2
          - \frac{5324}{27} L_{fr}^3
          + \frac{594058}{729} \gamma_E
          + 384 \gamma_E \zeta_5
          - \frac{24656}{27} \gamma_E \zeta_3
          + \frac{64784}{81} \gamma_E \zeta_2
          - \frac{2336}{3} \gamma_E \zeta_2 \zeta_3
          - \frac{176}{5} \gamma_E \zeta_2^2
          - \frac{122276}{81} \gamma_E L_{qr}
          + \frac{5632}{9} \gamma_E L_{qr} \zeta_3
          - \frac{15008}{9} \gamma_E L_{qr} \zeta_2
          - \frac{2752}{5} \gamma_E L_{qr} \zeta_2^2
          + \frac{12304}{27} \gamma_E L_{qr}^2
          - 192 \gamma_E L_{qr}^2 \zeta_3
          + \frac{1936}{3} \gamma_E L_{qr}^2 \zeta_2
          - \frac{968}{27} \gamma_E L_{qr}^3
          + \frac{51172}{81} \gamma_E L_{fr}
          + \frac{1760}{9} \gamma_E L_{fr} \zeta_3
          + \frac{15008}{9} \gamma_E L_{fr} \zeta_2
          + \frac{2752}{5} \gamma_E L_{fr} \zeta_2^2
          + \frac{11216}{27} \gamma_E L_{fr} L_{qr}
          + 384 \gamma_E L_{fr} L_{qr} \zeta_3
          + 352 \gamma_E L_{fr} L_{qr} \zeta_2
          - \frac{968}{9} \gamma_E L_{fr} L_{qr}^2
          - \frac{7840}{9} \gamma_E L_{fr}^2
          - 192 \gamma_E L_{fr}^2 \zeta_3
          - \frac{2992}{3} \gamma_E L_{fr}^2 \zeta_2
          - \frac{968}{3} \gamma_E L_{fr}^2 L_{qr}
          + \frac{12584}{27} \gamma_E L_{fr}^3
          + \frac{122276}{81} \gamma_E^2
          - \frac{5632}{9} \gamma_E^2 \zeta_3
          + \frac{15008}{9} \gamma_E^2 \zeta_2
          + \frac{2752}{5} \gamma_E^2 \zeta_2^2
          - \frac{10784}{9} \gamma_E^2 L_{qr}
          + 640 \gamma_E^2 L_{qr} \zeta_3
          - \frac{2464}{3} \gamma_E^2 L_{qr} \zeta_2
          + 584 \gamma_E^2 L_{qr}^2
          + 384 \gamma_E^2 L_{qr}^2 \zeta_2
          - \frac{352}{3} \gamma_E^2 L_{qr}^3
          - \frac{3472}{27} \gamma_E^2 L_{fr}
          - 640 \gamma_E^2 L_{fr} \zeta_3
          - \frac{2464}{3} \gamma_E^2 L_{fr} \zeta_2
          - \frac{6640}{9} \gamma_E^2 L_{fr} L_{qr}
          - 768 \gamma_E^2 L_{fr} L_{qr} \zeta_2
          - \frac{352}{3} \gamma_E^2 L_{fr} L_{qr}^2
          + \frac{7192}{9} \gamma_E^2 L_{fr}^2
          + 384 \gamma_E^2 L_{fr}^2 \zeta_2
          + \frac{1760}{3} \gamma_E^2 L_{fr}^2 L_{qr}
          - 352 \gamma_E^2 L_{fr}^3
          + \frac{67264}{81} \gamma_E^3
          - 448 \gamma_E^3 \zeta_3
          + \frac{704}{3} \gamma_E^3 \zeta_2
          - \frac{29600}{27} \gamma_E^3 L_{qr}
          - 768 \gamma_E^3 L_{qr} \zeta_2
          + 352 \gamma_E^3 L_{qr}^2
          - \frac{256}{3} \gamma_E^3 L_{qr}^3
          + \frac{21856}{27} \gamma_E^3 L_{fr}
          + 768 \gamma_E^3 L_{fr} \zeta_2
          + \frac{704}{3} \gamma_E^3 L_{fr} L_{qr}
          + 256 \gamma_E^3 L_{fr} L_{qr}^2
          - \frac{1760}{3} \gamma_E^3 L_{fr}^2
          - 256 \gamma_E^3 L_{fr}^2 L_{qr}
          + \frac{256}{3} \gamma_E^3 L_{fr}^3
          + \frac{14800}{27} \gamma_E^4
          + 384 \gamma_E^4 \zeta_2
          - \frac{3520}{9} \gamma_E^4 L_{qr}
          + 256 \gamma_E^4 L_{qr}^2
          - \frac{704}{9} \gamma_E^4 L_{fr}
          - 512 \gamma_E^4 L_{fr} L_{qr}
          + 256 \gamma_E^4 L_{fr}^2
          + \frac{1408}{9} \gamma_E^5
          - 256 \gamma_E^5 L_{qr}
          + 256 \gamma_E^5 L_{fr}
          + \frac{256}{3} \gamma_E^6
          \bigg\}
         \,,
\end{autobreak}
\end{align}

 \subsection{The SV resummed exponent $g^g_i$} \label{app:gN}
The resummation exponents $g^g_i$ in Eq. \ref{PsiSVN}, till three-loop, are given by   
\begin{align}
   
\begin{autobreak}
   g_1^g =

        \frac{1}{\beta_0} C_A   \bigg\{
           8
          + 8 \frac{L_\omega}{\omega}
          - 8 L_\omega
          \bigg\}\,,
\end{autobreak}\\
\begin{autobreak}
   g_2^g =

          \frac{\beta_1}{ \beta_0^3} C_A   \bigg\{
           4 \omega
          + 4 L_\omega
          + 2 L_\omega^2
          \bigg\}

       + \frac{1}{\beta_0^2} C_A n_f   \bigg\{
          \frac{40}{9} \omega
          + \frac{40}{9} L_\omega
          \bigg\}

       + \frac{1}{\beta_0^2} C_A^2   \bigg\{
          - \frac{268}{9} \omega
          + 8 \omega \zeta_2
          - \frac{268}{9} L_\omega
          + 8 L_\omega \zeta_2
          \bigg\}

       + \frac{1}{\beta_0} C_A   \bigg\{
            4 \omega L_{fr}
          + 4 L_\omega L_{qr}
          - 8 L_\omega \gamma_E
          \bigg\}\,,
\end{autobreak}\\
\begin{autobreak}
   g_3^g =
       \frac{1}{(1- \omega)} \bigg[  
         \frac{\beta_1^2}{\beta_0^4} C_A   \bigg\{
           2 \omega^2
          + 4 L_\omega \omega
          + 2 L_\omega^2
          \bigg\}

       + \frac{\beta_2}{\beta_0^3} C_A   \bigg\{
           4 \omega
          - 2 \omega^2
          \bigg\}

       + \frac{\beta_1}{\beta_0^3} C_A n_f   \bigg\{
           \frac{40}{9} \omega
          + \frac{20}{9} \omega^2
          + \frac{40}{9} L_\omega
          \bigg\}

       + \frac{\beta_1}{\beta_0^3} C_A^2   \bigg\{
          - \frac{268}{9} \omega
          + 8 \omega \zeta_2
          - \frac{134}{9} \omega^2
          + 4 \omega^2 \zeta_2
          - \frac{268}{9} L_\omega
          + 8 L_\omega \zeta_2
          \bigg\}

       + \frac{1}{\beta_0^2} C_A n_f^2   \bigg\{
          - \frac{8}{27}
          \bigg\}

       + \frac{1}{\beta_0^2} C_A C_F n_f   \bigg\{
          - \frac{55}{3}
          + 16 \zeta_3
          \bigg\}

       + \frac{1}{\beta_0^2} C_A^2 n_f   \bigg\{
          - \frac{418}{27}
          - \frac{56}{3} \zeta_3
          + \frac{80}{9} \zeta_2
          \bigg\}

       +  \frac{1}{\beta_0^2} C_A^3   \bigg\{
           \frac{245}{3}
          + \frac{44}{3} \zeta_3
          - \frac{536}{9} \zeta_2
          + \frac{88}{5} \zeta_2^2
          \bigg\}

       + \frac{\beta_1}{\beta_0^2} C_A   \bigg\{
           4 \omega L_{qr}
          - 8 \omega \gamma_E
          + 4 L_\omega L_{qr}
          - 8 L_\omega \gamma_E
          \bigg\}

       + \frac{1}{\beta_0} C_A n_f   \bigg\{
          - \frac{112}{27}
          + \frac{4}{3} \zeta_2
          + \frac{40}{9} L_{qr}
          - \frac{80}{9} \gamma_E
          \bigg\}

       + \frac{1}{\beta_0} C_A^2   \bigg\{
           \frac{808}{27}
          - 28 \zeta_3
          - \frac{22}{3} \zeta_2
          - \frac{268}{9} L_{qr}
          + 8 L_{qr} \zeta_2
          + \frac{536}{9} \gamma_E
          - 16 \gamma_E \zeta_2
          \bigg\}

       + C_A   \bigg\{
           2 \zeta_2
          + 2 L_{qr}^2
          - 8 \gamma_E L_{qr}
          + 8 \gamma_E^2
          \bigg\}  \bigg]
          
          + \frac{\beta_2}{\beta_0^3}  C_A   \bigg\{
          + 4 L_\omega
          \bigg\}

       + \frac{1}{\beta_0^2} C_A n_f^2   \bigg\{
           \frac{8}{27}
          + \frac{8}{27} \omega
          \bigg\}

       + \frac{1}{\beta_0^2} C_A C_F n_f   \bigg\{
           \frac{55}{3}
          - 16 \zeta_3
          + \frac{55}{3} \omega
          - 16 \omega \zeta_3
          \bigg\}

       + \frac{1}{\beta_0^2} C_A^2 n_f   \bigg\{
           \frac{418}{27}
          + \frac{56}{3} \zeta_3
          - \frac{80}{9} \zeta_2
          + \frac{418}{27} \omega
          + \frac{56}{3} \omega \zeta_3
          - \frac{80}{9} \omega \zeta_2
          \bigg\}

       + \frac{1}{\beta_0^2} C_A^3   \bigg\{
          - \frac{245}{3}
          - \frac{44}{3} \zeta_3
          + \frac{536}{9} \zeta_2
          - \frac{88}{5} \zeta_2^2
          - \frac{245}{3} \omega
          - \frac{44}{3} \omega \zeta_3
          + \frac{536}{9} \omega \zeta_2
          - \frac{88}{5} \omega \zeta_2^2
          \bigg\}

       + \frac{1}{\beta_0} C_A n_f   \bigg\{
           \frac{112}{27}
          - \frac{4}{3} \zeta_2
          - \frac{40}{9} L_{qr}
          + \frac{80}{9} \gamma_E
          - \frac{40}{9} \omega L_{fr}
          \bigg\}

       + \frac{1}{\beta_0} C_A^2   \bigg\{
          - \frac{808}{27}
          + 28 \zeta_3
          + \frac{22}{3} \zeta_2
          + \frac{268}{9} L_{qr}
          - 8 L_{qr} \zeta_2
          - \frac{536}{9} \gamma_E
          + 16 \gamma_E \zeta_2
          + \frac{268}{9} \omega L_{fr}
          - 8 \omega L_{fr} \zeta_2
          \bigg\}

       + C_A   \bigg\{
          - 2 \zeta_2
          - 2 L_{qr}^2
          + 8 \gamma_E L_{qr}
          - 8 \gamma_E^2
          - 2 \omega L_{fr}^2
          \bigg\}\,.
\end{autobreak}
\end{align}
Here, ${L}_{\omega}=\ln(1-\omega)$ and $\omega = 2 \beta_0 a_s(\mu_R^2) \ln N$. 
And $\beta_i$'s are the QCD $\beta$ functions which are given by
\begin{align}
  \beta_0&={11 \over 3 } C_A - {2 \over 3 } n_f \, ,
           \nonumber \\[0.5ex]
  \beta_1&={34 \over 3 } C_A^2- 2 n_f C_F -{10 \over 3} n_f C_A \, ,
           \nonumber \\[0.5ex]
  \beta_2&={2857 \over 54} C_A^3 
           -{1415 \over 54} C_A^2 n_f
           +{79 \over 54} C_A n_f^2
           +{11 \over 9} C_F n_f^2
           -{205 \over 18} C_F C_A n_f
           + C_F^2 n_f \,.
\end{align}

 \subsection{The NSV resummed exponents  $\overline{g}^g_i$} \label{app:gbN}
The resummation exponents  $\overline{g}^q_i$ in \ref{PsiNSVN}, till three-loop, are given by
\begin{align}
\begin{autobreak}

   \overline{g}^q_1 =

        \frac{1}{\beta_0} C_A   \bigg\{
           4 L_\omega
          \bigg\}\,,
\end{autobreak}    \\
\begin{autobreak}
    \overline{g}^q_2=

         \frac{1}{(1-\omega)}\bigg[ \frac{1}{\beta_0^2} C_A C_F n_f   \bigg\{
          - 8 \omega
          - 8 L_\omega
          \bigg\}

       + \frac{1}{\beta_0^2} C_A^2 n_f   \bigg\{
          - \frac{40}{3} \omega
          - \frac{40}{3} L_\omega
          \bigg\}

       + \frac{1}{\beta_0^2} C_A^3   \bigg\{
           \frac{136}{3} \omega
          + \frac{136}{3} L_\omega
          \bigg\}

       + \frac{1}{\beta_0} C_A n_f   \bigg\{
           \frac{40}{9} \omega
          \bigg\}

       + \frac{1}{\beta_0} C_A^2   \bigg\{
          - \frac{268}{9} \omega
          + 8 \omega \zeta_2
          \bigg\}

       + C_A   \bigg\{
          - 8
          + 4 L_{qr}
          - 4 L_{fr}
          + 4 L_{fr} \omega
          - 8 \gamma_E
          \bigg\} \bigg] \,,
\end{autobreak}    \\
\begin{autobreak}
    \overline{g}^q_3 =

         \frac{1}{(1-\omega)^2} \bigg[ \frac{1}{\beta_0^3} C_A C_F^2 n_f^2   \bigg\{
           8 \omega^2
          - 8 L_\omega^2
          \bigg\}

       + \frac{1}{\beta_0^3} C_A^2 C_F n_f^2   \bigg\{
           \frac{80}{3} \omega^2
          - \frac{80}{3} L_\omega^2
          \bigg\}

       + \frac{1}{\beta_0^3} C_A^3 n_f^2   \bigg\{
           \frac{200}{9} \omega^2
          - \frac{200}{9} L_\omega^2
          \bigg\}

       + \frac{1}{\beta_0^3} C_A^3 C_F n_f   \bigg\{
          - \frac{272}{3} \omega^2
          + \frac{272}{3} L_\omega^2
          \bigg\}

       + \frac{1}{\beta_0^3} C_A^4 n_f   \bigg\{
          - \frac{1360}{9} \omega^2
          + \frac{1360}{9} L_\omega^2
          \bigg\}

       + \frac{1}{\beta_0^3} C_A^5   \bigg\{
           \frac{2312}{9} \omega^2
          - \frac{2312}{9} L_\omega^2
          \bigg\}

       + \frac{1}{\beta_0^2} C_A C_F n_f^2   \bigg\{
           \frac{80}{9} \omega
          - \frac{62}{9} \omega^2
          + \frac{80}{9} L_\omega
          \bigg\}

       + \frac{1}{\beta_0^2} C_A C_F^2 n_f   \bigg\{
          - 2 \omega^2
          \bigg\}

       + \frac{1}{\beta_0^2} C_A^2 n_f^2   \bigg\{
           \frac{400}{27} \omega
          - \frac{31}{3} \omega^2
          + \frac{400}{27} L_\omega
          \bigg\}

       + \frac{1}{\beta_0^2} C_A^2 C_F n_f   \bigg\{
          - \frac{536}{9} \omega
          + 16 \omega \zeta_2
          + \frac{473}{9} \omega^2
          - 8 \omega^2 \zeta_2
          - \frac{536}{9} L_\omega
          + 16 L_\omega \zeta_2
          \bigg\}

       + \frac{1}{\beta_0^2} C_A^3 n_f   \bigg\{
          - \frac{4040}{27} \omega
          + \frac{80}{3} \omega \zeta_2
          + \frac{1145}{9} \omega^2
          - \frac{40}{3} \omega^2 \zeta_2
          - \frac{4040}{27} L_\omega
          + \frac{80}{3} L_\omega \zeta_2
          \bigg\}

       + \frac{1}{\beta_0^2} C_A^4   \bigg\{
           \frac{9112}{27} \omega
          - \frac{272}{3} \omega \zeta_2
          - \frac{2471}{9} \omega^2
          + \frac{136}{3} \omega^2 \zeta_2
          + \frac{9112}{27} L_\omega
          - \frac{272}{3} L_\omega \zeta_2
          \bigg\}

       + \frac{1}{\beta_0} C_A n_f^2   \bigg\{
           \frac{16}{27} \omega
          - \frac{8}{27} \omega^2
          \bigg\}

       + \frac{1}{\beta_0} C_A C_F n_f   \bigg\{
           \frac{110}{3} \omega
          - 32 \omega \zeta_3
          - \frac{55}{3} \omega^2
          + 16 \omega^2 \zeta_3
          - 16 L_\omega
          + 8 L_\omega L_{qr}
          - 16 L_\omega \gamma_E
          \bigg\}

       + \frac{1}{\beta_0} C_A^2 n_f   \bigg\{
           \frac{836}{27} \omega
          + \frac{112}{3} \omega \zeta_3
          - \frac{160}{9} \omega \zeta_2
          - \frac{418}{27} \omega^2
          - \frac{56}{3} \omega^2 \zeta_3
          + \frac{80}{9} \omega^2 \zeta_2
          - \frac{80}{3} L_\omega
          + \frac{40}{3} L_\omega L_{qr}
          - \frac{80}{3} L_\omega \gamma_E
          \bigg\}

       + \frac{1}{\beta_0} C_A^3   \bigg\{
          - \frac{490}{3} \omega
          - \frac{88}{3} \omega \zeta_3
          + \frac{1072}{9} \omega \zeta_2
          - \frac{176}{5} \omega \zeta_2^2
          + \frac{245}{3} \omega^2
          + \frac{44}{3} \omega^2 \zeta_3
          - \frac{536}{9} \omega^2 \zeta_2
          + \frac{88}{5} \omega^2 \zeta_2^2
          + \frac{272}{3} L_\omega
          - \frac{136}{3} L_\omega L_{qr}
          + \frac{272}{3} L_\omega \gamma_E
          \bigg\}

       + C_A n_f   \bigg\{
           \frac{352}{27}
          - \frac{88}{9} L_{qr}
          + \frac{4}{3} L_{qr}^2
          + \frac{40}{9} L_{fr}
          - \frac{80}{9} L_{fr} \omega
          + \frac{40}{9} L_{fr} \omega^2
          - \frac{4}{3} L_{fr}^2
          + \frac{8}{3} L_{fr}^2 \omega
          - \frac{4}{3} L_{fr}^2 \omega^2
          + \frac{176}{9} \gamma_E
          - \frac{16}{3} \gamma_E L_{qr}
          + \frac{16}{3} \gamma_E^2
          \bigg\}

       + C_A^2   \bigg\{
          - \frac{2416}{27}
          + 28 \zeta_3
          + 16 \zeta_2
          + \frac{532}{9} L_{qr}
          - 8 L_{qr} \zeta_2
          - \frac{22}{3} L_{qr}^2
          - \frac{268}{9} L_{fr}
          + 8 L_{fr} \zeta_2
          + \frac{536}{9} L_{fr} \omega
          - 16 L_{fr} \omega \zeta_2
          - \frac{268}{9} L_{fr} \omega^2
          + 8 L_{fr} \omega^2 \zeta_2
          + \frac{22}{3} L_{fr}^2
          - \frac{44}{3} L_{fr}^2 \omega
          + \frac{22}{3} L_{fr}^2 \omega^2
          - \frac{1064}{9} \gamma_E
          + 16 \gamma_E \zeta_2
          + \frac{88}{3} \gamma_E L_{qr}
          - \frac{88}{3} \gamma_E^2
          \bigg\} \bigg] \,.
\end{autobreak} 
\end{align}

 \subsection{The NSV resummed exponents $h^g_{ij}$} \label{app:hN}
The resummation constants $h^g_{ij}$ in Eq. \ref{hNSV}, till three-loop, are given by
\begin{align}
\begin{autobreak}
     h^g_{00}  =

       \frac{1}{\beta_0} C_A   \bigg\{
          - 8 L_\omega
          \bigg\}\,,
\end{autobreak}  \\
\begin{autobreak}
     h^g_{01}  = 
     0\,,
\end{autobreak}  \\
\begin{autobreak}
     h^g_{10}  =

         \frac{1}{(1-\omega)} \bigg[ \frac{\beta_1}{\beta_0^2} C_A   \bigg\{
          - 8 \omega
          - 8 L_\omega
          \bigg\}

       +\frac{1}{\beta_0} C_A n_f   \bigg\{
          - \frac{80}{9} \omega
          \bigg\}

       +\frac{1}{\beta_0} C_A^2   \bigg\{
          + \frac{536}{9} \omega
          - 16 \omega \zeta_2
          + 32 \gamma_E \omega
          \bigg\}

       + C_A   \bigg\{
          + 8
          - 8 L_{qr}
          + 8 L_{fr}
          - 8 L_{fr} \omega
          + 16 \gamma_E
          \bigg\} \bigg] \,,
\end{autobreak}  \\
\begin{autobreak}
     h^g_{11}  =

        \frac{1}{\beta_0} C_A^2   \bigg\{
           32 \omega
          \bigg\}

       + \frac{1}{(1-\omega)^2} \bigg[ \frac{1}{\beta_0} C_A^2   \bigg\{
          -  4 \omega
          \bigg\} \bigg] \,,
\end{autobreak}  \\
\begin{autobreak}
     h^g_{20}  =

         \frac{1}{(1-\omega)^2} \bigg[ \frac{\beta_1^2}{\beta_0^3} C_A   \bigg\{
          - 4 \omega^2
          + 4 L_\omega^2
          \bigg\}

       + \frac{\beta_2}{\beta_0^2} C_A   \bigg\{
          + 4 \omega^2
          \bigg\}

       + \frac{\beta_1}{\beta_0^2} C_A n_f   \bigg\{
          + \frac{80}{9} \omega
          - \frac{40}{9} \omega^2
          + \frac{80}{9} L_\omega
          \bigg\}

       + \frac{\beta_1}{\beta_0^2} C_A^2   \bigg\{
          - \frac{536}{9} \omega
          + 16 \omega \zeta_2
          + \frac{268}{9} \omega^2
          - 8 \omega^2 \zeta_2
          - 32 \gamma_E \omega
          + 16 \gamma_E \omega^2
          - \frac{536}{9} L_\omega
          + 16 L_\omega \zeta_2
          - 32 L_\omega \gamma_E
          \bigg\}

       +\frac{1}{\beta_0} C_A n_f^2   \bigg\{
          - \frac{32}{27} \omega
          + \frac{16}{27} \omega^2
          \bigg\}

       +\frac{1}{\beta_0} C_A C_F n_f   \bigg\{
          - \frac{172}{3} \omega
          + 64 \omega \zeta_3
          + \frac{86}{3} \omega^2
          - 32 \omega^2 \zeta_3
          \bigg\}

       +\frac{1}{\beta_0} C_A^2 n_f   \bigg\{
          - \frac{1096}{27} \omega
          - \frac{224}{3} \omega \zeta_3
          + \frac{320}{9} \omega \zeta_2
          + \frac{548}{27} \omega^2
          + \frac{112}{3} \omega^2 \zeta_3
          - \frac{160}{9} \omega^2 \zeta_2
          - \frac{640}{9} \gamma_E \omega
          + \frac{320}{9} \gamma_E \omega^2
          \bigg\}

       +\frac{1}{\beta_0} C_A^3   \bigg\{
          + \frac{724}{3} \omega
          - \frac{112}{3} \omega \zeta_3
          - \frac{2144}{9} \omega \zeta_2
          + \frac{352}{5} \omega \zeta_2^2
          - \frac{362}{3} \omega^2
          + \frac{56}{3} \omega^2 \zeta_3
          + \frac{1072}{9} \omega^2 \zeta_2
          - \frac{176}{5} \omega^2 \zeta_2^2
          + \frac{4288}{9} \gamma_E \omega
          - 128 \gamma_E \omega \zeta_2
          - \frac{2144}{9} \gamma_E \omega^2
          + 64 \gamma_E \omega^2 \zeta_2
          \bigg\}

       + \frac{\beta_1}{\beta_0} C_A   \bigg\{
          + 8 \omega
          - 4 \omega^2
          - 8 L_\omega
          + 8 L_\omega L_{qr}
          - 16 L_\omega \gamma_E
          \bigg\}

       + C_A n_f   \bigg\{
          - \frac{392}{27}
          + \frac{32}{3} \zeta_2
          + \frac{80}{9} L_{qr}
          - \frac{80}{9} L_{fr}
          + \frac{160}{9} L_{fr} \omega
          - \frac{80}{9} L_{fr} \omega^2
          - \frac{148}{9} \gamma_E
          \bigg\}

       + C_A^2   \bigg\{
          + \frac{2612}{27}
          - 56 \zeta_3
          - \frac{248}{3} \zeta_2
          - \frac{536}{9} L_{qr}
          + 16 L_{qr} \zeta_2
          + \frac{536}{9} L_{fr}
          - 16 L_{fr} \zeta_2
          - \frac{1072}{9} L_{fr} \omega
          + 32 L_{fr} \omega \zeta_2
          + \frac{536}{9} L_{fr} \omega^2
          - 16 L_{fr} \omega^2 \zeta_2
          + \frac{1060}{9} \gamma_E
          - 32 \gamma_E \zeta_2
          - 32 \gamma_E L_{qr}
          + 32 \gamma_E L_{fr}
          - 64 \gamma_E L_{fr} \omega
          + 32 \gamma_E L_{fr} \omega^2
          + 56 \gamma_E^2
          \bigg\}

       + \beta_0 C_A   \bigg\{
          + 16 \zeta_2
          - 8 L_{qr}
          + 4 L_{qr}^2
          - 4 L_{fr}^2
          + 8 L_{fr}^2 \omega
          - 4 L_{fr}^2 \omega^2
          + 16 \gamma_E
          - 16 \gamma_E L_{qr}
          + 16 \gamma_E^2
          \bigg\} \bigg] \,,
\end{autobreak}  \\
\begin{autobreak}
     h^g_{21}  =

         \frac{1}{(1-\omega)^2} \bigg[ \frac{\beta_1}{\beta_0^2} C_A^2   \bigg\{
          - 32 \omega
          + 16 \omega^2
          - 32 L_\omega
          \bigg\}

       +\frac{1}{\beta_0} C_A^2 n_f   \bigg\{
          - \frac{640}{9} \omega
          + \frac{320}{9} \omega^2
          \bigg\}

       +\frac{1}{\beta_0} C_A^3   \bigg\{
          + \frac{4288}{9} \omega
          - 128 \omega \zeta_2
          - \frac{2144}{9} \omega^2
          + 64 \omega^2 \zeta_2
          \bigg\}

       + C_A n_f   \bigg\{
          + \frac{4}{3}
          \bigg\}

       + C_A^2   \bigg\{
          - \frac{4}{3}
          - 32 L_{qr}
          + 32 L_{fr}
          - 64 L_{fr} \omega
          + 32 L_{fr} \omega^2
          + 48 \gamma_E
          \bigg\} \bigg] \,,
\end{autobreak}  \\
\begin{autobreak}
     h^g_{22}  =

        \frac{1}{(1-\omega)^3} \bigg[  \frac{1}{\beta_0} C_A^2 n_f   \bigg\{
          - \frac{32}{27} \omega
          \bigg\}

       +\frac{1}{\beta_0} C_A^3   \bigg\{
          + \frac{176}{27} \omega
          \bigg\} \bigg] \,.
\end{autobreak} 
\end{align}
 \section{Resummation Coefficients for the $\overline{N}$ exponentiation} \label{app:NBexp}
% %%%%%%%%%%%%%%%%%%%%%%%%%%%%%%%%%%%%%%%%%%%%%%%%%%%%%%%%%%%%%%%%%%%%%%%%%%%%%
 For the case of  $\overline{N}$ exponentiation, all the ${N}$-dependent resummed exponents namely $g^g_i(\overline{\omega})$, $\overline{g}^g_i(\overline{\omega})$, and $h^g_{i}(\overline{\omega})$ given in 
 Eqs \ref{eq:gnb} and \ref{hnb}, respectively can be obtained from the corresponding exponents in standard $N$-approach through setting all the $\gamma_E$ terms to zero and replacing all the $\ln N$ terms by $\ln \overline{N}$ as well as all the $\mathcal{O}(1)$ $\omega$ terms by $\overline{\omega}$ as mentioned in 
 Sec. \ref{resNSV}. The $N$-independent constants $\bar{\tilde {g}}^g_0$ given in Eq. (\ref{g0nb}) can be obtained from their counterparts in standard $N$-approach by simply putting the $\gamma_E$ terms equal to zero.

 % %-----------------------------------
 \section{Resummation Coefficients for the {\textbf {\textit {Soft exponentiation}}}} \label{app:softexp}
% %%%%%%%%%%%%%%%%%%%%%%%%%%%%%%%%%%%%%%%%%%%%%%%%%%%%%%%%%%%%%%%%%%%%%%%%%%%%%
 For the case of {\it soft  exponentiation}, all the terms coming from the soft-collinear function $\rm \Phi_g$ are exponentiated and hence this means all the contribution to the finite ($N$-independent) piece from the soft-collinear function is also being exponentiated. The resummation coefficients for the {\it soft  exponentiation} denoted by $\Tilde{g}_{0i}^{g, \rm Soft}$ and $g_i^{g,\rm Soft}$ in Eqs (\ref{eq:g0bsoft}) and (\ref{eq:gnbsoft}), respectively can be obtained from thier counterparts in standard $\Nb$ scheme as described below.

The $N$-independent constants $\Tilde{g}_{0i}^{g, \rm Soft}$ in Eq. (\ref{eq:g0bsoft}) can be put in the following form:
\begin{align}
\Tilde{g}_{01}^{g,\rm Soft} &= \bar{\Tilde{g}}_{01}^g +   \Delta^{g,\rm Soft}_{\Tilde{g}_{01}} \,,  \nn\\
\Tilde{g}_{02}^{g,\rm Soft} &= \bar{\Tilde{g}}_{02}^g +   \Delta^{g,\rm Soft}_{\Tilde{g}_{02}} \,,  \nn\\
\Tilde{g}_{03}^{g,\rm Soft} &= \bar{\Tilde{g}}_{03}^g +  \Delta^{g,\rm Soft}_{\Tilde{g}_{03}} \,,  
\end{align}
where the coefficients $ \Delta^{g,\rm Soft}_{\Tilde{g}_{0i}}$ are given by,
\begin{align}
    
\begin{autobreak}
 \Delta^{g,\rm Soft}_{\Tilde{g}_{01}} =
           n_f   \bigg\{
           \frac{4}{3} L_{fr}
          \bigg\}

       + C_A   \bigg\{
           14 \zeta_2
          - 2 L_{qr}^2
          - \frac{22}{3} L_{fr}
          \bigg\}
         \,,
   
\end{autobreak}\\
\begin{autobreak}
   \Delta^{g,\rm Soft}_{\Tilde{g}_{02}} =
           n_f^2   \bigg\{
           \frac{4}{3} L_{fr}^2
          \bigg\}

       + C_F n_f   \bigg\{
          - \frac{67}{3}
          + 16 \zeta_3
          + 4 L_{qr}
          + 4 L_{fr}
          \bigg\}

       + C_A n_f   \bigg\{
          - \frac{1832}{81}
          - \frac{92}{9} \zeta_3
          - \frac{50}{3} \zeta_2
          + \frac{104}{27} L_{qr}
          + \frac{32}{3} L_{qr} \zeta_2
          + \frac{20}{9} L_{qr}^2
          - \frac{4}{9} L_{qr}^3
          + \frac{16}{3} L_{fr}
          + \frac{56}{3} L_{fr} \zeta_2
          - \frac{8}{3} L_{fr} L_{qr}^2
          - \frac{44}{3} L_{fr}^2
          \bigg\}

       + C_A^2   \bigg\{
           \frac{5105}{81}
          - \frac{286}{9} \zeta_3
          + \frac{335}{3} \zeta_2
          + 74 \zeta_2^2
          + \frac{160}{27} L_{qr}
          - 4 L_{qr} \zeta_3
          - \frac{176}{3} L_{qr} \zeta_2
          - \frac{134}{9} L_{qr}^2
          - 24 L_{qr}^2 \zeta_2
          + \frac{22}{9} L_{qr}^3
          + 2 L_{qr}^4
          - \frac{64}{3} L_{fr}
          - 24 L_{fr} \zeta_3
          - \frac{308}{3} L_{fr} \zeta_2
          + \frac{44}{3} L_{fr} L_{qr}^2
          + \frac{121}{3} L_{fr}^2
          \bigg\}
         \,,
       
\end{autobreak}\\
\begin{autobreak}
    \Delta^{g,\rm Soft}_{\Tilde{g}_{03}}=
             n_f^3   \bigg\{
           \frac{32}{27} L_{fr}^3
          \bigg\}

       + C_F n_f^2   \bigg\{
           \frac{8962}{81}
          - \frac{224}{3} \zeta_3
          - \frac{184}{9} \zeta_2
          - \frac{32}{45} \zeta_2^2
          - \frac{104}{3} L_{qr}
          + \frac{64}{3} L_{qr} \zeta_3
          + \frac{8}{3} L_{qr}^2
          - \frac{290}{9} L_{fr}
          + \frac{64}{3} L_{fr} \zeta_3
          + \frac{16}{3} L_{fr} L_{qr}
          + \frac{28}{3} L_{fr}^2
          \bigg\}

       + C_F^2 n_f   \bigg\{
           \frac{608}{9}
          - 320 \zeta_5
          + \frac{592}{3} \zeta_3
          - 4 L_{qr}
          - 2 L_{fr}
          \bigg\}

       + C_A n_f^2   \bigg\{
           \frac{611401}{6561}
          + \frac{9152}{243} \zeta_3
          - \frac{440}{27} \zeta_2
          - \frac{424}{27} \zeta_2^2
          - \frac{28114}{729} L_{qr}
          - \frac{256}{27} L_{qr} \zeta_3
          - \frac{640}{27} L_{qr} \zeta_2
          + \frac{232}{81} L_{qr}^2
          + \frac{64}{9} L_{qr}^2 \zeta_2
          + \frac{80}{81} L_{qr}^3
          - \frac{4}{27} L_{qr}^4
          - \frac{8111}{243} L_{fr}
          - \frac{368}{27} L_{fr} \zeta_3
          - \frac{200}{9} L_{fr} \zeta_2
          + \frac{416}{81} L_{fr} L_{qr}
          + \frac{128}{9} L_{fr} L_{qr} \zeta_2
          + \frac{80}{27} L_{fr} L_{qr}^2
          - \frac{16}{27} L_{fr} L_{qr}^3
          + \frac{116}{9} L_{fr}^2
          + \frac{56}{3} L_{fr}^2 \zeta_2
          - \frac{8}{3} L_{fr}^2 L_{qr}^2
          - \frac{176}{9} L_{fr}^3
          \bigg\}

       + C_A C_F n_f   \bigg\{
          - \frac{341219}{486}
          + \frac{1216}{9} \zeta_5
          + \frac{29128}{81} \zeta_3
          - \frac{2947}{9} \zeta_2
          + \frac{1040}{3} \zeta_2 \zeta_3
          - \frac{128}{45} \zeta_2^2
          + \frac{3833}{27} L_{qr}
          - \frac{752}{9} L_{qr} \zeta_3
          + 88 L_{qr} \zeta_2
          + \frac{32}{5} L_{qr} \zeta_2^2
          + \frac{145}{3} L_{qr}^2
          - 48 L_{qr}^2 \zeta_3
          - \frac{28}{3} L_{qr}^3
          + \frac{1715}{9} L_{fr}
          - \frac{352}{3} L_{fr} \zeta_3
          + 56 L_{fr} \zeta_2
          - \frac{88}{3} L_{fr} L_{qr}
          - 8 L_{fr} L_{qr}^2
          - \frac{154}{3} L_{fr}^2
          \bigg\}

       + C_A^2 n_f   \bigg\{
          - \frac{7530014}{6561}
          + \frac{856}{9} \zeta_5
          - \frac{920}{81} \zeta_3
          - \frac{265942}{729} \zeta_2
          - \frac{2224}{9} \zeta_2 \zeta_3
          + \frac{116}{45} \zeta_2^2
          + \frac{266072}{729} L_{qr}
          + \frac{56}{9} L_{qr} \zeta_3
          + \frac{32504}{81} L_{qr} \zeta_2
          + \frac{1432}{15} L_{qr} \zeta_2^2
          + \frac{4094}{81} L_{qr}^2
          + \frac{328}{9} L_{qr}^2 \zeta_3
          - \frac{68}{3} L_{qr}^2 \zeta_2
          - \frac{1780}{81} L_{qr}^3
          - \frac{232}{9} L_{qr}^3 \zeta_2
          - \frac{76}{27} L_{qr}^4
          + \frac{8}{9} L_{qr}^5
          + \frac{67015}{243} L_{fr}
          + \frac{2320}{27} L_{fr} \zeta_3
          + \frac{3160}{9} L_{fr} \zeta_2
          + \frac{304}{3} L_{fr} \zeta_2^2
          - \frac{1648}{81} L_{fr} L_{qr}
          - \frac{16}{3} L_{fr} L_{qr} \zeta_3
          - \frac{1408}{9} L_{fr} L_{qr} \zeta_2
          - \frac{1264}{27} L_{fr} L_{qr}^2
          - 32 L_{fr} L_{qr}^2 \zeta_2
          + \frac{176}{27} L_{fr} L_{qr}^3
          + \frac{8}{3} L_{fr} L_{qr}^4
          - \frac{1090}{9} L_{fr}^2
          - 48 L_{fr}^2 \zeta_3
          - \frac{616}{3} L_{fr}^2 \zeta_2
          + \frac{88}{3} L_{fr}^2 L_{qr}^2
          + \frac{968}{9} L_{fr}^3
          \bigg\}

       + C_A^3   \bigg\{
           \frac{29639273}{13122}
          + \frac{4444}{9} \zeta_5
          - \frac{305432}{243} \zeta_3
          - \frac{208}{9} \zeta_3^2
          + \frac{1260769}{729} \zeta_2
          - \frac{4928}{9} \zeta_2 \zeta_3
          + \frac{92516}{135} \zeta_2^2
          + \frac{189748}{945} \zeta_2^3
          - \frac{373975}{729} L_{qr}
          + 32 L_{qr} \zeta_5
          + \frac{4096}{27} L_{qr} \zeta_3
          - \frac{84680}{81} L_{qr} \zeta_2
          - \frac{88}{3} L_{qr} \zeta_2 \zeta_3
          - \frac{8404}{15} L_{qr} \zeta_2^2
          - \frac{2065}{9} L_{qr}^2
          + \frac{572}{9} L_{qr}^2 \zeta_3
          - \frac{1414}{9} L_{qr}^2 \zeta_2
          - \frac{548}{5} L_{qr}^2 \zeta_2^2
          + \frac{2600}{81} L_{qr}^3
          + 8 L_{qr}^3 \zeta_3
          + \frac{1276}{9} L_{qr}^3 \zeta_2
          + \frac{683}{27} L_{qr}^4
          + 20 L_{qr}^4 \zeta_2
          - \frac{44}{9} L_{qr}^5
          - \frac{4}{3} L_{qr}^6
          - \frac{131507}{243} L_{fr}
          + 160 L_{fr} \zeta_5
          - \frac{3356}{27} L_{fr} \zeta_3
          - \frac{10106}{9} L_{fr} \zeta_2
          - 304 L_{fr} \zeta_2 \zeta_3
          - \frac{1672}{3} L_{fr} \zeta_2^2
          - \frac{3520}{81} L_{fr} L_{qr}
          + \frac{88}{3} L_{fr} L_{qr} \zeta_3
          + \frac{3872}{9} L_{fr} L_{qr} \zeta_2
          + \frac{4100}{27} L_{fr} L_{qr}^2
          + 48 L_{fr} L_{qr}^2 \zeta_3
          + 176 L_{fr} L_{qr}^2 \zeta_2
          - \frac{484}{27} L_{fr} L_{qr}^3
          - \frac{44}{3} L_{fr} L_{qr}^4
          + \frac{2486}{9} L_{fr}^2
          + 264 L_{fr}^2 \zeta_3
          + \frac{1694}{3} L_{fr}^2 \zeta_2
          - \frac{242}{3} L_{fr}^2 L_{qr}^2
          - \frac{5324}{27} L_{fr}^3
          \bigg\}
         \,.
          \end{autobreak}
          \end{align}

% %As a result, the $\Tilde{g}_0^g$ constants and the resummed exponents $g_i^g$ in the Standard $N$-approach change. We write the changes in $g_i^{g, \rm Soft}$ below in terms of the  exponents in Standard $N$, 
% %
 The $N$-dependent coefficients $g_i^{g,\rm Soft}$ in Eq. (\ref{eq:gnbsoft}) can be obtained as follows:

\begin{align}
g_1^{g,\rm Soft} &= \bar{g}_1^g \,,  \nn\\
g_2^{g,\rm Soft} &= \bar{g}_2^g + \as ~\Delta^{g,\rm Soft}_{g_2} \,,  \nn\\
g_3^{g,\rm Soft} &= \bar{g}_3^g + \as~ \Delta^{g,\rm Soft}_{g_3} \,,  \nn\\
%g_4^{g,\rm Soft} &= g_4^g + \as ~\Delta^{g,\rm Soft}_{g_4} \,,  \nn\\
\end{align}

where the coefficients $\Delta^{g,\rm Soft}_{g_i}$ are given as, 
 \begin{align}
    
\begin{autobreak}
   \Delta^{\rm Soft}_{g_2} =

         C_A   \bigg\{
           2 \zeta_2
          + 2 L_{qr}^2
          \bigg\}\,,
\end{autobreak}\\
\begin{autobreak}
   \Delta^{\rm Soft}_{g_3} =

            C_A   \bigg\{
           \frac{46}{3} \beta_0 \zeta_3
          - 2 L_{qr} \zeta_2 \beta_0
          - \frac{2}{3} L_{qr}^3 \beta_0
          \bigg\}

       + C_A n_f   \bigg\{
          - \frac{328}{81}
          + \frac{32}{3} \zeta_3
          - \frac{10}{9} \zeta_2
          + \frac{112}{27} L_{qr}
          - \frac{4}{3} L_{qr} \zeta_2
          - \frac{20}{9} L_{qr}^2
          \bigg\}

       + C_A^2   \bigg\{
           \frac{2428}{81}
          - \frac{176}{3} \zeta_3
          + \frac{67}{9} \zeta_2
          - 12 \zeta_2^2
          - \frac{808}{27} L_{qr}
          + 28 L_{qr} \zeta_3
          + \frac{22}{3} L_{qr} \zeta_2
          + \frac{134}{9} L_{qr}^2
          - 4 L_{qr}^2 \zeta_2 \bigg\}\,.
\end{autobreak}
\begin{autobreak}
\end{autobreak}
\end{align}

% %--------------------------------------------------------------------

% %%%%%%%%%%%%%%%%%%%%%%%%%%%%%%%%%%%%%%%%%%%%%%%%%%%%%%%%%%%%%%%%%%%%%%%%%%%%%
 \section{Resummation Coefficients for the {\textbf {\textit {All exponentiation}}}} \label{app:Allexp}
% %%%%%%%%%%%%%%%%%%%%%%%%%%%%%%%%%%%%%%%%%%%%%%%%%%%%%%%%%%%%%%%%%%%%%%%%%%%%%
 For the case of {\it All exponentiation}, the complete $\Tilde{g}_0^g$ is being exponentiated along with the large-$\Nb$ pieces. This brings into modification only for the resummed exponent compared to the `Standard ${\Nb}$ exponentiation'. We write the modified resummed exponents denoted by $g_i^{g, \rm All}$ in Eq. (\ref{gnall}) in terms of exponents in standard ${\Nb}$  as,
% %
\begin{align}
g_1^{g,\rm All} &= \bar{g}_1^g\,,  \nn\\
g_2^{g,\rm All} &= \bar{g}_2^g +  \as~  \Delta^{g,\rm All}_{g_2} \,,  \nn\\
g_3^{g,\rm All} &= \bar{g}_3^g +  \as~  \Delta^{g,\rm All}_{g_3} \,,  \nn\\
% %g_4^{g,\rm All} &= g_4^g +  \as ~ \Delta^{g,\rm All}_{g_4} \,,
\end{align}
% %
 where $ \Delta^{g,\rm All}_{g_i}$ terms are found from exponentiating the complete $\bar{\Tilde{g}}_0^g$ and they are given as,
\begin{align}
  \Delta^{g,\rm All}_{g_2} &= \bar{\Tilde{g}}_{01}^{g} \,, \nn \\
   \Delta^{g,\rm All}_{g_3} &=  \bigg(-\frac{(\bar{\Tilde{g}}_{01}^{g})^{2}}{2} +  \bar{\Tilde{g}}_{02}^{g} \bigg) \,, %\nn\\
%   \Delta^{g,\rm All}_{g_4} &= \bigg(\frac{(\Tilde{g}_{01}^{g})^{3}}{3} - \Tilde{g}_{01}^{g} \Tilde{g}_{02}^{g} +  \Tilde{g
%   }_{03}^{g} \bigg) \,,
\end{align}
where the coefficients $\bar{\Tilde{g}}_{0i}^g$ can be obtained as described in \ref{app:NBexp}.

%%%References
\bibliographystyle{JHEP}
\bibliography{main}

\end{document}